\documentclass[12pt,a4paper]{report}
\usepackage[dvips]{graphicx}
\usepackage{feynmf}
\usepackage{amsmath,amssymb}
\usepackage{epsfig}
\usepackage{subfigure}
\usepackage{amsmath}
\usepackage{amsfonts}
\usepackage{amssymb}
\usepackage{url}
\usepackage[hypertex]{hyperref}

\newcommand{\be}{\begin{equation}}
\newcommand{\ee}{\end{equation}}
\newcommand{\bea}{\begin{eqnarray}}
\newcommand{\eea}{\end{eqnarray}}
\newcommand{\bref}[1]{(\ref{#1})}
\newcommand{\BR}{{\rm BR}}
\newcommand{\la}{\langle}
\newcommand{\ra}{\rangle}
\newcommand{\bi}{\bibitem}
\newcommand{\rep}[1]{{\bf #1}}
\newcommand{\pa}{\partial}
\begin{document}
\begin{titlepage}
\begin{flushright}
\today
\end{flushright}
\vspace{4\baselineskip}
\begin{center}
{\Large\bf 
 SO(10) GUT in Four and Five Dimensions: A Review        
}
\end{center}
\vspace{1cm}
\begin{center}
{\large Takeshi Fukuyama 
\footnote{\tt E-mail:fukuyama@se.ritsumei.ac.jp}}
\end{center}
\vspace{0.2cm}
\begin{center}
{\small \it Department of Physics and R-GIRO, Ritsumeikan University,
Kusatsu, Shiga,525-8577, Japan} \\
\end{center}
\vskip 10mm
\begin{abstract}
We review SO(10) grand unified theories (GUTs) in four and five dimensions (4D and 5D). The renormalizable minimal SO(10) SUSY GUT is the central theme of this review. It is very predictive and makes it possible to construct all mass matrices including those of the Dirac and heavy right-handed Majorana neutrinos.
So it is not only able to reproduce all the low energy data, except for too larger $\theta_{13}$ in the lepton mixing angles (which can be evaded without spoiling the basic ingredients), but also predicts almost all new physics beyond the standard model (SM) like neutrinoless double beta decay, the electric and magnetic dipole moments of the quark-lepton, lepton flavour violation, leptogenesis etc. 
To be very predictive, on the other hand, implies that predictions are unambiguous and they are always exposed to severe compatibilities with observations  as well as to a conceptual consistency check. The explicit construction of the Higgs superpotential and the explicit display of a symmetry breaking pattern from GUT to the SM show that the naive desert from the SM to GUT in the minimal supersymmetric standard model (MSSM) is a too simplified concept and we have many definitely determined intermediate energy scales in general.
This situation destroys the naive gauge coupling unification in the MSSM scheme.
Also the precise measurements of neutrino oscillation data have revealed several small but manifest mismatches with our predictions. 
Also there are arguments that it is impossible to construct a GUT theory in 4D with a finite number of multiplets that leads to the MSSM with a residual R symmetry. Of course, there are several loopholes which solve these problems in 4D. However, if we try to solve all these pathologies comprehensively, it is very attractive for us to go into extra dimensions. Extra dimension may be either warped or flat. 
The fifth dimension, for simplicity, is compactified on the $S^1/(Z_2)$ (warped) or on the $S^1/(Z_2\times Z_2')$ (flat) orbifold with two inequivalent branes at the orbifold fixed points.  In the former warped case, intermediate energy scales are translated with the positions of Higgs fields in the bulk and the fundamental scheme of the MSSM is recovered. On the other hand, in the latter flat scenario, all matter and Higgs multiplets reside on the Pati-Salam (PS) brane where the PS symmetry is manifest. There the original renormalizablity in Yukawa coupling is broken but its essential structures of mass matrices in minimal SO(10) GUT in 4D is promoted to the PS invariant action in 4D. In the gaugino mediation mechanism, the SO(10) gauge multiplet is transmitted to the PS brane through the gaugino mediation with bulk gauge multiplet. 
Further breaking of the PS gauge group to the SM group is realized by VEVs of the Higgs multiplets 
 $({\bf 4},{\bf 1},{\bf 2}) \oplus (\overline{{\bf 4}},{\bf 1},{\bf 2})$. 
We show that this model not only cures all pathologies in SO(10) GUT in 4D but also provides the consistent inflation scenario and dark matter candidate and leptogenesis.
The gauge coupling unification is successfully realized 
 at $M_{\rm GUT} =4.6 \times 10^{17}$ GeV after incorporating 
 the threshold corrections of the Kaluza-Klein modes, 
 with the compactification scale (assumed to be the same as 
 the PS symmetry breaking scale) 
 $M_c = v_{\rm PS}= 1.2 \times 10^{16}$ GeV. This orbifold GUT model can naturally leads us to  the smooth hybrid inflation, which turns out to be consistent with the WMAP 5-years and 7-years data with the predicted $M_{\rm GUT}$ and $v_{PS}$. However, this simple model suffers from the stau LSP (the lightest SUSY particle) problem. Neutralino LSP can be realized when the compactification scale of the fifth dimension is higher than the PS symmetry breaking scale, keeping the gauge coupling unification. We reanalyze all new physics beyond the SM in 5D,  the leptogenesis, LFV etc. So our predictions range over all particle physics. The recent discoveries of a Higgs-like object and null search of LHC give very clear and important inmpacts on the above mentioned scenario and we discuss to where they drive our theory.
Finally we add some comments on the impacts of the discovery of a Higgs-like particle by the Large Hadronic Collider at CERN (LHC).
\end{abstract}
\end{titlepage}
\chapter{Introduction}
\setcounter{page}{3}
SUSY GUT is the most promising model beyond the SM \cite{GUT}.  
The SM is a very powerful theory but it has validity limits like the other great theories.
There are discrepancies with observations like neutrino mass as well as the other non affirmative ones like muon g-2 \cite{Bennett},an anomalous like-sign dimuon charge asymmetry \cite{Abazov}, and $h\rightarrow 2\gamma$. The CKM phase only is insufficient for baryogenesis \cite{baryogenesis}. The SM does not predict dark matter (DM) candidates. 
These facts strongly suggest a more comprehensive theory which improve the deficits of the SM.\\
Above the above mentioned observational deficiencies, the SM has conceptual problems which do not allow it to be complete.
Indeed, the SM+$\nu$ has too many parameters, 19 + 9. It does not explain quark-lepton mass hierarchy, mixing angles, CP phases, Higgs mass stability against quantum corrections, three different gauge couplings and their unification. They are all independent given or unknown objects and have no mutual relationships in the SM framework.\\
There are the motivations for us to consider the theory beyond the SM from the bottom up approach.
The top-down approach from string theory is also necessary and complimentary to the bottom-up approaches \cite{JEKim} \cite{Raby2}.

Even if we restrict ourselves in the Higgs hierarchy problem, there are many approaches like little Higgs \cite{little}, composite particle models of technicolor \cite{technicolor} etc.  However, it seems to be only SUSY GUT which
may solve all these problems comprehensively.  \\
MSSM explains hierarchy problems and gauge coupling unification. However, it gives rather few definite predictions beyond the SM since it gives no information on mass matrices of quarks-leptons and on the physical structures at the unification scale. This drives us to SUSY GUT.
Even if we accept a scheme of SUSY GUT, there are so many options.
What is the gauge group, SU(5), SO(10), $E_6$, $E_7$, $E_8$ or their products ?
So we need the additional criteria to select the gauge symmetry.
An anomaly free condition may be a good candidate for it since chiral symmetry must be preserved under quantum correction. SO(10) is the smallest group which is free from anomalies in a single multiplet.
Such an anomaly free condition is meaningful if the theory is renormalizable. 
Even if we fixed a gauge group, we need other criteria to make a definitive model. Renormalizability will be one such criteria.
Before proceeding to the detailed discussion, let us consider the structure of the existing theories. For $SU(3)\times U(1)$ theory at $\mu < {\cal O}(10^2$GeV), it has the following form 
\be
L=L_{ren}+\frac{L_1}{\Lambda_1}
\label{cutoff1}
\ee
in general.
Here the first $L_{ren}$ denotes renormalizable Lagrangian and the second term an unrenormalizable effective Lagrangian. 
 $L_1$ implies the fermi coupling $\propto J^\mu J_\mu$ and $\Lambda_1^{1/2}={\cal O}(10^2\mbox{GeV})$ in the effective weak interaction before the SM.
For $SU(3)\times SU(2)\times U(1)$ SM+$\nu$, this $L_1$ becomes the renormalizable
$g_2J^\mu W_\mu$ term but a new effective term appears in the (type I) seesaw mechanism \cite{seesaw},
\be
L=L_{ren}'+Y_\nu^T\frac{1}{M_R}Y_\nu(LH)^2\equiv L_{ren}'+\frac{L_2}{\Lambda_2}.
\label{cutoff2}
\ee
Here $L_{ren}'\supset L_{ren}+\frac{L_1}{\Lambda_1}$ and $\Lambda_2(=O(10^{13}GeV))\gg \Lambda_1^{1/2}$. 
Thus the theory is expressed as the sum of a renormalizable theory plus effective action with cut off, and the renormalizable Lagrangian becomes more involved as the energy scale goes up.  When the energy scale $\mu$ goes up and $\mu>M_R$, this Lagrangian is expected to be renormalizable. This is the rough sketch of GUT schemes.
This scenario, of course, is not a unique one. For the other types of seesaw mechanism \cite{Rabi}, \cite{Valle}, and  \cite{Rabi2} (Type II, III, and Inverse seesaw, respectively) give the corresponding phase transitions at each energy scale. However, even in these cases, GUT scheme that the renormalizable part becomes enlarged as energy scale goes up seems to be universally valid.
Especially, minimal SO(10) GUT includes $M_R$ in the Yukawa coupling with the other quark-lepton for the case of $\overline{\bf 126}$ Higgs field, and this field is suitable for the Yukawa coupling. Since after GUT all local field contents becomes massless, it is very natural to consider renormalizability as a guiding principle for constructing GUT theory.

Quarks and leptons have analogous structures. They have both three families of left-handed doublets and three right-handed singlets.
However, they have quite different mass hierarchies and mixing angles.
The successful gauge coupling unification of the minimal supersymmetric standard model (MSSM)
strongly supports the emergence of a supersymmetric (SUSY) GUT 
 around $M_{\rm GUT} \simeq 2 \times 10^{16}$ GeV. 

However, the MSSM does not specify the gauge group at $M_{GUT}$, and therefore the symmetry breaking pattern to the SM gauge group.

Group theoretical properties give strong constraints on seemingly quite different objects of quark-lepton.  The minimal SU(5) model \cite{Georgi} is ruled out from the wrong mass matrix relation and fast proton decay \cite{flipped}, and, among possible gauge groups, SO(10) GUT seems to be most promising.  SO(10) \cite{Fritzsch}
is the smallest simple gauge group 
 under which the entire SM matter contents of each generation are unified into a single anomaly-free irreducible representation: 
${\bf 16}$ representation. 
This ${\bf 16}$ representation includes the right-handed neutrino and no other exotic matter particles.
 Among several models based on the gauge group SO(10), 
 the renormalizable minimal SUSY SO(10) model (minimal SO(10) GUT) has been paid  a particular attention, where two Higgs multiplets 
 $\{{\bf 10} \oplus {\bf \overline{126}}\}$ 
 are utilized for the Yukawa couplings with matter 
 ${\bf 16}_i~(i=\mbox{generation})$ \cite{Babu} \cite{Fukuyama1} \cite{Fukuyama2}.

Before discussing the details of this model, let us explain the general structure of SUSY GUT.
\noindent
It consists of three ingredients:\\
(A) To build the Yukawa coupling based on gauge symmetry. \\
(B) To construct the gauge invariant Higgs superpotential and exhibit the gauge symmetry breaking pattern from GUT to the SM.\\
(C) To show the supersymmetry breaking mechanism and the initial conditions of soft SUSY breaking terms at GUT .\\
The SM discusses only (A).
The MSSM and its variations like the NMSSM and nMSSM etc. deal with (A) and (C).
So far minimal SO(10) GUT can uncover (B) since it can fix the gauge invariant Higgs superpotential.

Let us proceed to the detailed arguments of the model.
Concerning (A), a remarkable feature of the minimal SO(10) GUT model is its high predictivity on neutrino oscillations as well as reproducing charged fermion masses and mixing angles.
In this review we consider the renormalizability as one of the principles.  
If we relax the renormalizability, different SO(10) models are also possible \cite{Raby1} \cite{Frogatt} \cite{Barr2} etc., which will be discussed at subsection \ref{perturbative} breifly.
Minimal SO(10) GUT includes the additional CP phases which solves the too smaller baryogenesis problem for the Cabibbo-Kobayashi-Maskawa (CKM) CP phase \cite{Yoshimura}. 
However, after KamLAND data \cite{Eguchi:2002dm} had been released, 
 it entered to the stage of precision measurements, and some mismatches with our predictions were revealed in $\theta_{13}$ in the Maki-Nakagawa-Sakata (MNS) matrix and neutrino mass square ratios. Recently its value has been measured precisely \cite{Daya-Bay}
\be
\mbox{sin}^22\theta_{13}=0.092\pm 0.016(stat)\pm 0.005(syst).
\label{Daya-Bay}
\ee

Many authors performed data-fitting analysis to match up these new data.
Some models have seeked the solution of mismatches in the threshold corrections, though no one has yet performed these tremendous calculations.
The SO(10) model has, except for the SUSY partner, no exotic matter but has many guge and scalar particles other than the SM. Such heavy particles or would-be Nambu-Goldstone bosons are studied in (B). The Higgs superpotential in the minimal SO(10) GUT has been constructed and a detailed analysis of symmetry breaking patterns 
 have been extensively studied by us \cite{Fukuyama:2004xs, Bajc:2004xe, Aulakh:2004hm}. 
This construction gives the vacuum expectation values (VEVs)
 at intermediate energy scales explicitly, which gives rise to a trouble in gauge coupling unification as well as the necessary scales of the seesaw mechanism \cite{seesaw} and fast proton decay.

Probably, we have no desert between electroweak scale and GUT which MSSM has been assumed so far for simplicity.   
At each stage of the intermediate energy scales, there enters new massless particle into the renormalization group equation, changing the naive behaviour of gauge couplings and spoiling the unification.
This mismatch of gauge couplings has been explicitly shown 
 in Ref.~\cite{Bertolini}, where the couplings are not unified any more 
 and even the SU(2) gauge coupling blows up far below the GUT scale.

Even if this problem was solved, 
 the renormalizable minimal SO(10) GUT potentially suffers from the problem 
 that the gauge coupling blows up above GUT and below the Planck scale.
The minimal SO(10) GUT also predicts too faster proton decay \cite{Fukuyama:2004xs}\cite{Fukuyama3}.

So far we have discussed mainly on the observational conflicts. In order that a realistic GUT model works well, it must not involve the internal inconsistency concerned with (C).  Supersymmetry and gauge symmetry are related via $U(1)_R$ symmetry. The no-go theorem on SUSY breaking \cite{Nelson} \cite{Ratz} says that consistent spontaneous SUSY breaking for gauge group equal and higher than SU(5) is naively impossible in 4D.

Thus, to solve these observational and conceptual problems comprehensively, we consider an orbifold GUT \cite{Kawamura} preserving the merit of the minimal SO(10) GUT. 

One of the demerits to go beyond 4D is to break the renormalizability of the preceding theory. However, the extra dimension opens up only in the neighbourhood of the GUT scale and the deviation from the renormalizable theory give only small corrections.
On the other hand, there appear many merits by considering an extra dimension. 
First we can solve the problems mentioned above. Moreover, the problems involved in the minimal SO(10), fast proton decay and coupling blow up,
can also be evaded.
We have a new geometrical SUSY breaking mechanism.

This paper is organized as follows.

In Chapter 2 we review the minimal SO(10) GUT in 4D.
In the first two sections we overview the historical background before we proposed SO(10) GUT model. In section 2.3 we give a setup of the minimal SO(10) GUT model and its data fitting is given in section 2.4. Section 2.5 is devoted to leptogenesis and electric dipole moments (EDM). The latter is an important signal of new physics beyond the SM (BSM) since minimal SO(10) model has the definite additional CP phases, giving larger EDMs than those in the SM.
Some mismatches with observations are $\theta_{13}$ and the neutrino mass square ratio. However, they may not be serious since we can improve the situation by extending to incorporate also type II seesaw, preserving the minimal SO(10) model.  Leptogenesis is discussed in section 2.6. In section 2.7 we go further to the analyses of the Higgs superpotential and reveal the concrete structure of intermediate energy scales between GUT and the SM, which leads us to the ruin of gauge coupling unigfication of the MSSSM. Another deficit of minimal SO(10), fast proton decay, is discussed in section 2.8. So we need some modifications of the minimal SO(10) model. Some solutions are discussed in section 2.9 but other bad news of GUT in 4D (No-Go theorem on SUSY breaking) is explained.  Consequently, in the susequent part, we consider a class of SO(10) models with 5D orbifold \cite{Raby}. In Chapter 3 we consider an SO(10) model in 5D and will explain how this model rescues the problems of minimal SO(10) GUT in 4D.
In section 3.2, we explain the setup of our model, \cite{F-O1}, where
all matters and Higgs multiplets reside only 
 on a Pati-Salam brane (PS brane) where the PS gauge symmetry is manifest, 
 so that a low energy effective description of this model 
 is nothing but the PS model in 4D with a special set of 
 matter and Higgs multiplets. 
At energies higher than the compactification scale, 
 the Kaluza-Klein (KK) modes of the bulk SO(10) gauge multiplet 
 are involved in the particle content. In section 3.3, the gauge coupling 
 unification is shown to be successfully realized 
 by incorporating the KK mode threshold corrections 
 into the gauge coupling running.
The unification scale ($M_{\rm GUT}$) and 
 the compactification scale ($M_c$) which was set to be the same 
 as the PS symmetry breaking scale ($v_{\rm PS}$) is found to be 
 $M_{\rm GUT}=4.6 \times 10^{17}$ GeV and 
 $M_c=v_{\rm PS}=1.2 \times 10^{16}$ GeV. The improvements of mass spectra data fitting will be discussed. This is rather trivial fact. 
In section 3.4, we apply this SO(10) model to the inflationary scenario \cite{F-O2}. 
The idea of inflation \cite{InflationRev} has been strongly favored 
 from the view point of not only providing the solutions 
 to the horizon and flatness problems of the standard big bang cosmology 
 but also recent precise cosmological observations on the cosmic microwave background radiation and the large 
 scale structure in the Universe. 
Therefore, it is an important task to construct a realistic 
 inflation model based on some well-motivated particle physics model. 
Single-field inflation is disfavored by its requirement of tiny parameter value to reproduce the results of COBE \cite{COBE} and WMAP \cite{WMAP}. 
Hybrid inflation 
 \cite{Linde} \cite{HIRev} \cite{Hybrid1} solves the above problem but gives rise to monopole problem in the original form.
 Variants of hybrid inflation, in particular, 
 applicable to SUSY GUT models have been proposed: 
  standard \cite{StdHI}, shifted \cite{ShiftedHI} 
 and smooth \cite{SmoothHI} hybrid inflation models. 
Some of these models are based on the SUSY PS model 
 with one singlet and Higgs multiplets 
 $({\bf 4},{\bf 1},{\bf 2}) \oplus (\overline{{\bf 4}},{\bf 1},{\bf 2})$ 
 whose VEVs break the PS symmetry to the SM one. 
Interestingly, except for the singlet field, the orbifold GUT model 
 of Ref.~\cite{F-O1}, which we are interested in, has the same particle content. 
Therefore, the GUT model can naturally incorporate hybrid inflation and is constrained from the inflation model. 
So far we have assumed for simplicity $M_c=v_{PS}$.
However in this case LSP becomes stau. In section 3.5, we show that
 the neutralino becomes the LSP by generalizing $M_c\neq v_{PS}$ and DM candidate \cite{F-O3}.
In sections 3.6 and 3.7 we reanalyze leptogenesis \cite{F-O4} and LFV, which were discussed in the minimal SO(10) model in 4D, respectively in the scheme of SO(10) in 5D. According to the recent discovery of Higgs (like) particle, we have added a subsection on the impact of the LHC results to GUT.

The last section is devoted to discussion.
Appendix is served for a compact mathematical resume of SO(10) group property.
\chapter{Renormalizable Minimal SO(10) GUT in Four Dimensions}
\section{Why Do We Need GUT ?}
It is natural to start our review with this title since GUT is not necessarily indispensable for new physics BSM to all model builders.
In going beyond the SM, we have rather serious constraints on matter contents, whereas we have no definite criteria for Higgs sector.
One of the reasons for more Higgs than the SM is concerned with electro-weak baryogenesis \cite{Shaposhnikov}.
We need strong first-order phase transition at the sphaleron transition (its rate $\Gamma$),
\be
H_c>\Gamma_c.
\label{sphaleron}
\ee
It requires in the SM
\be
\frac{2E}{\lambda_c}>1
\ee
with
\be
E=\frac{1}{2\pi v^3}(6m_W^3+3m_Z^3)
\ee
and $\lambda$ is the coefficient of $\frac{\lambda}{4}\phi^4$.
This leads us to $m_H<56$ GeV. If we add additional scalar particles, $E$ can get larger and $m_H$ can go beyond the LEP or LHC bound \cite{Kastening}.

So it is rather natural to consider more Higgs than one SM Higgs.
In this case we have two major constraints: One is the $\rho$ parameter,
\be
\rho\equiv \frac{m_W^2}{m_Z^2\mbox{cos}^2\theta_W}
\ee
and the other is FCNC. In two and more Higgs doublet model, FCNC appears at tree level in general.
Let us consider $\rho$ parameter first.  It is famous that this parameter first predicted correctly large top quark mass before its discovery.
Indeed, one loop quark correction to W propagator gives \cite{Veltman}
\be
\rho=1+\frac{3G_F}{8\pi^2\sqrt{2}}\left[m_t^2+m_b^2-\frac{2m_t^2m_b^2}{m_t^2-m_b^2}\mbox{ln}\left(\frac{m_t^2}{m_b^2}\right)\right].
\ee
Two Higgs Doublet Model (2HDM) \cite{2HDM} contributes at one loop level and Higgs Triplet Model (HTM) \cite{HTM} \cite{Schechter} does at tree level.

So first we consider on the HTM. The HTM is the minimum extension to the SM. It adds only one 
Higgs triplet and no matter field even right-handed neutrino.
In the Higgs Triplet Model,
we introduce a SU(2) triplet $Y=2$ scalar as
\begin{eqnarray}
\Delta
\equiv
 \left(
  \begin{array}{cc}
   \Delta^+/\sqrt{2} & \Delta^{++}\\
   \Delta^0 & -\Delta^+/\sqrt{2}
  \end{array}
 \right), \ \ \
{\mathcal L}_{triplet~Yukawa}
= - h_{\alpha\beta} \overline{L_\alpha^C} i\sigma^2 \Delta P_L L_\beta
  + h.c.
\end{eqnarray}
 This model generates neutrino masses without right-handed neutrinos
with the triplet vacuum expectation value $v_\Delta$
which is given by the explicit breaking of the lepton number.
 This model is very predictive
because of a clear relation
\begin{eqnarray}
m_{\alpha\beta} = \sqrt{2} v_\Delta h_{\alpha\beta},
\end{eqnarray}
where $m_{\alpha\beta}$ denotes the Majorana mass matrix for neutrinos.

The experimental limit of $\rho=1.0004^{+0.0027}_{-0.0007}$ at $2\sigma$ \cite{Amsler} gives
\be
\frac{v_\Delta}{v_H}\leq 0.01,
\ee
where $v=246$ GeV.

 Next simple model is two Higgs doublet model (2HDM), which add, in addition to the SM Higgs doublet $H_1$, another Higgs doublet $H_2$.
 There are several types of the model
depending on which doublet couples with which fermion:
\begin{eqnarray}
\text{type I (SM-like)}
&:&
 \text{$H_1$ couples with all fermions}\nonumber\\
&&
 \text{$H_2$ decouples with fermions}\nonumber\\
\text{type II (MSSM-like)}
&:&
 \text{$H_1$ couples with down-type quarks and charged leptons}\nonumber\\
&&
 \text{$H_2$ couples with up-type quarks}\nonumber\\
\text{type III (general)}
&:&
 \text{both of Higgs doublets couple with all fermions}\nonumber\\
etc.\ etc. && \nonumber
\end{eqnarray}

In the 2HDM as a whole, there is no criteria to classify matter content and it is the reason why there are several types mentioned above. Its strategy seems to let observations tell a story without model prejudice. In other words, in the 2HDM, it is rather difficult to specify the model
and predict new phenomena. Also the masses of neutral and charged Higgses and phases are tightly constrained from
 $R_b\equiv \frac{\Gamma(Z\rightarrow b\overline{b})}{\Gamma(Z\rightarrow hadrons)},~
\Gamma(b\rightarrow s\gamma),~\overline{B}^0-B$ mixing, $\rho$ parameter etc.,
and we should take those constraints all into account.

Both the HTM and 2HDM are very simple and useful towards the final theory. However, they are phenomenological models and can not be comprehensive new
theories BSM.

\section{What and Why Is SO(10) Group ?}

Chiral (left-handed) fermions in the SM with right-handed $\nu_R$ are composed of left-handed $SU(2)_L$ doublets ($Q$ and $L$) and the charge conjugates of right-handed singlets $\left((u_R)^c,(d_R)^c,(e_R)^c,(\nu_R)^c\right)$ (hereafter we abbreviate them $u^c,..$ for simplicty),
\bea
Q &=&\left(\begin{array}{ccc}
u_r &u_y &u_b  \\
d_r &d_y &d_b 
\end{array}\right)=\left(3,2,\frac{1}{6}\right),\nonumber\\
u^c&=&\left(u^{rc},u^{yc},u^{bc}\right)=\left(\overline{3},1,-\frac{2}{3}\right),~~d^c=\left(d^{rc},d^{yc},d^{bc}\right)=\left(\overline{3},1,\frac{1}{3}\right),\\
L&=&\left(\begin{array}{c} 
\nu\\
e^-
\end{array}\right)=\left(1,2,-\frac{1}{2}\right),\nonumber\\
e^c&= &(1,1,1),~~\nu^c=(1,1,0).\nonumber
\label{qnumber}
\eea
The first unfication scheme began with partial unification of quark-leptons under $SU(4)_C\times SU(2)_L\times SU(2)_R$ \cite{Pati}
\bea
({\bf 4},{\bf 2},{\bf 1})=\left(\begin{array}{cccc}
u_r &u_y &u_b &\nu_e \\
d_r &d_y &d_b & e
\end{array}
\right)_L\equiv F_{L1}.
\label{PS}
\eea
Likewise, $({\bf \overline{4}},{\bf 1},{\bf 2})$ is the charge conjugation of their right-handed partners.
Here, $L(R)$ indicates left-handed (right-handed) fermions and $1$ indicates the first family. We have, of course, the second, third families.
Thus lepton number was considered as fourth color.

One of the great aqchievments of the PS model is the realization of charge quantization via
\be
Q=T_{L3}+\frac{Y}{2}=T_{L3}+T_{R3}+\frac{B-L}{2}.
\ee
SU(5) model unifies strong, electromagnetic, and weak forces but does not in single multiplet for matters
matters,
\be
{\bf \overline{5}}=(d^{rc},d^{yc},d^{bc},e^-,-\nu_e)_L^T
\ee
and
\bea
{\bf 10}=\frac{1}{\sqrt{2}}\left(\begin{array}{ccccc}
0& u^{cr}&-u^{cy}&-u^{r}&-d^{r}\\
-u^{cb}&0&u^{cr}&-u^y&-d^y\\
u^{cy}&-u^{cr}&0&-u^b&-d^b\\
u^r&u^y&u^b&0&-e^+\\
d^r&d^y&d^b&e^+&0
\end{array}\right)_L
\label{SU5}
\eea
and
\be
{\bf 1}=\nu_e^c.
\ee
Minimal SU(5) model gives problematic mass relation
\be
M_e=M_d
\label{SU5mass}
\ee
and fast proton decay. The flipped SU(5) \cite{flipped} circumvents this pathology and gives a good instrument, missing partner mechanism, for doublet-triplet problem (See subsection {\bf 2.9.2}).
Unfortunately, $\nu^c$ moves into {\bf 10}-plet from SU(5) singlet and heavy Majorana mass term appears as unrenormalizable term.

On the other hand, the fundamental representation ${\bf 16}$ in SO(10) is an anomaly free and includes all fermions in a single multiplet 
\be
{\bf 16}_L=(Q, u^c,d^c,L,\nu^c,e^c)_L.
\ee
${\bf 16}$ is decomposed as
\be
{\bf 16}_L=({\bf 4},{\bf 2},{\bf 1})+(({\bf \overline{4}},{\bf 1},{\bf 2})
\ee
under PS and as
\be
{\bf 16}_L={\bf \overline{5}}+{\bf 10}+{\bf 1}
\ee
under SU(5), respectively.   Of course, we need another Higgs to cancell the contribution of the fermion partners of the Higgs of SM if SUSY is involved.

Let us go to further extension of gauge group.
Next large gauge group is of rank 6, $E_6$, whose most conventional mass assignment is \cite{Hewett}
\bea
{\bf 27}=\left(\begin{array}{cccccc}
Q, &u^c, &e^c, &L, &d^c, &\nu^c \\
H, & g^c, & H^c, & g, &S^c
\label{27}
\end{array}
\right).
\label{E6}
\eea
Here
\bea
H=\left(\begin{array}{c}
N\\
E
\end{array}\right),~~H^c=\left(\begin{array}{c}
E\\
N
\end{array}\right)^c
\eea
and $g$ is colored SU(2) singlet.
In \bref{27} the contents of upper line are matters of ${\bf 16}={\bf 10}+\overline{\bf 5}+{\bf 1}$ and those of lower are exotics of
${\bf 10}=\overline{\bf 5}+{\bf 5}$ and ${\bf 1}$, respectively.
$\tilde{N}$ and $\tilde{N}^c$ play the role of Higgs fields, $H_d$ and $H_u$, respectively.

Unlike the case of SO(10), it includes exotic particles. One of advantages was that Higgs fields are assigned in matter multiplet as $\tilde{H}$ and $\tilde{H}^c$ ($\tilde{H}$ implies the scalar partner of $H$).
However, we must have three copies of Higgs multiplets corresponding to three generations. 
So some one consider that only $H_3$ and $H^c_3$ have vev and the other $H_i~(i=1,2)$ and $H_i^c~(i=1,2)$ are non-Higgs \cite{King}.
So we must consider an additional scheme for the mass generation of the remaining first and second families.
Also, if we set fermions and osons in the same multiplet we can not define R-parity,
\be
R=(-1)^{3(B-L)+2S}.
\label{R-parity}
\ee

R-parity is $Z_2$ subgroup of $U(1)_{B-L}$ which is carried out by $\overline{{\bf 126}}$ for SO(10)
and ${\bf 351}_s={\bf 1(-8)}+{\bf 10(-2)}+{\bf 16(-5)}+{\bf 54(4)}+{\bf 126(-2)}+{\bf 144(1)}$ for $E_6$.

Also general $E_6$ invariant Yukawa coupling ${\bf 27}\times {\bf 27}\times \overline{{\bf 27}}$ leads us to B-L violation 
at low energy scale.
To circumvent the troubles, one method is to prepare ${\bf 27}$ Higgs additionally and separate matters and Higgs sectors.
In that case we must add another Higgs to realize non diagonal mixing angle of lepton mass matrix.
For the renormalized case it may be ${\bf 351}_H$.
As for unrenormalized effective action we do not adopt such strategy as we have repeated several times.
Anyhow, this is far from being minimal mentioned above, and hereafter we concentrate on SO(10) GUT.

First, we give a brief review of the minimal SUSY SO(10) model.

There are many excellent textbooks for GUT. Here we list some of them emphasizing neutrino physics, \cite{F-Y} and \cite{M-P}.
\section{Yukawa Coupling}
A major part of great successes of the SM comes from its perturbative predictions. From the bottom-up approach, we have seen in the introduction that theoretical developments overlap with the extension of renormalizable area.
So at least the main part of GUT should be guided from renormalizability.
If we respect renormalizability, the dimension of the coupling constant (we symbolically write it $Y$ here) must have zero or negative index in length units.
So if fermion masses are generated by Higgs mechanism as in the SM, their interaction must be Yukawa coupling.
\bea
Y^{(u)}_{ij}\overline{u}_iH_uQ_j+Y^{(d)}_{ij}\overline{d}_iH_dQ_j+\nonumber\\
+Y^{(\nu)}\overline{\nu}_iH^uL_j+Y^{(e)}_{ij}\overline{e}_iH_dL_j+h.c.
\eea
We have written here for the case of MSSM. In the case of the SM+massive (Dirac) neutrino, we may change
\be
H_u\rightarrow \Phi=(\phi^0, \phi^-)^T,~~H_d\rightarrow i\tau_2\Phi^*.
\ee
So matter multiplets products in Yukawa coupling become group theoretically, 
\be
{\bf 16}\otimes {\bf 16}={\bf 10}\oplus {\bf 120}\oplus {\bf 126}.
\ee
The details of group theoretical arguments of SO(10) are given in \cite{fuku1} \cite{Mohapatra}.

So the Higgs fields which can construct SO(10) singlet with bi-product of above
fermions are ${\bf 10},~{\bf 120},$ and ${\bf \overline{126}}$.
Obviously single Higgs is incompatible with the observed mixing matrices, CKM and MNS.
Since ${\bf 10}$ is the simplest and inevitable, so the set of \{${\bf 10}$ and ${\bf 120}\}$ or \{${\bf 10}$ and ${\bf \overline{126}}\}$ is a minimal model.
${\bf \overline{126}}$ is more essential than ${\bf 120}\}$ in connection with neutrino mass since the former includes $({\bf 10},{\bf 3},{\bf 1})$ and $({\bf \overline{10}},{\bf 1},{\bf 3})$ under $SU(4)_c\otimes SU(2)_L\otimes SU(2)_R$.
These two subgroups, if they have vevs, induce the right-handed Majorana and left-handed Majorana neutrino, respectively.
It also automatically conserves R-parity \cite{M-L}. (Whereas SU(5) GUT also induces R-parity violation term via ${\bf 10}\overline{\bf 5}~\overline{\bf 5}$ term like $LLe^c,~LQd^c,~u^cd^cd^c$.)

So we select \{${\bf 10}$ and ${\bf \overline{126}}\}$ in the Yukawa coupling.
This model is called the renormalizable minimal SUSY SO(10) GUT (the minimal SO(10) GUT).

This model was first applied to neutrino oscillation in \cite{Babu}. 
However, it did not reproduce the large mixing angles.
It has been pointed out that 
 CP-phases in the Yukawa sector play important roles 
 to reproduce the neutrino oscillation data \cite{Fukuyama1}. 
More detailed analysis incorporating the renormalization group (RG) 
 effects in the context of MSSM 
 has explicitly shown that the model is consistent with the neutrino
 oscillation data at that time, Thus the minimal SO(10) GUT became a realistic model \cite{Fukuyama2}. 
We give a brief review of the minimal SO(10) model.\\
Yukawa coupling is given by
\begin{eqnarray}
 W_Y = Y_{10}^{ij} {\bf 16}_i H_{10} {\bf 16}_j 
           +Y_{126}^{ij} {\bf 16}_i H_{126} {\bf 16}_j \; , 
\label{Yukawa1}
\end{eqnarray} 
where ${\bf 16}_i$ is the matter multiplet of the $i$-th generation,  
 $H_{10}$ and $H_{126}$ are the Higgs multiplet 
 of {\bf 10} and $\overline{\bf 126} $ representations 
 under SO(10), respectively. 
Note that, by virtue of the gauge symmetry, 
 the Yukawa couplings, $Y_{10}$ and $Y_{126}$, 
 are, in general, complex symmetric $3 \times 3$ matrices. 
After the symmetry breaking of SO(10) to 
${\rm SU}(3)_c \times {\rm SU}(2)_L \times {\rm U}(1)_Y$
via ${\rm SU}(4)_c \times {\rm SU}(2)_L \times {\rm SU}(2)_R$
or ${\rm SU}(5) \times {\rm U}(1)$,
 we find that two pairs of Higgs doublets 
 in the same representation appear as the pair in the MSSM. 
One pair comes from $({\bf 1},{\bf 2},{\bf 2}) \subset {\bf 10}$ 
 and the other comes from 
 $(\overline{\bf 15}, {\bf 2}, {\bf 2}) \subset \overline{\bf 126}$. 
Using these two pairs of the Higgs doublets, 
 the Yukawa couplings of Eq.~(\ref{Yukawa1}) are rewritten as 
\begin{eqnarray}
W_Y &=& (U^c)_i  \left(
Y_{10}^{ij}  H^u_{10} + Y_{126}^{ij}  H^u_{126} \right) Q
+ (D^c)_i  \left(
Y_{10}^{ij}  H^d_{10} + Y_{126}^{ij}  H^d_{126} \right) Q_j  
\nonumber \\ 
&+& (N^c)_i \left( 
Y_{10}^{ij}  H^u_{10} - 3 Y_{126}^{ij} H^u_{126} \right) L_j 
+ (E^c)_i  \left(
Y_{10}^{ij}  H^d_{10}  - 3 Y_{126}^{ij} H^d_{126} \right) L_j   
\nonumber \\
&+&
 L_i \left( Y_{126}^{ij} \; v_L \right) L_j +
(N^c)_i \left( Y_{126}^{ij} \; v_R \right) (N^c)_j \;  . 
\label{Yukawa2}
\
\end{eqnarray} 
Here $u_R$, $d_R$, $\nu_R$ and 
 $e_R$ are the right-handed ${\rm SU}(2)_L$ 
 singlet quark and lepton superfields, $Q$ and $L$ 
 are the left-handed ${\rm SU}(2)_L$ doublet quark and lepton superfields, 
 $H_{10}^{u,d}$ and $H_{126}^{u,d}$ 
 are up-type and down-type Higgs doublet superfields 
 originated from $H_{10}$ and $H_{126}$, respectively, 
 and the last two terms are the Majorana mass terms 
 of the left- and right-handed neutrinos developed 
 by the VEVs ($v_L$ and $v_R$) of the $({\bf 10}, {\bf 3}, {\bf 1})$ and $(\overline{\bf 10}, {\bf 1}, {\bf 3})$ Higgs. 
The factor $-3$ in the lepton sector 
 is the Clebsch-Gordan coefficient. 

In order to preserve a successful gauge coupling unification, 
 suppose that one pair of Higgs doublets 
 given by a linear combination $H_{10}^{u,d}$ and $H_{126}^{u,d}$ 
 is light while the other pair is  heavy ($\geq M_{\rm GUT}$).  
The light Higgs doublets are identified as 
 the MSSM Higgs doublets ($H_u$ and $H_d$) 
 and given by 
\begin{eqnarray} 
 H_u &=& \tilde{\alpha}_u  H_{10}^u  
      + \tilde{\beta}_u  H_{126}^u \;,
 \nonumber \\
 H_d &=& \tilde{\alpha}_d  H_{10}^d  
      + \tilde{\beta}_d  H_{126}^d  \; , 
 \label{mix}
\end{eqnarray} 
where $\tilde{\alpha}_{u,d}$ and $\tilde{\beta}_{u,d}$ 
 denote elements of the unitary matrix  
 which rotate the flavor basis in the original model 
 into the (SUSY) mass eigenstates (see \bref{UV} in detail). 
Omitting the heavy Higgs mass eigenstates, 
 the low energy superpotential is described 
 by only the light Higgs doublets $H_u$ and $H_d$ such that 
\begin{eqnarray}\label{Yukawa3}
W_Y &=& 
(U^c) _i \left( \alpha^u  Y_{10}^{ij} + 
\beta^u Y_{126}^{ij} \right)  H_u \, Q_j 
+ (D^c)_i  
\left( \alpha^d  Y_{10}^{ij} + 
\beta^d Y_{126}^{ij}  \right) H_d \,Q_j  \nonumber \\ 
&+& (N^c)_i  
\left( \alpha^u  Y_{10}^{ij} -3 
\beta^u Y_{126}^{ij} \right)  H_u \,L_j 
+ (E^c)_i  
\left( \alpha^d  Y_{10}^{ij} -3 
\beta^d  Y_{126}^{ij}  \right) H_d \,L_j  \\ 
&+& 
  L_i \left( Y_{126}^{ij} \; v_L \right) L_j + 
 (N^c)_i  
  \left( Y_{126}^{ij} v_R \right)  (N^c)_j \nonumber \; ,  
\end{eqnarray}
where the formulas of the inverse unitary transformation 
 of Eq.~(\ref{mix}), 
 $H_{10}^{u,d} = \alpha^{u,d} H_{u,d} + \cdots $ and 
 $H_{126}^{u,d} = \beta^{u,d} H_{u,d} + \cdots $, 
 have been used. 
Note that the elements of the unitary matrix, 
 $\alpha^{u,d}$ and $\beta^{u,d}$,   
 are in general complex parameters, 
 through which CP-violating phases are introduced 
 into the fermion mass matrices. 

Providing the Higgs VEVs, 
 $H_u = v \sin \beta$ and $H_d = v \cos \beta$ 
 with $v=174 \mbox{GeV}$, 
 the quark and lepton mass matrices can be read off as%
\begin{eqnarray}
\label{massmatrix}
  M_u &=& c_{10} M_{10} + c_{126} M_{126}\; ,   \nonumber \\
  M_d &=&     M_{10} +     M_{126}\; ,   \nonumber \\
  M_D &=& c_{10} M_{10} -3 c_{126} M_{126}\; ,    \\
  M_e &=&     M_{10} -3     M_{126}\; ,   \nonumber \\
  M_L &=& c_L M_{126}\; ,\nonumber \\  
  M_R &=& c_R M_{126}  \nonumber \; . 
\end{eqnarray} 
Here $M_u$, $M_d$, $M_D$, $M_e$, $M_L$, and $M_R$ 
 denote the mass matrices of up-type quark, down-type quark, 
 Dirac neutrino, charged-lepton, left-handed Majorana, and 
 right-handed Majorana neutrino, respectively. 
Note that all the quark and lepton mass matrices 
 are characterized by only two basic mass matrices, $M_{10}$ and $M_{126}$,   
 and four complex coefficients 
 $c_{10}$, $c_{126}$, $c_L$ and $c_R$, 
 which are defined as 
 $M_{10}= Y_{10} \alpha^d v \cos\beta$, 
 $M_{126} = Y_{126} \beta^d v \cos\beta$, 
 $c_{10}= (\alpha^u/\alpha^d) \tan \beta$, 
 $c_{126}= (\beta^u/\beta^d) \tan \beta $, 
 $c_L = v_L/( \beta^d  v  \cos \beta)$) and 
 $c_R = v_R/( \beta^d  v  \cos \beta)$), respectively.  
These are the mass matrix relations required by 
 the minimal SO(10) model. 

You should remark especially the relation between $M_d$ and $M_e$ different from \bref{SU5mass}. In other word, $Y_{10}$ and $Y_{126}$ must be of the same order at least in the first $2\times 2$ matrix components. 
Another essential point is that the heavy right-handed is represented by $M_{126}$ (also a part of the mass matrices of charged fermions).
This is the main reason of its high predictivity of this theory.

In the following in Part I, we set $c_L=0$ as the first approximation.
Except for $c_R$, 
  which is used to determine the overall neutrino mass scale, 
 this system has fourteen free parameters in total if we resrict ourselves $M_{10}$ and $M_{126}$ are real \cite{Fukuyama1}.
This few parameters in all mass matrices assures the strong predictability of the minimal SO(10) Model.
\section{Data Fitting}
Data fitting is performed as follows:
Firstly we fit the data of charged fermions, masses of quarks and charged leptons and
CKM mixing angles. Using the parameters fixed by this process we proceed to fit the neutrino data.
\subsection{Charged fermions}
Eliminating $M_{10}$ and $M_{126}$ from Eq.\bref{massmatrix}, we obtain
\begin{equation}
M_e=c_dM_d+c_uM_u, \label{M2}
\end{equation}
where
\begin{equation}
c_d=-
{\frac{3c_{10}+c_{126}}{c_{10}-c_{126}}} \ ,\ \ 
c_u={\frac{4}{c_{10}-c_{126}}} \ .
\end{equation}
Since $M_u$, $M_d$, and $M_e$ are complex symmetric matrices, they are 
diagonalized
by unitary matrices $U_u$, $U_d$, and $U_e$, respectively, as 
\begin{equation}
U_u^{T}M_uU_u=D_u \ , \ \
U_d^{T}M_dU_d=D_d \ , \ \
U_e^{T}M_eU_e=D_e \ , \label{diag1}
\end{equation}
where $D_u$, $D_d$, and $D_e$ are diagonal matrices given by
\begin{eqnarray}
&&D_u \equiv \mbox{diag}(m_u,m_c,m_t) \ , \ \
  D_d \equiv \mbox{diag}(m_d,m_s,m_b) \ , \nonumber \\
&&D_e \equiv \mbox{diag}(m_e,m_\mu,m_\tau) \ . \label{diag2}
\end{eqnarray}
Since the CKM matrix $V_q$ is given by 
\begin{equation}
V_q=U_u^{T} U_d^* \ ,
\end{equation}
the relation (\ref{M2}) is re-written as follows: 
\begin{equation}
(U_e^{\dagger}U_u)^T D_e(U_e^{\dagger}U_u)
=c_{d}V_q D_{d}V_q^{T}
+c_uD_u .\label{eq071703}
\end{equation}
Therefore, we obtain the independent three equations:
\begin{eqnarray}
{\rm Tr}D_e D_e^\dagger &=& |c_d|^2 \, {\rm Tr}
\Bigl[(V_q D_{d}V_q^{T}+\kappa D_u)
(V_q D_{d}V_q^{T}+\kappa D_u)^\dagger\Bigr],
 \label{eq82501}\\
{\rm Tr}(D_e D_e^\dagger)^2 &=&|c_d|^4 \,
{\rm Tr}\Bigl[
((V_q D_{d}V_q^{T}+\kappa D_u)
(V_q D_{d}V_q^{T}+\kappa D_u)^\dagger)^2\Bigr],
 \label{eq82502}\\
{\rm det}D_eD_e^\dagger &=& |c_d|^6 \, {\rm det}
\Bigl[(V_q D_{d}V_q^{T}+\kappa D_u)
(V_q D_{d}V_q^{T}+\kappa D_u)^\dagger\Bigr]
\label{eq82503}
\end{eqnarray}
with $\kappa=c_u/c_d$.
\begin{figure}
\begin{center}
\epsfig{file=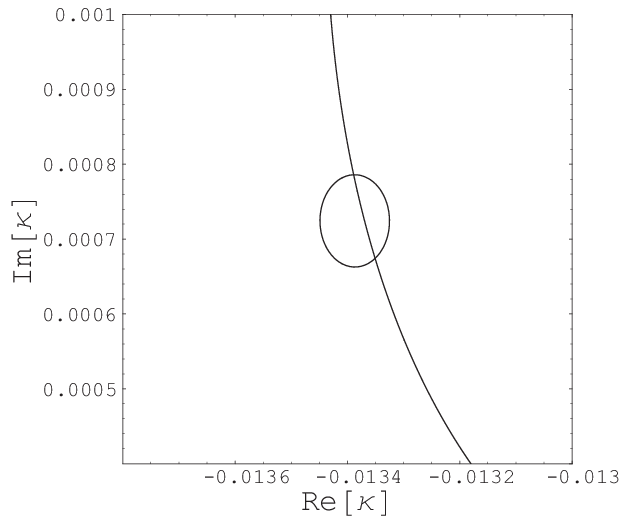, width=10cm}
\caption{
Contour plot on complex $\kappa$-plane.
The vertical line and the circle correspond to the solutions 
 of Eqs.~(\ref{eq82511})  and (\ref{eq82512}), respectively. 
}
\label{GCU}
\end{center}
\end{figure}
By eliminating the parameter \(c_d\), we have two equations 
for the parameter \(\kappa\):
\begin{eqnarray}
\frac{(m_e^2+m_\mu^2+m_\tau^2)^3}{m_e^2 m_\mu^2 m_\tau^2}&=&
  \frac{(\ref{eq82501})^3}{(\ref{eq82503})}, \label{eq82511}\\
\frac{(m_e^2+m_\mu^2+m_\tau^2)^2}
     {2(m_e^2 m_\mu^2+m_\mu^2 m_\tau^2+m_\tau^2 m_e^2)}&=&
\frac{(\ref{eq82501})^2}{(\ref{eq82501})^2-(\ref{eq82502})}, 
\label{eq82512}
\end{eqnarray}
Here \((\ref{eq82501})^3\), for instance, 
means the right-hand side of Eq.(\ref{eq82501}) to the third power.
Let us denote the parameter values of $\kappa$ evaluated from 
Eqs.(\ref{eq82511}) and (\ref{eq82512}) as ${\kappa}_A$ and ${\kappa}_B$, respectively.  
If ${\kappa}_A$ and ${\kappa}_B$ coincide with each other, 
then we have a possibility 
that the SO(10) GUT model can reproduce the observed quark and charged 
lepton mass spectra.  
If ${\kappa}_A$ and ${\kappa}_B$ do not so, the SO(10) model with 
one {\bf 10} and one {\bf 126} Higgs scalars is ruled out, 
and we must bring 
more Higgs scalars into the model.  The result is depicted in Fig.\ref{GCU}.

\par

\subsection{The number of parameters in the minimal SO(10) model \label{parameter}}
As we have discussed in the previous section, 
among four freedoms of complex $\{c_{10},~c_{126}\}$ or $\{c_d ,~\kappa\}$, 
we have been able to fix the three of them, $\kappa$ and $|c_d|$.  
This is not accidental.
Let us discuss the situation in details in the SO(10) two Higgs model. 

By using the relation (\ref{eq071703}),
we have investigated whether there is a set of parameters which
can give the 13 observable quantities $D_e$, $D_u$, $D_d$, and $V_q$
or not.
We can rewrite Eq.(\ref{eq071703}) as
\begin{equation}
A_e^T D_eA_e=c_d (V_q D_{d}V_q^{T}+\kappa D_u), \label{eq01080701}
\end{equation}
where 
\begin{equation}
A_e = U_e^\dagger U_u,
\end{equation}
\begin{equation}
c_d = |c_d| e^{i\sigma} .
\end{equation}
The quantities $D_e$, $D_u$, $D_d$, and $V_q$
are inputs, and the quantities $|c_d|$, $\kappa$, and $A_e$ are
the parameters which should be fixed from those observed quantities.
In general, an $n\times n$ unitary matrix for \(n\) generations has $n^2$ parameters.
Therefore, the number of the parameters is
\begin{equation}
N(\mbox{pmt}) = N(A_e) +N(c_d)+N(\kappa) = n^2+2+2.
\end{equation}
On the other hand, the number of equations is
\begin{equation}
N(\mbox{eqs}) = n(n+1) ,
\end{equation}
because Eq.(\ref{eq01080701}) is symmetric.
Therefore, the number of the unfixed parameters is given by
\begin{equation}
N_{{\rm free}}= N(\mbox{pmt}) -N(\mbox{eqs}) = 4-n = 1 ,
\end{equation}
for $n=3$, i.e., the 13 observed quantities fix the parameters
$|c_d|$, $\kappa$, and $A_e$, but 1 parameter \(\sigma\) 
remains as an unknown parameter \cite{Fukuyama1} and \cite{Fukuyama2}.  

Let us reconsider this number counting in more detail, going back to \bref{massmatrix}.
We first set the base as $M_u=D_u$ (3). $M_d$ is $3\times 3$ complex symmetric matrix (12)
which is diagonalized by a general unitary matrix V as
\begin{equation}
M_d=V^*D_dV^\dagger.
\end{equation}
Here nine parameters of V are divided into the CKM matrix (4) and five phases,
\begin{equation}
V=e^{i\alpha}e^{i\beta T_3}e^{i\gamma T_8}~ V_{CKM}~ e^{i\beta'T_3}e^{i\gamma 'T_8}.
\label{phases}
\end{equation}
However, the first three phases are delieted by rephasing. Thus we have 3+6=9 in $M_d$.
Furthermore we have 2 complex number $c_d,~\kappa$ (4). Thus totally 3+9+4=16.
In this numbers we have set the above two phases ($\beta ',~\gamma '$) as 0 or $\pi$ for simplicity. Thus we have 14 parameters. \footnote{This is in contrast with 28 parameters in the SM+$\nu$.} Using the remaining one parameter and one $c_R$ (in \bref{massmatrix}) we can fit all mass matrices including $M_D$ and $M_R$, that is, all low energy phenomena. This is a miraculous predictivity.


\subsection{Application to neutrino sector}
Next, we proceed to study how to predict neutrino masses
\begin{equation}
D_\nu = U_\nu^T M_\nu U_\nu 
\end{equation}
and the Maki-Nakagawa-Sakata (MNS) mixing matrix
\begin{equation}
U_\ell = U_e^T U_\nu^*\equiv A_e^*A_\nu^T
\end{equation}
by using the observed quantities $D_e$, $D_u$, $D_d$, and $V_q$
and the parameter values $|c_d|$, $\kappa$, and $A_e$ fixed by
Eq.(\ref{eq01080701}).

SO(10) GUT asserts that the Dirac neutrino mass matrix $M_D$
is given by the form 
\begin{equation}
M_D = c_{10} M_{10}- 3 c_{126} M_{126}
\end{equation}
and 
Majorana mass matrices of the left-handed and right-handed neutrinos,
$M_L$ and $M_R$, are proportional to the matrix $M_{126}$:
\begin{equation}
M_L = c_L M_{126}, \ \ \ M_R = c_R  M_{126},
\end{equation}
where $M_{10}$ and $M_{126}$ are related to the quark and charged lepton
mass matrices $M_u$, $M_d$, and $M_e$ as follows:
\begin{eqnarray}
M_{10}&=&\frac{3M_d+ M_e}{4},\label{eq081401}\\
M_{126}&=&\frac{M_d-M_e}{4}.\label{eq071701}
\end{eqnarray}
Then the neutrino mass matrix derived form the seesaw mechanism becomes
\begin{eqnarray}
M_\nu&=&M_L-M_DM_R^{-1}M_D^T\nonumber\\
 &=&c_LM_{126} \nonumber\\
 & &-c_R^{-1}(c_{10}M_{10}-3c_{126}M_{126})M_{126}^{-1}(c_{10}M_{10}-3c_{126}M_{126})^T. 
\label{eq071702}
\end{eqnarray}
In the present paper we adopt $c_L=0$.
Also we may ignore the phase of \(c_R\) 
which does not affect the observed values.
Therefore, we can rewrite Eq.(\ref{eq071702}) as
\begin{equation}
|c_R| A_\nu^T D_\nu A_\nu = \widetilde{M}_D \widetilde{M}_1^{-1}
                            \widetilde{M}_D^T, \label{eq081004}
\end{equation}
similarly to Eq.(\ref{eq01080701}), where
\begin{eqnarray}
\widetilde{M}_D&=&c_{10} \widetilde{M}_0- 3 c_{126} \widetilde{M}_1,\\
\widetilde{M}_{10}&=&\frac{1}{4} (3\widetilde{M}_d+ \widetilde{M}_e ),\\
\widetilde{M}_{126}&=&\frac{1}{4} ( \widetilde{M}_d-\widetilde{M}_e ), 
\label{eq0717xx}
\end{eqnarray}
with
\begin{eqnarray}
\widetilde{M}_d &=& U_u^T M_d U_u = V_q D_d V_q^T , \\
\widetilde{M}_e &=& U_u^T M_e U_u = A_e^T D_e A_e 
\nonumber \\
                &=& c_d ( V_q D_d V_q^T +\kappa D_u).
\end{eqnarray}
The reasonable results we found at that time are listed in Table \ref{Table:neutrino}.
The shortening of almost all the minimal SO(10) models is large value of $\theta_{13}$ (see Eq.\bref{Daya-Bay} for the recent result).

\begin{table}[pt]
\caption{The input values of $\tan \beta$, $m_s(M_Z)$
and $\delta$ in the CKM matrix and the outputs for the neutrino
oscillation parameters.
\label{Table:neutrino}}
{\begin{tabular}{c|cc|c|ccc|c}
\hline \hline
 $\tan \beta $ & $m_s(M_Z)$ & $\delta$  & $\sigma $
 & $\sin^2 2 \theta_{1 2}$
 & $\sin^2 2 \theta_{2 3}$
 & $\sin^2 2 \theta_{1 3} $
 & $\Delta m_{\odot}^2/\Delta m_{\oplus}^2$ \\ \hline
40 & 0.0718 & $ 93.6^\circ $ & 3.190& 
0.738 & 0.900 & 0.163 & 0.205 \\
45 & 0.0729 & $ 86.4^\circ $ & 3.198& 
0.723 & 0.895 & 0.164 & 0.188 \\
50 & 0.0747 & $ 77.4^\circ $ & 3.200& 
0.683 & 0.901 & 0.164 & 0.200 \\
55 & 0.0800 & $ 57.6^\circ $ & 3.201& 
0.638 & 0.878 & 0.152 & 0.198 \\
\hline \hline
\end{tabular}}
\end{table}
\begin{figure}
  \includegraphics[height=.3\textheight]{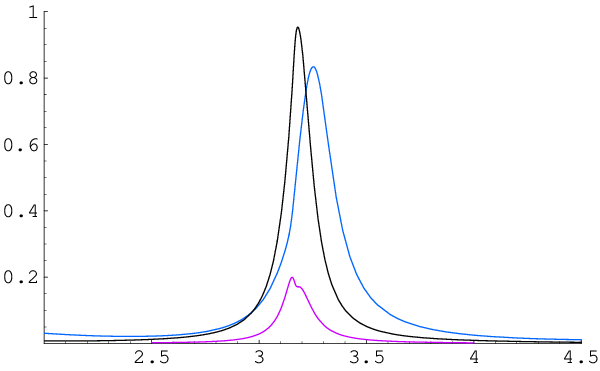}
  \caption{Three mixing angles in the MNS matrix as functions of $\sigma \mbox{[rad]}$. 
The graphs with the highest, middle and lowest peaks
correspond to $\sin^2 2 \theta_{2 3}$, 
$\sin^2 2 \theta_{1 2}$ and $\sin^2 2 \theta_{1 3}$, respectively. 
The plots of $\sin^2 2 \theta_{23}$ and $\sin^2 2 \theta_{13}$ have 
the sharp peaks at $\sigma \sim 3.2 [\mbox{rad}]$, 
while $\sin^2 2 \theta_{12}$ has the sharp peak at 
$\sigma \sim 3.3 [\mbox{rad}]$ cited from \cite{Fukuyama2}.}
\end{figure}

As mentioned above, 
 our resultant neutrino oscillation parameters 
 are sensitive to all the input parameters. 
In other words, if we use the neutrino oscillation data 
 as the input parameters, 
 the other input, for example, the CP-phase in the CKM matrix 
 can be regarded as the prediction of our model. 
It is a very interesting observation 
 that the CP-phases listed above are 
 in the region consistent with experiments.
The CP-violation in the lepton sector
 is characterized by the Jarlskog parameter
 defined as 
\begin{eqnarray}
 J_{CP} = \mbox{Im}\left[ 
     U_{e2} U_{\mu 2}^*  U_{e 3}^*  U_{\mu 3} \right] \; , 
\end{eqnarray}  
where $U_{f i}$ is the MNS matrix element. 

\section{Lepton Flavour Violation and Dipole Moments}

Lepton flavour violation (LFV), anomalous magnetic dipole moment (anomalous MDM), and electric dipole moment (EDM) are discussed in the unified way. These phenomena are very sensitive to New Physics BSM via new CP phases and new particles.

The SM gives negligibly small LFV probability in charged leptons, even taking into account the neutrino oscillation, 
\be
Br(l\rightarrow l'\gamma)\propto \left(\frac{\delta m_\nu^2}{m_W^2}\right)<10^{-54},
\ee
and observation of LFV process becomes a clear signature of BSM.

At tree level in the SM,
the interaction of fermion $\psi$
(of mass $m_\psi$ and electro-magnetic charge $\text{e} Q_\psi$)
with photon is given by
\begin{eqnarray}
- \text{e} Q_\psi \overline{\psi} \gamma^\mu \psi A_\mu
&=&
 - \frac{e Q_\psi}{2m_\psi}\,
   \overline{\psi}
    ( i\partial^\mu - i\overleftarrow{\partial}^\mu )
   \psi\, A_\mu
 - \frac{e Q_\psi}{4m_\psi}\,
   \overline{\psi}
   \sigma^{\mu\nu}
   \psi\, F_{\mu\nu}\\
&=&
 - \frac{e Q_\psi}{2m_\psi}\,
   \overline{\psi}
    ( i\partial^\mu - i\overleftarrow{\partial}^\mu )
   \psi\, A_\mu
 -i\,\frac{e Q_\psi}{2m_\psi}\,
   \overline{\psi}
   \left(
    \begin{array}{cc}
     \vec{\sigma}\cdot\vec{B} & 0\\
     0 & \vec{\sigma}\cdot\vec{B}
    \end{array}
   \right)
   \psi\,.
\label{eq:SM_tree}
\end{eqnarray} 
%
It is clear in eq.~(\ref{eq:SM_tree}) that
a fermion has a MDM with $g = 2$
at tree level in the SM\@.

 In loop level in the SM and/or models BSM,
following effective interaction of gauge invariant form
can be obtained:
\begin{eqnarray}
&&
 -i\overline{\psi}_i
 \left(
  A_L^{ij} P_L + A_R^{ij} P_R
 \right)
 \sigma^{\mu\nu}
 \psi_j
 F_{\mu\nu}\nonumber\\
&&\hspace*{5mm}
=
 \frac{-i}{\,2\,}
 (A_L^{ij} + A_R^{ij})
 \overline{\psi}
 \sigma^{\mu\nu}
 \psi
 F_{\mu\nu}
 +
 \frac{1}{\,2\,}
 (A_R^{ij} - A_L^{ij})
 \overline{\psi}
 \sigma^{\mu\nu} \gamma^5
 \psi
 F_{\mu\nu}\nonumber\\
&&\hspace*{5mm}
=
 (A_L^{ij} + A_R^{ij})
 \overline{\psi}
 \left(
  \begin{array}{cc}
   \vec{\sigma}\cdot\vec{B} & 0\\
   0 & \vec{\sigma}\cdot\vec{B}
  \end{array}
 \right)
 \psi
 +i
 (A_R^{ij} - A_L^{ij})
 \overline{\psi}
 \left(
  \begin{array}{cc}
   0 & \vec{\sigma}\cdot\vec{E}\\
    \vec{\sigma}\cdot\vec{E} & 0
  \end{array}
 \right)
 \psi.
\label{eff_dipole}
\end{eqnarray}
 For the EDM and MDM,
we take zero momentum of the photon.
 Then imaginary part of coefficients
of the effective interaction vanishes
because of the optical theorem
(imaginary part of the forward scattering amplitude
is given by the sum of possible cuts of intermediate states).
 We have an anomalous MDM $a_\psi$
and EDM $d_\psi$ as
\begin{eqnarray}
a_\psi
&=&
 \frac{g-2}{2}
=
 - \frac{2 m_\psi}{ \text{e} Q_\psi }\,
   \text{Re}( A_R^{ii} + A_L^{ii} ),\\
d_\psi
&=&
 2\,\text{Im}( A_R^{ii} - A_L^{ii} ).
\label{imaginary}
\end{eqnarray}
 Note that $A_L$ and $A_R$ must include
a fermion mass ($m_\psi$ or fermion mass in the loop)
because the effective interaction
$\overline{\psi} \sigma^{\mu\nu} \psi$
changes the chirality which can be done
by the mass term in the fundamental Lagrangian.
 If one of particles in the loop is
much heavier than others,
$A_L$ and $A_R$ are suppressed by the mass.
 Thus,
for large $A_L$ and/or $A_R$,
it is preferred that masses of particles in the loop
are similar to each other.

The explicit formulas of $A_{L,R}$ etc. used in our analysis 
 are summarized in \cite{Hisano-etal} \cite{Okada-etal}. 
According to these papers, hereafter we renormalize $A_L$ and $A_R$ as
\be
{\cal L}_{eff}=-i\frac{e}{2}m_{l_i}\overline{l}_j\sigma_{\mu\nu}F^{\mu\nu}(A_L^{ji}P_L+A_R^{ji}P_R)l_i.
\ee
The decay rate of the LFV of charged lepton is given by 
\begin{eqnarray}
\Gamma (\ell_i \rightarrow \ell_j \gamma) 
= \frac{e^2}{16 \pi} m_{\ell_i}^5 
 \left( |A_L^{j i}|^2  +  |A_R^{j i}|^2  \right) \; , 
\end{eqnarray}
while the real diagonal components of $A_{L,R}$ 
 contribute to the anomalous MDMs and EDMs of 
 the charged-leptons such as 
\begin{eqnarray}
 a_{\ell_i}^{\rm SUSY} &=& \frac{g_{\ell_i}-2}{2} 
  = -  m_{\ell_i}^2 
  \mbox{Re} \left[ A_L^{i i}  +  A_R^{i i}  \right]  \\
d_{l_i}/e&=&-m_{l_i}\mbox{Im}(A_L^{ii}-A_R^{ii})
\end{eqnarray}
Let us consider first
an anomalous MDM of muon,
$a_\mu \equiv (g-2)/2$.
 The muon anomalous MDM
has been measured very precisely%
~\cite{Bennett:2006fi} as
\begin{eqnarray}
a_\mu^{\text{exp}} = 11659208.0(6.3)\times 10^{-10},
\label{aMDMexp}
\end{eqnarray}
where the number in parentheses shows $1\sigma$ uncertainty.
 On the other hand,
the SM predicts
\begin{eqnarray}
a_\mu^{\text{SM}}[\tau] &=& 11659193.2(5.2)\times 10^{-10},\\
a_\mu^{\text{SM}}[e^+e^-] &=& 11659177.7(5.1)\times 10^{-10},
\label{aMDMSNM}
\end{eqnarray}
where the hadronic contributions
to $a_\mu^{\text{SM}}[\tau]$ and $a_\mu^{\text{SM}}[e^+e^-]$
were calculated~\cite{Davier:2009ag} by using data of
hadronic $\tau$ decay and $e^+e^-$ annihilation to hadrons,
respectively (see also
\cite{de Troconiz:2004tr,Hagiwara:2006jt,
DEHZ,Jegerlehner:2007xe,Prades:2009qp}).
 The deviations of the SM predictions from the experimental result
are given by
\begin{eqnarray}
\Delta a_\mu[\tau]
&\equiv& a_\mu^{\text{exp}} - a_\mu^{\text{SM}}[\tau]
 = 14.8(8.2)\times 10^{-10},\\
\Delta a_\mu[e^+e^-]
&\equiv& a_\mu^{\text{exp}} - a_\mu^{\text{SM}}[e^+e^-]
 = 30.3(8.1)\times 10^{-10}.
\end{eqnarray}
 These values of $\Delta a_\mu[\tau]$ and $\Delta a_\mu[e^+e^-]$
correspond to $1.8\sigma$ and $3.7\sigma$ deviations
from the SM predictions, respectively.
In order to clarify the parameter dependence 
 of the decay amplitude, 
 we give here an approximate formula of the LFV decay rate 
 \cite{Hisano-etal}, 
\begin{eqnarray}
\Gamma (\ell_i \rightarrow \ell_j \gamma) 
 \sim  \frac{e^2}{16 \pi} m_{\ell_i}^5 
 \times  \frac{\alpha_2}{16 \pi^2} 
 \frac{\left| \left(\Delta  m^2_{\tilde{\ell}} \right)_{ij}\right|^2}{M_S^8} 
 \tan^2 \beta \; , 
 \label{LFVrough}
\end{eqnarray}
where $M_S$ is the average slepton mass at the electroweak scale, 
 and $ \left(\Delta  m^2_{\tilde{\ell}} \right)_{ij}$ 
 is the slepton mass estimated in Eq.~(\ref{leading}). 
We can see that the neutrino Dirac Yukawa coupling matrix 
 plays the crucial role in calculations of the LFV processes. 
We use the neutrino Dirac Yukawa coupling matrix of Eq.~(\ref{Ynu})
 in our numerical calculations. 


These quantities are evaluated by using the outputs 
 presented in Table \ref{Table:neutrino}, 
 and the results are listed in Table \ref{Table:tanb} \cite{F-K-O}\cite{F-I-K}.  
\begin{table}[pt]
\caption{The input values of $\tan \beta$ and the outputs 
for the CP-violating observables}
{\begin{tabular}{c|ccc}
\hline \hline
 $\tan \beta $ & 
 $ \langle m_\nu \rangle_{ee}~ (\mbox{eV})$ & 
 $J_{CP}$ &   $\epsilon$  \\   \hline
40 & 0.00122  & $~~0.00110$ & $ 7.39 \times 10^{-5} $ \\
45 & 0.00118  & $-0.00429$  & $ 6.80 \times 10^{-5} $ \\
50 & 0.00119  & $-0.00631$  & $ 6.50 \times 10^{-5} $ \\
55 & 0.00117  & $-0.00612$  & $ 11.2 \times 10^{-5} $ \\ 
\hline \hline
\end{tabular}}
\label{Table:tanb}
\end{table}

Here some comments on the rate of the LFV processes
and the muon $g-2$ are added.
The evidence of the neutrino flavor mixing implies that 
 the lepton flavor of each generation is not individually conserved. 
Therefore the lepton flavor violating (LFV) processes 
 in the charged lepton sector such as 
 $\mu \rightarrow e \gamma$, $\tau \rightarrow \mu \gamma$ 
 are allowed. 
In simply extended models 
 so as to incorporate massive neutrinos into the SM, 
 the rate of the LFV processes is accompanied 
 by a highly suppression factor, 
 the ratio of neutrino mass to the weak boson mass, 
 because of the GIM mechanism,  
 and is far out of the reach of the experimental detection. 
However, in supersymmetric models, the situation is quite different. 
In this case, soft SUSY breaking parameters can be new LFV sources, 
 and the rate of the LFV processes 
 are suppressed by only the scale 
 of the soft SUSY breaking parameters, which is assumed to be the electroweak scale. 
Thus the huge enhancement occurs compared to the previous case. 
In fact, the LFV processes can be one of the most important processes 
 as the low-energy SUSY search. 
However, this needs another assumption of universal boundary condition of supersymmetry-breaking parameters independent on the GUT framework.
\paragraph{Universal SUSY-breaking}
\bea
{\cal L}_{soft}&=&-\frac{1}{2}\left(M_3\tilde{g}\tilde{g}+M_2\tilde{W}\tilde{W}+M_1\tilde{B}\tilde{B}+c.c.\right)\nonumber\\
&-&\left(\tilde{\overline{u}}{\bf a}_u\tilde{Q}H_u-\tilde{\overline{d}}{\bf a}_d\tilde{Q}H_d-\tilde{\overline{e}}{\bf a}_e\tilde{L}H_d+c.c.\right)\\
&-&\tilde{Q}^\dagger {\bf m}_Q^2\tilde{Q}-\tilde{L}^\dagger {\bf m}_L^2\tilde{L}-\tilde{\overline{u}}^\dagger {\bf m}_u^2\tilde{\overline{u}}^\dagger-\tilde{\overline{d}}{\bf m}_d^2\tilde{\overline{d}}^\dagger-\tilde{\overline{e}}{\bf m}_e^2\tilde{\overline{e}}^\dagger\nonumber\\
&-&m_{H_u}^2H_u^*H_u-m_{H_d}^2H_d^*H_d-\left(bH_uH_d+c.c.\right).\nonumber
\eea
So generically we have 19 parameters (3 gaugino masses + tan$\beta$ + $\mu$ + $m_A$ + 10 sfermion masses + 3 trilinear terms), called phenomenological MSSM (pMSSM).
Universal SUSY-breaking is a very strong assumpotion that it requires not only flavour-blindness but also universality over quarks and leptons at GUT scale,
\bea
&&{\bf m}_Q^2={\bf m}_{\overline{u}}^2={\bf m}_{\overline{d}}^2={\bf m}_{\overline{L}}^2={\bf m}_{\overline{e}}^2=m_0^2{\bf 1}_3,\label{USB1}\\
&&m_{H_u}=m_{H_d}=m_0\label{NUHM}\\
&&\frac{M_3}{g_3^2}=\frac{M_2}{g_2^2}=\frac{M_1}{g_1^2}=\frac{M_{1/2}}{g_u^2}\label{USB2}\\
&&{\bf a}_u=A_0{\bf Y}_u,~~{\bf a}_d=A_0{\bf Y}_d~~{\bf a}_u=A_0{\bf Y}_e.\label{USB3}
\eea
This MSSM+universal SUSY breaking is called constrained MSSM (CMSSM) \cite{Martin}.
If, in place of \bref{NUHM}, we set $m_{H_u},~m_{H_d}$ as free parameters, it is called non-universal Higgs masses (NUHM2) model \cite{E-O-S} or called NUHM1 for $m_{H_u}=\pm m_{H_d}$ \cite{Baer}.
 From the SO(10) view point, there is some reason to set the universality between squarks and sleptons at GUT scale since all matters belong to a single {\bf 16} but there is no definite reason to extend it to Higgs masses. (We will argue on the problems on CMSSM or its alternatives at the last part of this review in connection with Higgs-like boson around 125 GeV discovered at the LHC.). Hereafter, we will discuss in the CMSSM framework.
We evaluate the rate of the LFV processes 
 in the minimal SUSY SO(10) model, 
 where the neutrino Dirac Yukawa couplings 
 are the primary LFV sources. 
Although in Ref.~\cite{Fukuyama2} 
 various cases with given $\tan \beta = 40-55$ have been analyzed, 
 we consider only the case $\tan \beta =45$ in the following. 
Our final result is almost insensitive 
 to $\tan \beta$ values in the above range. 
The predictions of the minimal SUSY SO(10) model 
 necessary for the LFV processes are as follows \cite{Fukuyama2}: 
 with $\sigma=3.198$ fixed, 
 the right-handed Majorana neutrino mass eigenvalues 
 are found to be (in GeV) 
\be
 M_{R_1}=1.64 \times 10^{11},~  
 M_{R_2}=2.50 \times 10^{12}~\mbox{and}~ 
 M_{R_3}=8.22 \times 10^{12}, 
\label{M_R}
\ee
 where $c_R$ is fixed so that 
 $\Delta m_\oplus^2 = 2 \times 10^{-3} \mbox{eV}^2$. 
In the basis where both of the charged-lepton 
 and right-handed Majorana neutrino mass matrices 
 are diagonal with real and positive eigenvalues, 
 the neutrino Dirac Yukawa coupling matrix at the GUT scale 
 is found to be \footnote{We are now reconsidering data fitting with
the update experimental data and new RGE results. It gives the differen values from \bref{Ynu} but the LFV results are not essentially changed.}
\begin{eqnarray}
 Y_{\nu} = 
\left( 
 \begin{array}{ccc}
-0.000135 - 0.00273 i & 0.00113  + 0.0136 i  & 0.0339   + 0.0580 i  \\ 
 0.00759  + 0.0119 i  & -0.0270   - 0.00419  i  & -0.272    - 0.175   i  \\ 
-0.0280   + 0.00397 i & 0.0635   - 0.0119 i  &  0.491  - 0.526 i 
 \end{array}   \right) \; .  
\label{Ynu}
\end{eqnarray}     
LFV effect most directly emerges 
 in the left-handed slepton mass matrix 
 through the RGEs such as \cite{Hisano-etal}
\begin{eqnarray}
\mu \frac{d}{d \mu} 
  \left( m^2_{\tilde{\ell}} \right)_{ij}
&=&  \mu \frac{d}{d \mu} 
  \left( m^2_{\tilde{\ell}} \right)_{ij} \Big|_{\mbox{MSSM}} 
 \nonumber \\
&+& \frac{1}{16 \pi^2} 
\left( m^2_{\tilde{\ell}} Y_{\nu}^{\dagger} Y_{\nu}
 + Y_{\nu}^{\dagger} Y_{\nu} m^2_{\tilde{\ell}} 
 + 2  Y_{\nu}^{\dagger} m^2_{\tilde{\nu}} Y_{\nu}
 + 2 m_{H_u}^2 Y_{\nu}^{\dagger} Y_{\nu} 
 + 2  A_{\nu}^{\dagger} A_{\nu} \right)_{ij}  \; ,
 \label{RGE} 
\nonumber\\
\end{eqnarray}
where the first term in the right-hand side denotes 
 the normal MSSM term with no LFV. 
We have found $Y_\nu$ explicitly and we can calculate LFV and
 related phenomena unambiguously \cite{Fukuyama2}. In the leading-logarithmic approximation, 
 the off-diagonal components ($i \neq j$)
 of the left-handed slepton mass matrix are estimated as 
\begin{eqnarray}
 \left(\Delta  m^2_{\tilde{\ell}} \right)_{ij}
 \sim - \frac{3 m_0^2 + A_0^2}{8 \pi^2} 
 \left( Y_{\nu}^{\dagger} L Y_{\nu} \right)_{ij} \; ,  
 \label{leading}
\end{eqnarray}
where the distinct thresholds of the right-handed 
 Majorana neutrinos are taken into account 
 by the matrix $ L = \log [M_{\rm GUT}/M_{R_i}] \delta_{ij}$.

\begin{figure}[t]
\includegraphics[scale=0.7]{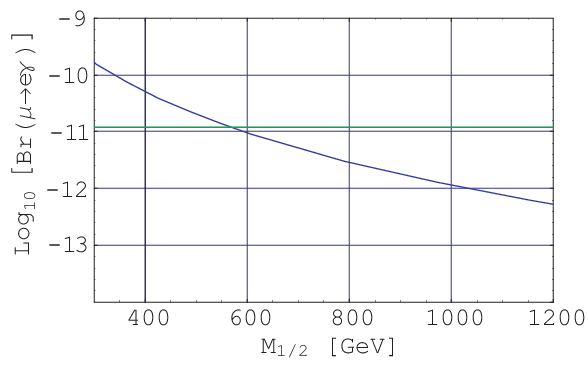}
\includegraphics[scale=0.7]{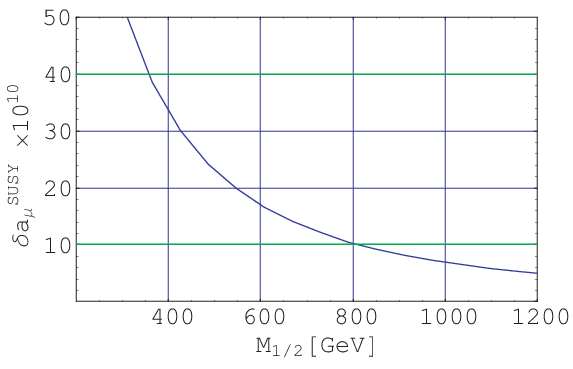}
\includegraphics[scale=0.7]{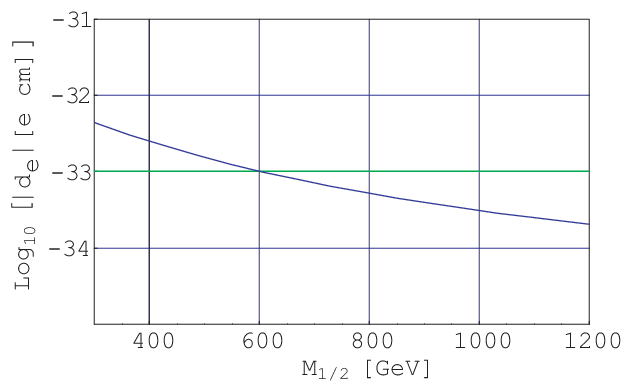}\label{Fig2c}
\hspace{2.0cm}
\includegraphics[scale=0.7]{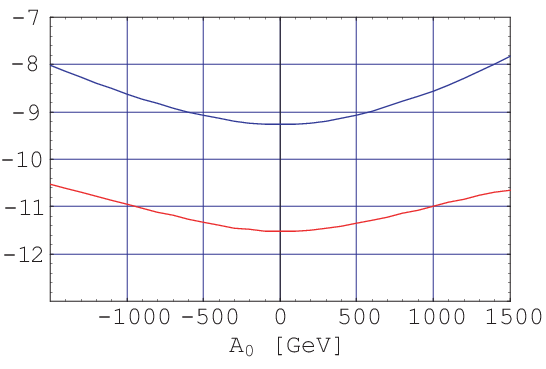}
\vspace*{8pt}
\caption{
The branching ratio, a:
$\mbox{Log}_{10} \left[{\rm Br} (\mu \rightarrow e \gamma) \right]$, b:
the SUSY contribution to the muon $g-2$ in units of $10^{-10}$, 
$\delta a_{\ell_i}^{\rm SUSY} = \frac{g_{\ell_i}-2}{2}$,
and c:
the electron EDM, 
$\mbox{Log}_{10} \left[ | d_e | [ \mbox{e cm}] \right]$. 
All these figures are plotted as a function of $M_{1/2}$ (GeV) 
 along the cosmological constraint of Eq.~(\ref{relation}). Trilinear term $A_0$ is assumed to be zero except for the last panel. The last panel is added for the reference to see the behaviour of non zero $A_0$, where
the branching ratios,  
$\mbox{Br}(\tau \rightarrow \mu \gamma)$ (top)
and 
$\mbox{Br}(\mu \rightarrow e \gamma)$ (bottom) are given
 as functions of $A_0$ (GeV) 
 for $m_0 =600$ GeV and $M_{1/2}=800$ GeV.  All are cited from \protect\cite{F-K-O}.
\label{Fig2a}}
\end{figure}

The recent Wilkinson Microwave Anisotropy Probe (WMAP) satellite data 
 \cite{WMAP}  
 provide estimations of various cosmological parameters 
 with greater accuracy. 
The current density of the universe is composed of 
 about 73\% of dark energy and 27\% of matter. 
Most of the matter density is in the form of 
 the CDM, and its density is estimated to be (in 2$\sigma$ range) 
\begin{eqnarray}
\Omega_{\rm CDM} h^2 = 0.1126^{+0.0161}_{-0.0181} \; . 
 \label{WMAP} 
\end{eqnarray}
The parameter space of the CMSSM 
 which allows the neutralino relic density 
 suitable for the cold dark matter 
 has been recently re-examined 
 in the light of the WMAP data \cite{CDM}. 
It has been shown that the resultant parameter space 
 is dramatically reduced into the narrow stripe 
 due to the great accuracy of the WMAP data. 
It is interesting to combine this result 
 with our analysis of the LFV processes and the muon $g-2$. 
In the case relevant to our analysis, 
 $\tan \beta=45$, $\mu>0$ and $A_0=0$, 
 we can read off the approximate relation 
 between $m_0$ and $M_{1/2}$ 
 such as (see Figure 1 in the second paper of Ref.~\cite{CDM}) 
\begin{eqnarray}
m_0(\mbox{GeV}) = \frac{9}{28} M_{1/2}(\mbox{GeV}) + 150(\mbox{GeV}) \;,  
 \label{relation} 
\end{eqnarray} 
along which the neutralino CDM is realized. 
$M_{1/2}$ parameter space is constrained within the range 
 $300 \mbox{GeV} \leq M_{1/2} \leq 1000 \mbox{GeV}$ 
 due to the experimental bound on the SUSY contribution 
 to the $ b \rightarrow s \gamma$ branching ratio 
 and the unwanted stau LSP parameter region. 
We show $\mbox{Br}(\mu \rightarrow e \gamma)$ and 
 the muon $g-2$ as functions of $M_{1/2}$ 
 in Fig.~\ref{Fig2a} 
 along the neutralino CDM condition of Eq.~(\ref{relation}). 
We find the parameter region, 
 $560 \mbox{GeV} \leq M_{1/2} \leq 800 \mbox{GeV}$, 
 being consistent with all the experimental data. 

%
%

The semileptonic flavor violation processes were also considered
in \cite{F-I-K}, for instance, for 
$\tau^- \to e^- (\mu^-)\, \pi^0$,
$\tau^- \to e^- (\mu^-)\, \eta$,
$\tau^- \to e^- (\mu^-)\, \eta^\prime$,
$\tau^- \to e^- (\mu^-)\, \rho^0$,
$\tau^- \to e^- (\mu^-)\, \phi$,
$\tau^- \to e^- (\mu^-)\, \omega$, {\it etc.}

Similarly we can estimate EDM from \bref{imaginary}.
Electron EDM is depicted also in Fig.\ref{Fig2a}.

When the KamLAND data \cite{Eguchi:2002dm} was released, 
 the results in \cite{Fukuyama2} were found to be deviated 
 by 3$\sigma$ from the observations. 
Afterward this minimal SO(10) was modified by many authors, 
 using the so-called type-II seesaw mechanism \cite{Bajc:2002iw} 
 and/or considering a ${\bf 120}$ Higgs coupling to the matter
 in addition to the ${\bf \overline{126}}$ Higgs \cite{Goh:2003hf}. 
Based on an elaborate input data scan \cite{Babu:2005ia}, 
\cite{Bertolini:2006pe},
 it has been shown that the minimal SO(10) is essentially consistent 
 with low energy data of fermion masses and mixing angles. \footnote{See \bref{MEGsignal2} for the recent result of $\mu\rightarrow e\gamma$.}
The importance of the threshold
 corrections was also discussed in \cite{Parida}
\section{Leptogenesis \label{leptogenesis}}

Cosmic baryon asymmetry is one of the most important subjects for new physics BSM. Sakharov pointed out \cite{Sakharov} that we need three conditions for baryogenesis: \\
1. Bayon number violation, \\
2. CP violation,\\
3. Out of equilibrium condition.

As we emphasized in the previous section, CP violation process in the SM is parametrized by the Jarlskog parameter. Its magnitude is too small to generate the observed baryon asymmetry $Y_{\Delta B}={\cal O}(10^{-10})$.

One of the reasons that we adopted $\overline{{\bf 126}}$ was that it includes $({\bf 10},{\bf 1},{\bf 3})$ which generates heavy right-handed Majorana mass term and induces $\Delta (B-L)=2$ with additional CP-violating (CPV) phases.

The minimal SO(10) GUT has many scalar fields and many CPV phases, heavy right-handed neutrino $N_R$ is not the unique parent for baryon asymmetry \footnote{Indeed, there are alternative approaches \cite{Babu-Mohapatra}.}. However, it seems to be most natural to accept  leptogenesis  via $N_R$  \cite{Fuku-Yanagida}.
Supersymmetry requires the leptogenesis by the decays of both the lightest heavy right-handed neutrino $Y_{L_f}$ and sneutrino $Y_{L_s}$ equal and total lepton asymmetry is \cite{plumacher}
\be
Y_L=Y_{L_f}+Y_{L_s}.
\ee
The processes of neutrino and sneutrino decays are essentially same and sneutrino case will be discussed at section \ref{leptogenesis5D}. 
\paragraph{Thermal leptogenesis}
$N_R$ has a large mass and no gauge interaction, and is out of equilibrium.
The effective Lagrangian at energies lower than the right-handed neutrino masses is
\begin{eqnarray}
{\cal L}_{\rm eff}=- 
\int d^2 \theta 
\left(Y_\nu^{ij} N_i^c L_j H_u
+\frac{1}{2} \sum_{i} M_{Ri} N^c_i N_i^c \right)
+ h.c. \; , 
\end{eqnarray} 
where $i,j=1,2,3$ denote the generation indices and $Y_\nu$ the Yukawa coupling, $L$ and $H_u$ are 
the lepton and the Higgs doublets chiral supermultiplets, respectively, 
and $M_{Ri}$ is the lepton-number-violating mass term of the right-handed 
neutrino $N_i$. The peculiar properties of the minimal SO(10) are that we 
can fix $Y_\nu^{ij}$ and $M_{Ri}$ unambiguously 
from the low-energy phenomenologies of quarks and leptons.
The lepton asymmetry in the universe is generated by CP-violating 
out-of-equilibrium decay of the heavy neutrinos,  
$N \rightarrow \ell_L H_u^*$ and $N \rightarrow \overline{\ell_L} H_u$. 
The leading contribution is given by the interference between 
the tree level and one-loop level decay amplitudes, 
and the CP-violating parameter is found to be \cite{epsilon}
\begin{eqnarray}
\epsilon = 
\frac{1}{8 \pi (Y_\nu Y_\nu^\dag)_{11}}
\sum_{j=2,3}\mbox{Im} \left[ (Y_\nu Y_\nu^\dag)_{1j}^2 \right]
\left\{ f(M_{Rj}^2/M_{R1}^2)
+ 2 g(M_{Rj}^2/M_{R1}^2) \right\} \; .
\label{epsilon}
\end{eqnarray}
Here $f(x)$ and $g(x)$ correspond to 
the vertex and the wave function corrections, 
\begin{eqnarray}
f(x)&\equiv& \sqrt{x} \left[
1-(1+x)\mbox{ln} \left(\frac{1+x}{x} \right) \right] \;,
\nonumber\\
g(x)&\equiv& \frac{\sqrt{x}}{2(1-x)}   \; ,  
\end{eqnarray}
respectively, and both are reduced to 
$\sim -\frac{1}{2 \sqrt{x}}$ for $ x \gg 1$. 
So, in this approximation, $\epsilon$ becomes 
\begin{equation}
\epsilon = - 
\frac{3}{16 \pi (Y_\nu Y_\nu^\dag)_{11}}
\sum_{j=2,3} \mbox{Im} \left[(Y_\nu Y_\nu^\dag)_{1j}^2 \right]
\frac{M_{R1}}{M_{Rj}}\;.
\label{epsilon}
\end{equation}
Using the mass of neutrino via Type-I see-saw mechanism, 
$M_\nu = - Y_\nu^T M_R^{-1} Y_\nu \left<H_u \right>^2$, 
$\epsilon$ is further written as \cite{b-y}
\begin{eqnarray}
\epsilon &=& 
\frac{3}{16 \pi} \frac{M_{R1}}{\left<H_u \right>^2}
\frac{\mbox{Im}\left[(Y_\nu M_\nu^* Y_\nu^T)_{11} \right]}
{(Y_\nu Y_\nu^\dag)_{11}}
\nonumber\\
&\equiv& 
\frac{3}{16 \pi}\frac{m_{\nu 3} M_{R1} 
\delta_{\rm eff}}{\left<H_u \right>^2} \;. 
\label{epsilon2}
\end{eqnarray}
In the minimal SO(10) model we have the definite form of $Y_\nu$ and 
estimate these values unambiguously.
We have assumed that the lightest $N_1$ decay dominantly contributes 
to the resultant lepton asymmetry. In fact, this is confirmed by numerical 
analysis in the case of hierarchical right-handed neutrino masses 
\cite{plumacher}. Using the above $\epsilon$, 
generated $Y_{B}$ is described as
\begin{eqnarray}
Y_{B}  \sim  \frac{\epsilon}{g_*}  d \; , 
\end{eqnarray}
where $g_* \sim 100$ is the effective degrees of freedom 
in the universe at $T \sim M_{R1}$, and $ d \leq 1 $ 
is so-called the dilution factor. 
This factor parameterizes how the naively expected value 
$Y_B \sim \epsilon/g_*$ is reduced by washing-out processes. 
We can classify the washing-out processes into two cases with 
and without the external leg of the heavy right-handed neutrinos, 
respectively. The former includes the inverse-decay process 
and the lepton-number-violating scatterings 
mediated by the Higgs boson \cite{luty} such as 
$N + \overline{\ell_L} \leftrightarrow \overline{q_R} + q_L$,  
where $q_L$ and $q_R$ are quark doublet and singlet, respectively. 
The latter case is the one induced 
by the effective dimension five interaction, 
\begin{eqnarray} 
{\cal L}_N = \frac{1}{2} \left(Y_\nu^T M_R^{-1} Y_\nu \right)_{ij}
(L_i H_u)^{T} C^{-1} (L_j H_u) \; , 
\label{4point}
\end{eqnarray}
after integrating out the heavy right-handed neutrinos. 
This term is nothing but the one 
providing the see-saw mechanism \cite{seesaw}. 
The importance of this interaction was discussed in \cite{kolb},  
where the interaction was shown to be necessary 
to avoid the false  generation of the lepton asymmetry 
in thermal equilibrium. While numerical calculations 
\cite{plumacher} \cite{luty} are necessary in order to evaluate 
the dilution factor precisely, $Y_B \sim \epsilon/g_{*}$ roughly 
gives a correct answer, and the washing-out process is mostly not 
so effective. Note that this is the consequence from the current neutrino 
oscillation data as explained in \cite{Kolb}. 

These quantities are evaluated by using the outputs 
 presented in Table \ref{Table:neutrino}, 
 and the results are listed in Table \ref{Table:tanb} \cite{F-K-O}\cite{F-I-K}.  

\begin{table}[pt]
\caption{The input values of $\tan \beta$ and the outputs 
for the CPV observables}
{\begin{tabular}{c|ccc}
\hline \hline
 $\tan \beta $ & 
 $ \langle m_\nu \rangle_{ee}~ (\mbox{eV})$ & 
 $J_{CP}$ &   $\epsilon$  \\   \hline
40 & 0.00122  & $~~0.00110$ & $ 7.39 \times 10^{-5} $ \\
45 & 0.00118  & $-0.00429$  & $ 6.80 \times 10^{-5} $ \\
50 & 0.00119  & $-0.00631$  & $ 6.50 \times 10^{-5} $ \\
55 & 0.00117  & $-0.00612$  & $ 11.2 \times 10^{-5} $ \\ 
\hline \hline
\end{tabular}}
\label{Table:tanb}
\end{table}
$\la m_\nu\ra_{ee}$ is the averaged neutrino mass appearing in the $0\nu\beta\beta$ process.
Unfortunately, the CP parameter $\epsilon$ is too large 
to be consistent with the observed baryon asymmetry. In order to 
circumvent this trouble we made use of another pair of SU(2) doublets 
appearing in the minimal SO(10) model. We solved the Boltzman equation and 
obtained the consistent $Y_B$ \cite{okada2}. However, in this case 
we need the extra Higgs other than those in the MSSM, which may raise 
the other problems. So it is deserved to consider an alternative 
solution to this overproduction. 
On the other hand the gravitino problem forces us low reheating temperature 
less than the mass of $M_{R1}$. If we believe it, the above problem becomes 
fake since thermal $N_R$ are not generated in the reheating era. 
So the minimal SO(10) model itself drives us to the other approaches 
such as non-thermal leptogenesis scenario \cite{non-thermal} 
or the Affleck-Dine mechanism \cite{Affleck-Dine}. 
In the next section, we discuss on the non-thermal leptogenesis 
scenario in the minimal SO(10) model. 
\paragraph{Non-Thermal leptogenesis}
Now we turn to the discussions of the non-thermal leptogenesis scenario 
\cite{non-thermal}. In the non-thermal leptogenesis scenario, 
the right-handed neutrinos are produced through the direct 
non-thermal decay of the inflaton. 

Here we give a concrete model to specify the inflaton. 
We add a singlet chiral supermultiplet which plays a role of inflaton $S$
The interaction Lagrangian relevant to the inflaton 
and the right-handed neutrinos is given by 
\be
{\cal L}_S = - \frac{1}{2} \int d^2 \theta
\left(M_I S^2 +\sum_i \lambda_i S N^c_i N^c_i \right)\;.
\label{int}
\ee
When inflaton gets a VEV, it gives rise to the Majorana masses for 
the right-handed neutrinos in addition to the VEV of 
$({\bf 10,1,3})$ in $\overline{\bf 126}$ 
under $SU(4)_{PS} \times SU(2)_L \times SU(2)_R$ \cite{Fukuyama1, Fukuyama2}. 
However, the VEV $\left<S \right>$ is posted around the GUT scale and 
$\lambda_i$ is found to be $10^{-8}$ later, and this contribution gives 
a tiny correction to $M_R$. Also the first term in Eq.~(\ref{int}) 
dominates over the second, and is reduced to the chaotic inflationary 
model \cite{chaotic}. 

In such a superpotential, the inflaton decay rate is given by
\be
\Gamma(S \to N_i N_i) \simeq \frac{|\lambda_i|^2}{4 \pi} M_I \;.
\label{decay}
\ee
Then the consequently produced reheating temperature is obtained by
\be
T_R \ =\ \left(\frac{45}{2 \pi^2 g_*} \right)^{1/4}
(\Gamma M_P)^{1/2} \;.
\label{reheat}
\ee
If the inflaton dominantly couples to $N$, the branching ratio of this 
decay process is, of course, ${\rm BR} \sim 1$. 
Then the produced baryon asymmetry of the universe can be calculated 
by using the following formula, 
\bea
\left(\frac{n_B}{s} \right)
&=& - 0.35 \times \left(\frac{n_{N_1}}{s} \right)
\times \left(\frac{n_L}{n_{N_1}} \right)
\nonumber\\
&=& - 0.35 \times \frac{3}{2} ~{\rm BR}(S \to N_1 N_1)
\left(\frac{T_R}{M_I} \right) \times \epsilon\;.
\label{YB}
\eea
With the hierarchical mass spectra for the right-handed neutrinos, 
it can be approximated as 
\be
\left(\frac{n_B}{s} \right)
= - 1.95 \times 10^{-10} \times {\rm BR} \times
\left(\frac{T_R}{10^6~{\rm GeV}} \right)
\left(\frac{M_{R1}}{M_I}\right) 
\left(\frac{m_{\nu 3}}{0.065~{\rm eV}}\right) \times 
\delta_{\rm eff} \;,
\label{YB2}
\ee
where $\delta_{\rm eff} \equiv 
\mbox{Im}\left[(Y_\nu M_\nu^* Y_\nu^T)_{11} \right]/
\left[m_{\nu 3} (Y_\nu Y_\nu^\dag)_{11} \right]$ 
denotes the effective value of the CP violating phase parameter 
relevant to the leptogenesis and it can be estimated as 
$\delta_{\rm eff} = - 0.166$ in our model. 
As it can easily be seen that it is possible to produce the baryon asymmetry 
of the universe by using the reheating temperature as low as, 
$T_R \lesssim 10^6 ~{\rm GeV}$. Hence, a very wide range of the gravitino mass 
can be allowed, $m_{3/2} \gtrsim 10^{6} ~{\rm MeV}$. 
The result of the detailed numerical calculation based on Eq.~(\ref{YB}) 
is shown in Figs.2.4 and 2.5 \cite{nonthermal}. 
%

\begin{figure}[h]
\begin{center}
{\includegraphics*[width=.9\linewidth]{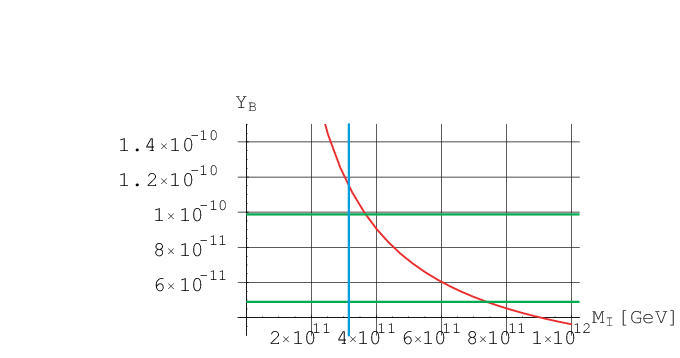}
\label{Fig1a}}
\caption{
The predicted baryon asymmetry of the universe $Y_B = n_B/s$ 
as a function of the inflaton mass $M_I ~\mbox{[GeV]}$ with the reheating 
temperature $T_R = 10^6 ~{\rm GeV}$ and ${\rm BR} = 1$. The vertical line represents the kinematical cut for the inflaton to 
have enough energy to decay, $E \geq 2 M_{R1}$, {\it i.e.}, 
the right side of this line is allowed from the kinematics. 
Two horizontal lines represent the upper and the lower bounds on 
the observed value of the baryon asymmetry at 95 \% C.L.  The lepton asymmetry parameter $\epsilon$ has been taken from 
\cite{Fukuyama2}.}
\end{center}
\end{figure}
\begin{figure}[h]
\begin{center}
{\includegraphics[width=.8\linewidth]{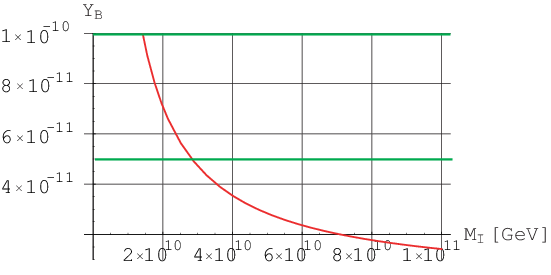}
\label{Fig1b}}
\caption{The same diagram as Fig.2.4 except for that the lepton asymmetry parameter $\epsilon$ has been taken from 
\cite{mohapatra}}
\end{center}
\end{figure}
As shown in Fig.2.4 that the predicted inflaton mass is 
heavier than the lightest right-handed neutrino 
($M_{R1} = 1.6 \times 10^{11}$ GeV) in our model \cite{Fukuyama2}. 
Hence the non-thermal leptogenesis is well workable. But in model 
\cite{mohapatra}, you can see from Fig.2.5 that 
the calculated inflaton mass is lighter than the lightest right-handed 
neutrino mass ($M_{R1} = 2.7 \times 10^{13}$ GeV), and the non-thermal 
leptogenesis scenario is prohibited by the kinematics. 
We hasten to add that this conclusion is valid under the non-thermal 
leptogenesis under the gravity mediated SUSY breaking scenario. 

It can be read from Fig.2.4 that the observed value of 
the baryon asymmetry leads to the inflaton mass around 
$M_I \sim 5 \times 10^{11}$ GeV. 
This corresponds to the coupling constant of the inflaton to 
the right-handed neutrinos as $\lambda_i \sim 10^{-8}$. 
Such a small coupling indicates that the model can naturally fit into
the chaotic inflationary model \cite{chaotic} 
based on a minimal supersymmetric SO(10) model. 

\section{Higgs Superpotential-\\
Symmetry Breaking Flows from GUT to the SM}
On the other hand, it has been long expected to construct 
 a concrete Higgs sector of the minimal SO(10) model. 
\paragraph{The simplest Higgs superpotential}
The simplest Higgs superpotential at the renormalizable level
is given by \cite{clark}, \cite{lee}, \cite{aulakh}
\be
W=m_1 \Phi^2 + m_2 \Delta \overline{\Delta} 
+m_3 H^2
+\lambda_1 \Phi^3 + \lambda_2 \Phi \Delta \overline{\Delta}
+\lambda_3 \Phi \Delta H + \lambda_4 \Phi \overline{\Delta} H \;,
\label{lee}
\ee
where $\Phi ={\bf 210}$, $\Delta ={\bf 126}$, 
$\overline{\Delta} ={\bf \overline{126}}$ and $H={\bf 10}$.
The interactions of ${\bf 210}$, ${\bf \overline{126}}$, 
${\bf 126}$ and ${\bf 10}$ lead to some complexities 
in decomposing the GUT representations to the MSSM 
and in getting the low energy mass spectra.  
Particularly, the CG coefficients 
corresponding to the decompositions of 
${\rm SO}(10) \to 
{\rm SU}(3)_C \times {\rm SU}(2)_L \times {\rm U}(1)_Y$ 
have to be found.
This problem was first attacked by X.~G.~He and S.~Meljanac
\cite{he} and further by D.~G.~Lee \cite{lee} and by J.~Sato \cite{Joe}. 
But they did not present the explicit form of mass matrices for 
a variety of Higgs fields and also did not perform a formulation 
of the proton life time analysis.  This is very laborious work and it is indispensable for the data fit of low energy physics.
We completed this program in \cite{Fukuyama:2004xs} (see also \cite{Bajc:2004xe},\cite{Aulakh:2004hm}).
This construction gives some constraints among the vacuum expectation
 values (VEVs) of several Higgs multiplets, 
 which gives rise to a trouble in the gauge coupling unification 
 \cite{Bertolini}. 
The trouble comes from the fact that the observed neutrino oscillation
 data suggests the right-handed neutrino mass around $10^{13-14}$ GeV,
 which is far below the GUT scale. Indeed \bref{lee} contains 
five directions which are singlets 
under $SU(3)_C \times SU(2)_L \times U(1)_Y$.  
Three of them are included in {\bf 210}, 
\bea
\hat{\phi}_1 &=& (1234), \nonumber
\\
\hat{\phi}_2 &=& (5678+5690+7890), 
\\
\label{210vacua}
\hat{\phi}_3 &=& (12+34)(56+78+90).  \nonumber
\eea
one in {\bf 126} and 
\be
\hat{v}_R = (13579), 
\label{126vec}
\ee
and one in ${\bf \overline{126}}$
\be
\hat{\overline{v_R}} = (24680). 
\label{126barvec} 
\ee

Due to the D-flatness condition the VEVs 
$v_R$ and $\overline{v_R}$ are equal, 
\be
v_R= \overline{v_R}. 
\ee

This intermediate scale is provided by Higgs field VEV, 
 and several Higgs multiplets are expected to have their masses 
 around the intermediate scale and contribute to 
 the running of the gauge couplings. 

We write down the VEV conditions which preserve supersymmetry, 
with respect to the directions 
$\hat{\phi}_1$, $\hat{\phi}_2$, $\hat{\phi}_3$, and $\hat{v}_R$, 
respectively.  
\bea
\label{VEV1}
2 m_1 \phi_1 + 3 \lambda_1 \frac{ \phi_3^2}{6 \sqrt{6}} 
+ \lambda_2 \frac{v_R^2}{10 \sqrt{6}} &=& 0,
\\
\label{VEV2}
2 m_1 \phi_2 + 3 \lambda_1 \left(\frac{\phi_2^2 +\phi_3^2}
{9 \sqrt{2}} \right)+ \lambda_2 \frac{v_R^2}{10 \sqrt{2}} &=& 0,
\\
\label{VEV3}
2 m_1 \phi_3 + 3 \lambda_1 \left(\frac{\phi_1 \phi_3}{3 \sqrt{6}}
+\frac{\sqrt{2} \phi_2 \phi_3}{9}\right)
+ \lambda_2 \frac{v_R^2}{10} &=& 0,
\\
\label{VEV4}
m_2 + \lambda_2 \left( \frac{\phi_1}{10 \sqrt{6}}
+ \frac{\phi_2}{10 \sqrt{2}}
+ \frac{\phi_3}{10} \right)&=& 0.
\label{VEV}
\eea

Eliminating $v_R^2$, $\phi_1$ and $\phi_3$ from 
Eqs. (\ref{VEV1})--(\ref{VEV4}), 
one obtains a fourth-order equation in $\phi_2$.  
The corresponding fourth-order polynomial in $\phi_2$ 
factorizes into a linear and a cubic term 
in $\phi_2$.  
Linear term gives the solution of the fourth-order equation 
which is very simple, $\phi_2 = -3 \sqrt{2} \, m_2/\lambda_2$, 
but it preserves the $SU(5)$ symmetry.  
Therefore, it is physically not 
interesting.  The cubic term solutions lead to the true 
$SU(3)_C \times SU(2)_L \times U(1)_Y$ symmetry.  
Here we consider only the solutions with $|v_R| \neq 0$.
Eliminating $v_R \cdot \overline{v_R}$, 
$\phi_1$ and $\phi_2$ from 
Eqs. (\ref{VEV1})--(\ref{VEV4}), 
one obtains a fourth-order equation in $\phi_3$,  
\be
\label{phi3eq}
\left(\phi_3 + \frac{{\cal M}_2}{10} \right)
\left\{8 \, \phi_3^3 - 15 \, {\cal M}_1 \phi_3^2 
+ 14 \, {\cal M}_1^2 \phi_3 - 3 \, {\cal M}_1 ^3 
+ \left(\phi_3 - {\cal M}_1 \right)^2 {\cal M}_2 \right\} 
= 0,  
\ee
where 
\be
{\cal M}_1 \equiv 12 \, \left(\frac{m_1}{\lambda_1} \right), \, 
{\cal M}_2 \equiv 60 \, \left(\frac{m_2}{\lambda_2} \right).  
\ee
Any solution of the cubic equation in $\phi_3$ is 
accompanied by the solutions 
\bea
\phi_1 &=& - \frac{\phi_3}{\sqrt{6}} 
\frac{\left({\cal M}_1^2 - 5 \, \phi_3^2 \right)}{({\cal M}_1 - \phi_3)^2}, 
\label{vev1}
\\
\phi_2 &=& - \frac{1}{\sqrt{2}} 
\frac{\left({\cal M}_1^2 - 2 \, {\cal M}_1 \phi_3 - \phi_3^2 \right)}{({\cal M}_1 - \phi_3)}, 
\label{vev2}
\\
v_R \cdot \overline{v_R} &=&  \frac{5}{3} \, 
\left(\frac{\lambda_1}{\lambda_2} \right) 
\frac{\phi_3 \left({\cal M}_1 - 3 \, \phi_3 \right) \left({\cal M}_1^2 + \phi_3^2 \right)}
{({\cal M}_1 - \phi_3)^2}.  
\label{nuR}
\eea
The linear term gives the solution of the fourth-order equation (\ref{phi3eq})
which is very simple, 
$\phi_3 = - 6 \, \left(\frac{m_2}{\lambda_2} \right)$.
It leads to $\phi_1 = -\sqrt{6} \, \left(\frac{m_2}{\lambda_2} \right)$, 
$\phi_2 = -3 \sqrt{2} \, \left(\frac{m_2}{\lambda_2} \right)$    and 
$\sqrt{\left( v_R \cdot \overline{v_R} \right)} = \sqrt{60} \, 
\left(\frac{m_2}{\lambda_2} \right) \sqrt{2 \left(\frac{m_1}{m_2} \right) 
- 3 \left(\frac{\lambda_1}{\lambda_2} \right)}$.  
This solution preserves the $SU(5)$ symmetry.  
Therefore, it is physically not interesting.  
Then we proceed to the most important part of the SO(10) GUT.
We can not show the details of the scenario but only show the essential part of it \cite{Fukuyama:2004xs}. 
\paragraph{Would-be Nambu-Goldstone bosons}

At first, we list the quantum numbers of 
the would-be Nambu-Goldstone (NG) modes under $SU(3)_C \times SU(2)_L \times U(1)_Y$.  

\begin{itemize}
\item
$
\left[
\left({\bf \overline{3}, 2},\frac{5}{6} \right)
\oplus
\left({\bf 3, 2}, -\frac{5}{6} \right)
\right],
$
\item 
$
\left[
\left({\bf \overline{3}, 2},-\frac{1}{6} \right)
\oplus
\left({\bf 3, 2},\frac{1}{6} \right)
\right],
$
\item
$
\left[
\left({\bf \overline{3}, 1},-\frac{2}{3} \right)
\oplus
\left({\bf 3, 1},\frac{2}{3} \right)
\right], 
$
\item
$
\left[
\left({\bf 1, 1},1 \right)
\oplus
\left({\bf 1, 1},-1 \right)
\right],
$
\item
$\left[\left({\bf 1, 1},0 \right)\right].$
\end{itemize}
Total number of the NG degrees of freedom is :  
${\bf 12 + 12 + 6 + 2 + 1 =  33}$. 
The cubic term solutions lead to the true 
$SU(3)_C \times SU(2)_L \times U(1)_Y$ symmetry.  
\bref{nuR} gives heavy right-handed neutrino, and the coefficient of \bref{massmatrix} is also written in terms of $\phi_3$.
\paragraph{Electroweak Higgs doublet}
In the standard picture of the electroweak symmetry breaking, 
we have the Higgs doublets which give masses to the matter.  
These masses should be less than or equal to the electroweak scale.  
Since we approximate the electroweak scale as zero, 
we must impose a constraint that the mass matrix should 
have one zero eigenvalue.  

We define 
\be
H_u^{10} \equiv H_{\bf(1,2,2)}^{({\bf 1,2},\frac{1}{2})}, \,
\overline{\Delta}_u \equiv 
\overline{\Delta}_{\bf(15,2,2)}^{({\bf 1,2},\frac{1}{2})}, \, 
\Delta_u \equiv 
\Delta_{\bf(15,2,2)}^{({\bf 1,2},\frac{1}{2})}, \,
\Phi_u \equiv 
\Phi_{\bf(\overline{10},2,2)}^{({\bf 1,2},\frac{1}{2})}
\ee
and 
\be
H_d^{10} \equiv H_{\bf(1,2,2)}^{({\bf 1,2},-\frac{1}{2})}, \,
\overline{\Delta}_d \equiv 
\overline{\Delta}_{\bf(15,2,2)}^{({\bf 1,2},-\frac{1}{2})}, \,
\Delta_d \equiv \Delta_{\bf(15,2,2)}^{({\bf 1,2},-\frac{1}{2})}, \, 
\Phi_d \equiv 
\Phi_{\bf(10,2,2)}^{({\bf 1,2},-\frac{1}{2})}.
\ee
In the basis 
$\left\{ H_u^{10}, \overline{\Delta}_u, \Delta_u, \Phi_u \right\}$, 
the mass matrix is written as  
\be
\label{Mdoublet}
M_{\mathsf{doublet}} \equiv 
\left(
\begin{array}{cccc}
\begin{array}{c}
2 m_3 \\
\frac{\lambda_4 \phi_2}{\sqrt{10}} 
-\frac{\lambda_4 \phi_3}{2 \sqrt{5}} \\
-\frac{\lambda_3 \phi_2}{\sqrt{10}} 
-\frac{\lambda_3 \phi_3}{2 \sqrt{5}} \\
\frac{\lambda_3 v_R}{\sqrt{5}}
\end{array}
&\begin{array}{c}
\frac{\lambda_3 \phi_2}{\sqrt{10}} 
-\frac{\lambda_3 \phi_3}{2 \sqrt{5}} \\
m_2 +\frac{\lambda_2 \phi_2}{15 \sqrt{2}} 
-\frac{\lambda_2 \phi_3}{30} \\
0 \\
0
\end{array}
&\begin{array}{c}
-\frac{\lambda_4 \phi_2}{\sqrt{10}} 
-\frac{\lambda_4 \phi_3}{2 \sqrt{5}} \\
0 \\
m_2 +\frac{\lambda_2 \phi_2}{15 \sqrt{2}} 
+\frac{\lambda_2 \phi_3}{30}\\
-\frac{\lambda_2 v_R}{10}
\end{array}
&\begin{array}{c}
\frac{\lambda_4 \overline{v_R}}{\sqrt{5}} \\
0 \\
-\frac{\lambda_2 \overline{v_R}}{10} \\
2 m_1 + \frac{\lambda_1 \phi_2}{\sqrt{2}} 
+ \frac{\lambda_1 \phi_3}{2}
\end{array}
\end{array}
\right).  
\ee
The corresponding mass terms of the superpotential read 
\be
W_m = \left(H_u^{10}, \overline{\Delta}_u, \Delta_u, \Phi_u \right) 
\, M_{\mathsf{doublet}} \,
\left(H_d^{10}, \Delta_d, \overline{\Delta}_d, \Phi_d \right)^{\mathsf{T}}.  
\label{Wm}
\ee
The requirement of the existence of a zero mode leads to the 
following condition.  
\be 
\det{M_{\mathsf{doublet}}} = 0.  
\label{splittings}
\ee
For instance, in case of $\lambda_3 = 0$, 
$m_2 + \frac{\lambda_2 \phi_2}{15 \sqrt{2}} 
- \frac{\lambda_2 \phi_3}{30}=0$, 
we obtain a special solution to Eq. (\ref{splittings}), 
while it keeps a desirable vacuum and it does not produce 
any additional massless fields.  However, we proceed our 
arguments hereafter without using this special solution.  

We can diagonalize the mass matrix, $M_{\mathsf{doublet}}$ 
by a bi-unitary transformation.   
\be
U^{\ast} \,M_{\mathsf{doublet}} \,V^{\dagger}
= {\mathrm{diag}}(0, M_{126}, M_2, M_3).    
\ee
Then the mass eigenstates are written as  

\bea
\left(H_u, 
\, {\mathsf{h}}_u^1, \, {\mathsf{h}}_u^2, \, {\mathsf{h}}_u^3 \right)
&=& 
\left(H_u^{10}, {\overline{\Delta}}_u, \Delta_u, 
\Phi_u \right) \, U^{\mathsf{T}}, 
\nonumber\\
\left(H_d, 
\, {\mathsf{h}}_d^1, \, {\mathsf{h}}_d^2, \, {\mathsf{h}}_d^3 \right)
&=&
\left(H_d^{10}, \Delta_d, {\overline{\Delta}}_d, 
\Phi_d \right) \, V^{\mathsf{T}}. 
\label{UV}
\eea
Here $H_u,~H_d$ are MSSM light Higgs doublets. 
We get the explicit form of $U$ and $V$ from \bref{Mdoublet}, and thus we can connect the oscillation data with the GUT Yukawa coupling.
Thus the intermediate energy scales are severely constrained from the low energy neutrino data, and the gauge coupling unification 
 at the GUT scale may be spoiled. \\
This fact has been explicitly shown in \cite{Bertolini}, 
 where the gauge couplings are not unified any more 
 and even the ${\rm SU}(2)$ gauge coupling blows up below the GUT scale (Fig.3). 
\begin{figure}[t]
\centering
\includegraphics[width=0.49\textwidth]{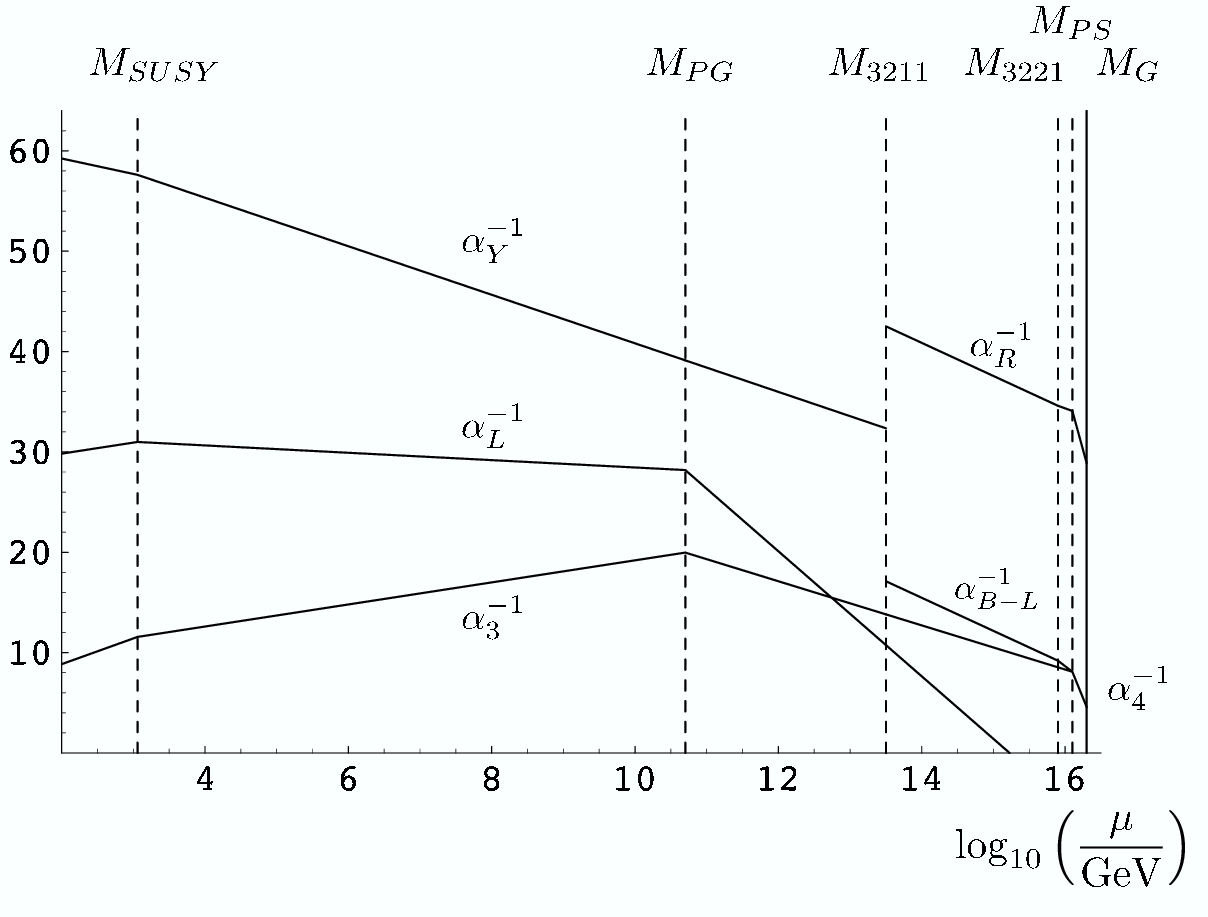}
  \caption{Running gauge coupling constants.
$M_\mathrm{SUSY} = 1$~TeV,
  $M_{PG}=M_{(8,3,0)}$, $M_{3211} \equiv <10,1,3>$, $M_{3221}
  \equiv <15,1,1>$, $M_{PS} \equiv <1,1,1>$, and
  $M_\mathrm{GUT} \equiv M_{210} = 2\times 10^{16}$~GeV cited from \cite{Bertolini}}
\label{fig:running}
\end{figure}
Thus the detailed analyses of superpotential was the great progress but it reveals unambiguously the details of structure, which also uncovers pathologies.\\
However, this is easily remedied by the addition of ${\bf 120}$ Higgs in Yukawa coupling \cite{mohapatra}. We mean that the dominant part may be governed by the minimal SO(10) but
such generalization does not spoil the renormalizable SO(10) GUT yet. 
\section{Proton Decay \label{protondecay}}

One of the problems we encountered in the minimal SO(10) GUT is the fast proton decay.
After the symmetry breaking from SO(10) to 
$SU(3)_C \times SU(2)_L \times U(1)_Y$, 
the generic Yukawa interactions between the matter fields 
and the color triplet Higgs fields are given by  
\bea
W_{Y}&=&
Y_{10}^{ij}\,H_{\overline{T}}
\left(q_i \ell_j + u^c_i d^c_j \right)
+Y_{126}^{ij}\,\overline{\Delta}_{\overline{T}}
\left(q_i \ell_j + u^c_i d^c_j \right)
\nonumber\\
&+&
Y_{10}^{ij}\,H_T 
\left(\frac{1}{2} q_i q_j 
+u^c_i e^c_j 
+d^c_i \nu^c_j \right)
\nonumber\\
&+&Y_{126}^{ij}\,\overline{\Delta}_T
\left(\frac{1}{2} q_i q_j 
+u^c_i e^c_j + d^c_i \nu^c_{j} 
\right)
\nonumber\\
&+&
Y_{126}^{ij}\,{\overline{\Delta}}_T^{\prime}
\left(u^c_i e^c_j + d^c_i \nu^c_{j} 
\right).
\label{WY}
\eea
Here we have defined  
\be
H_{\overline{T}} 
\equiv
H^{{\bf(\overline{3},1},\frac{1}{3})}_{\bf(6,1,1)}\,,\,\,
\equiv
H^{{\bf(3,1},-\frac{1}{3})}_{\bf(6,1,1)}\,,\,\,
\overline{\Delta}_{\overline{T}} 
\equiv
\overline{\Delta}^{{\bf(\overline{3},1},\frac{1}{3})}_{\bf(6,1,1)}\,,\,\,
\overline{\Delta}_T 
\equiv
\overline{\Delta}^{{\bf(3,1},-\frac{1}{3})}_{\bf(6,1,1)}\,,\,\,
{\overline{\Delta}}_T^{\prime} 
\equiv
\overline{\Delta}^{{\bf(3,1},-\frac{1}{3})}_{\bf(10,1,3)}\,. 
\label{triplet1}
\ee
For later use we define
\be
{\Delta}_{\overline{T}} 
\equiv
\Delta^{{\bf(\overline{3},1},\frac{1}{3})}_{\bf(6,1,1)}\,,\,\, 
{\Delta}_{T} 
\equiv
\Delta^{{\bf(3,1},-\frac{1}{3})}_{\bf(6,1,1)}\,,\,\, 
{\Delta}_{\overline{T}}^{\prime} 
\equiv
\Delta^{{\bf(\overline{3},1},\frac{1}{3})}_{\bf(\overline{10},1,3)}\,,\,\,
\Phi_{\overline T} 
\equiv
\Phi^{{\bf(\overline{3},1},\frac{1}{3})}_{\bf(15,1,3)}\,,\,\,
\Phi_T 
\equiv
\Phi^{{\bf(3,1},-\frac{1}{3})}_{\bf(15,1,3)}\,. 
\label{triplet2}
\ee

In the basis $
\left\{
H_{\overline T}, \Delta_{\overline T}, 
{\overline \Delta}_{\overline T}, 
\Phi_{\overline T}, 
\Delta_{\overline T}^{\prime} \right\}$, 
the mass matrix reads
\be
M_{\mathsf{triplet}} \equiv
\left(
\begin{array}{ccccc}
\begin{array}{c}
2 m_3 \\
-\frac{\lambda_3 \phi_1}{\sqrt{10}} - \frac{\lambda_3 \phi_2}{\sqrt{30}} \\
-\frac{\lambda_4 \phi_1}{\sqrt{10}} + \frac{\lambda_4 \phi_2}{\sqrt{30}} \\
\frac{\lambda_3 v_R}{\sqrt{5}} \\
-\frac{\sqrt{2}\lambda_3 \phi_3}{\sqrt{15}}
\end{array}
&\begin{array}{c}
-\frac{\lambda_4 \phi_1}{\sqrt{10}} - \frac{\lambda_4 \phi_2}{\sqrt{30}} \\
m_2 \\
0 \\
-\frac{\lambda_2 v_R}{10 \sqrt{3}} \\
\frac{\lambda_2 \phi_3}{15 \sqrt{2}} 
\end{array}
&\begin{array}{c}
-\frac{\lambda_3 \phi_1}{\sqrt{10}} + \frac{\lambda_3 \phi_2}{\sqrt{30}} \\
0 \\
m_2 \\
0 \\
0
\end{array}
&\begin{array}{c}
\frac{\lambda_4 \overline{v_R}}{\sqrt{5}} \\
-\frac{\lambda_2 \overline{v_R}}{10\sqrt{3}} \\
0 \\
\overline{m}_{44}  \\
-\frac{\lambda_2 \overline{v_R}}{5 \sqrt{6}}
\end{array}
&\begin{array}{c}
-\frac{\sqrt{2}\lambda_4 \phi_3}{\sqrt{15}} \\
\frac{\lambda_2 \phi_3}{15 \sqrt{2}} \\
0 \\
-\frac{\lambda_2 v_R}{5 \sqrt{6}} \\
\overline{m}_{55}
\end{array}
\end{array}
\right),  
\label{triplet3}
\ee

where $\overline{m}_{44} 
\equiv 
2 m_1 + \frac{\lambda_1 \phi_1}{\sqrt{6}}
+ \frac{\lambda_1 \phi_2 }{3 \sqrt{2}}
+ \frac{2 \lambda_1 \phi_3}{3}$ 
and 
$\overline{m}_{55} 
\equiv 
m_2 +\frac{\lambda_2 \phi_1}{10 \sqrt{6}}
+\frac{\lambda_2 \phi_2}{30 \sqrt{2}}$.  

The corresponding mass terms of the superpotential read 
\be
W_m = \left(
H_{\overline T}, \Delta_{\overline T}, 
{\overline \Delta}_{\overline T}, \Phi_{\overline T}, 
\Delta_{\overline T}^{\prime}
\right) \,
M_{\mathsf{triplet}} \,
\left(
H_T, {\overline \Delta}_T, 
\Delta_T, \Phi_T, 
{\overline{\Delta}}_T^{\prime}
\right)^{\mathsf{T}}.  
\label{WmT}
\ee
Integrating out $\Delta_T$, $\Phi_T$, and $\Delta_{\overline{T}}'$, 
we obtain the effective Yukawa interactions between 
the matter fields and the color triplet Higgs fields as  
\bea
W_{Y} &=& 
Y_{10}^{ij} \, 
H_{\overline{T}} \, 
\left(q_i \ell_j + u^c_i d^c_j \right) 
+ Y_{126}^{ij} \, 
\overline{\Delta}_{\overline{T}} \, 
\left(q_i \ell_j + u^c_i d^c_j \right)
\nonumber\\
&+&
Y_{10}^{ij} \, 
H_T \,
\frac{1}{2} q_i q_j 
\nonumber\\
&+&
\left(Y_{10}^{ij} - \frac{m_{31}}{m_{33}} \, Y_{126}^{ij} \right) \, 
H_T \, 
\left( u^c_i e^c_j + d^c_i \nu^c_j \right)
\nonumber\\
&+&Y_{126}^{ij} \, 
\overline{\Delta}_T \,
\frac{1}{2} q_i q_j 
\nonumber\\
&+& \left(1 -\frac{m_{32}}{m_{33}} \right) 
Y_{126}^{ij} \, 
\overline{\Delta}_T \,
\left(u^c_i e^c_j + d^c_i \nu^c_j \right). 
\label{yukawa2}
\eea

Then the effective mass terms for 
the remaining color triplet Higgs fields are written as  
\bea
W_{m}^{\mathrm{eff}}&=&
H_{\overline{T}}\left(a\, H_T +b \, \overline{\Delta}_T \right)
\nonumber\\
&+&
\overline{\Delta}_{\overline{T}}\left(c \,H_T +d\, \overline{\Delta}_T \right)
\nonumber\\
&\equiv&
\left(H_{\overline{T}},\,\overline{\Delta}_{\overline{T}} \right)
\,M_T \,
\left(
\begin{array}{c}
H_T \\
\overline{\Delta}_T 
\end{array}
\right).
\label{mass}
\eea
Eqs. (\ref{yukawa2}) and (\ref{mass}) leads us to 
the effective dimension-five interactions after integrating out 
the remaining color triplet Higgs fields \cite{Sakai},  
\be
-W_5 =  C_L^{ijkl} \, \frac{1}{2} q_i q_j q_k \ell_l
+ C_R^{ijkl} \, u^c_i e^c_j u^c_k d^c_l ,
\label{dim5}
\ee
inducing the dangerous proton decay.  
Here, $C_L$ and $C_R$ are given by the Yukawa coupling 
matrices at the GUT scale, $M_G$,    
\bea
C_L^{ijkl} (M_{G})
&=&
\left(Y_{10}^{ij},\, Y_{126}^{ij} \right)
M_T^{-1}
\left(
\begin{array}{c}
Y_{10}^{kl} \\
Y_{126}^{kl}
\end{array} 
\right),
\nonumber\\
C_R^{ijkl} (M_{G})
&=&
\left( Y_{10}^{ij} - \frac{m_{13}}{m_{33}} \,Y_{126}^{ij}, \, 
\left(1 -\frac{m_{32}}{m_{33}} \right) \,Y_{126}^{ij} \right)
M_T^{-1}
\left(
\begin{array}{c}
Y_{10}^{kl} \\
Y_{126}^{kl}
\end{array} 
\right).
\label{Wilson}
\eea
Note that
\bea
\left(
\begin{array}{c}
Y_{10} \\
Y_{126}
\end{array}
\right)
&=&
\left(
\begin{array}{cc}
\begin{array}{c}
\alpha_u \quad \\
\alpha_d \quad
\end{array}
\begin{array}{c}
\beta_u \\
\beta_d
\end{array}
\end{array}
\right)^{-1}
\left(
\begin{array}{c}
Y_{u} \\
Y_{d}
\end{array}
\right)
\nonumber\\
&\equiv&
A^{-1} \,\left(
\begin{array}{c}
Y_{u} \\
Y_{d}
\end{array}
\right).  
\label{A}
\eea
Thus we have 
\bea
C_L^{ijkl} 
&=& 
\left(Y_{u}^{ij},~Y_{d}^{ij} \right)
\left(
A\,
M_T
\,
A^{\mathsf{T}}
\right)^{-1}
\left(
\begin{array}{c}
Y_{u}^{kl} \\
Y_{d}^{kl}
\end{array} 
\right).
\label{Wilson2}
\eea
The Yukawa coupling matrices, $Y_{10}$ and $Y_{126}$, 
 are related to the corresponding mass matrices 
 $M_{10}$ and $M_{126}$ such that 
\bea
Y_{10} &=& \frac{c_{10}}{\alpha^u  v \sin{\beta}} M_{10}, 
\nonumber\\
Y_{126} &=& \frac{c_{126}}{\beta^u  v \sin{\beta}} M_{126}, 
\label{M10}
\eea
with $v \simeq 174.1 \, \mathrm{[GeV]}$.  
Here $\alpha^u$ and $\beta^u$ are the Higgs doublet mixing parameters
 introduced in \bref{Yukawa3}, 
 which are restricted in the range 
 $|\alpha^u|^2 +|\beta^u|^2 \leq 1$. 
Although these parameters are irrelevant 
 to fit the low energy experimental data 
 of the fermion mass matrices, 
 there are theoretical lower bound on them 
 in order for the resultant Yukawa coupling constant 
 not to exceed the perturbative regime. 
Since $c_{10}$, $c_{126}$, $M_{10}$ and $M_{126}$ 
 are the functions of only $\sigma$, 
 we can completely determine the Yukawa coupling matrices 
 once $\sigma$, $\alpha^u$ and $\beta^u$ are fixed. 
In order to obtain the most conservative values of 
 the proton decay rate, 
 we make a choice of the Yukawa coupling matrices 
 as small as possible. 
In the following analysis, we restrict the region 
 of the parameters in the range 
 $(\alpha^u)^2 +(\beta^u)^2 = 1$ 
 (we assume $\alpha^u$ and $\beta^u$ real for simplicity). 
Here we present examples of the Yukawa coupling matrices 
 with fixed $\sigma = \pi$.  
For $\tan\beta = 2.5 $ with $\alpha^u = 0.031$, we find  
\be
Y_{10}
=
\left(
\begin{array}{ccc}
\begin{array}{c}
0.000839 + 2.79 \times {10}^{-6} \,i \\
0.00151 - 0.0000265 \, i \\
0.000692 - 0.000818 \, i \\
\end{array}
\begin{array}{c}
0.00151 - 0.0000265 \, i \\
0.00479 + 0.0000811 \, i \\
-0.0128 + 3.17 \times {10}^{-6} \, i 
\end{array}
\begin{array}{c}
0.000692 - 0.000818 \, i \\
-0.0128 + 3.17 \times {10}^{-6} \, i \\
0.525 - 0.0420 \, i 
\end{array}
\end{array}
\right),
\label{Y10}
\ee
\be
Y_{126}
=
\left(
\begin{array}{ccc}
\begin{array}{c}
-0.0000613 - 2.17 \times {10}^{-7} \, i \\
-0.000111 + 1.94 \times {10}^{-6} \, i \\
-0.0000508 + 0.0000600 \, i 
\end{array}
\begin{array}{c}
-0.000111 + 1.94 \times {10}^{-6} \, i \\
-0.000428 - 2.46 \times {10}^{-6} \, i \\
0.000941 - 2.33 \times {10}^{-7} \, i 
\end{array}
\begin{array}{c}
-0.0000508 + 0.0000600 \, i \\
0.000941 - 2.33 \times {10}^{-7} \, i \\
1.42 \times {10}^{-7} + 0.00132 \, i
\end{array}
\end{array}
\right),  
\label{Y126}
\ee
and for $\tan\beta = 10$ with $\alpha^u = 0.111$,  
\be
Y_{10}
=
\left(
\begin{array}{ccc}
\begin{array}{c}
0.00101 + 1.87 \times {10}^{-6} \, i \\
0.00179 - 0.0000439 \, i \\
0.000348 - 0.00125 \, i 
\end{array}
\begin{array}{c}
0.00179 - 0.0000439 \, i \\
0.00541 + 0.000141 \, i \\
-0.0154 + 5.21 \times {10}^{-6} \, i 
\end{array}
\begin{array}{c}
0.000348 - 0.00125 \, i \\
-0.0154 + 5.21 \times {10}^{-6} \, i \\
0.530 - 0.0567 \, i 
\end{array}
\end{array}
\right),
\label{Y10'}
\ee
\be
Y_{126}
=
\left(
\begin{array}{ccc}
\begin{array}{c}
-0.000244 - 5.21 \times {10}^{-7} \, i \\
-0.000436 + 0.0000107 \, i \\
-0.0000847 + 0.000305 \, i 
\end{array}
\begin{array}{c}
-0.000436 + 0.0000107 \, i \\
-0.00164 - 0.0000154 \, i \\
0.00375 - 1.27 \times {10}^{-6} \, i 
\end{array}
\begin{array}{c}
-0.0000847 + 0.000305 \, i \\
0.00375 - 1.27 \times {10}^{-6} \, i \\
-0.000684 + 0.00626 \, i 
\end{array}
\end{array}
\right).  
\label{Y126'}
\ee

For the effective color triplet Higgsino mass matrix, 
 we assume the eigenvalues being the GUT scale, 
 $M_G = 2 \times 10^{16} \, \mathrm{[GeV]}$, 
 which is necessary to keep the successful gauge 
 coupling unification. 
Then, in general, we can parameterize the $2 \times 2$ 
 mass matrix as 
\be
M_T  = M_G I_2 \times U, 
\label{MT}
\ee
 with the unitary matrix, 
\be
U = e^{i \varphi \sigma_3} 
 \left(
 \begin{array}{cc}
 \begin{array}{c}
\cos{\theta} \\
- \sin{\theta}
\end{array}
\begin{array}{c}
\sin{\theta} \\
\cos{\theta}
\end{array} 
\end{array}
  \right) 
e^{i \varphi^\prime \sigma_3}.  
\label{U}
\ee
Here we omit an over all phase since it is irrelevant 
 to calculations of the proton decay rate.  
Now there are five free parameters in total 
 involved in the coefficient $C_L^{ i j k l}$, 
 namely, $\sigma$, $\alpha_u$, $\theta$, 
 $\varphi$ and $\varphi^\prime$. 
Once these parameters are fixed, $C_L^{i j k l}$ 
 is completely determined. 

The proton decay mode via the dimension five operator 
 in Eq.~(\ref{dim5}) 
 with the Wino dressing diagram 
 is found to be dominant, and leads to 
 the proton decay process, $p \to K^{+} \overline{\nu}$. 
The decay rate for this process 
 is approximately estimated as 
 (in the leading order of the Cabibbo angle $\lambda \sim 0.22$) 
\bea
\Gamma(p \to K^{+} \overline{\nu})
& \simeq &
\Gamma(p \to K^{+} \overline{\nu_\tau})
=
\frac{m_p}{32 \pi f_{\pi}^2} \,
\left|\beta_{H} \right|^2 \times
\left|A_L A_S \right|^2 \times
\left(\frac{\alpha_2}{4 \pi} \right)^2 
\frac{1}{m_{S}^2}
\nonumber\\
&\times&
 \Big|C_L^{2311} - C_L^{1312} 
  + \lambda \left( C_L^{2312} - C_L^{1322} \right)\Big|^2 
\nonumber\\
&\times& 
 5.0 \times 10^{31} \, [{\mathrm{years^{-1}/GeV}}].  
\label{rate1}
\eea
Here the first term denotes 
 the phase factor and the hadronic factor, 
\be
\beta_H u_L({\bf k})\equiv \la 0| d_Lu_Lu_R |p({\bf k})\ra,
\ee
and $\beta_{H} = 0.0096 \,[{\mathrm GeV^3}]$ 
is given by lattice calculations \cite{lattice}. 
$A_L \simeq 0.32$, $A_S \simeq 0.93$ are 
 the long-distance and the short-distance 
 renormalization factors about the coefficient $C_L^{ijkl}$, 
 respectively.  
\footnote{As suggested in Ref. \cite{turzynski}, 
it might be proper to use the renormalization factors 
$\tilde{A}_L$, $\tilde{A}_S$ in \cite{turzynski}, which directly treats 
the renormalization of the Wilson coefficients itself.  
But here, we adopt the use of the conventional factors $A_L$, $A_S$ 
to compare our results to the previous ones.  }
The third term in the first line comes from 
 the Wino dressing diagram, and 
 $m_S$ is a typical sparticle mass scale 
 multiplied by the ratio of a sfermion and Wino. 
In the case with the mass hierarchy 
 between the sfermions and the Wino 
 $(\widetilde{m}_f \gg M_2)$, 
 we find $m_S \sim \widetilde{m}_f \times ({\widetilde{m}_f}/M_2)$. 
In the following numerical analysis, 
 we take $\widetilde{m}_f  = 1 \,{\mathrm{[TeV]}}$ 
 and $M_2  = 100 \,{\mathrm{[GeV]}}$. \footnote{See the comments on the recent LHC results in the last section.}

\begin{figure}[h]
\begin{center}
\includegraphics{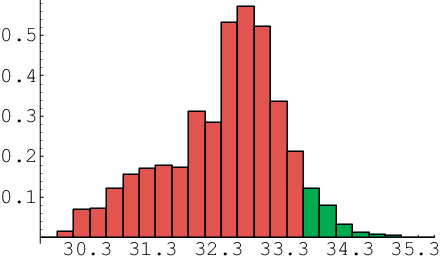}
\end{center}
\caption{
The distributions of the proton lifetime (log years) 
 for $\tan \beta =2.5$ 
 in arbitrary five parameter choices (normalized by 1).  
The green region (right side from 33.3 point) is consistent with experiment. 
}
\label{f3}
\end{figure}

\begin{figure}
\begin{center}
\includegraphics{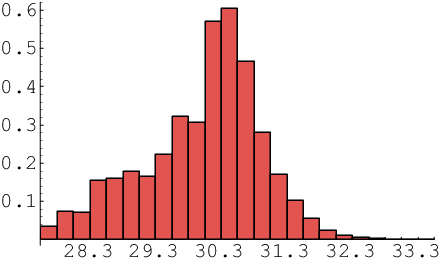}
\end{center}
\caption{Same as in Figure \ref{f3}, but for $\tan\beta=10$. There is very few region (right side from 35.3 ) allowed. }
\label{f4}
\end{figure}

Now we perform numerical analysis. 
Note that because of the very constrained flavor structure 
 of the minimal SUSY $SO(10)$ model 
 we can give definite predictions for the proton decay rate 
 once the five parameters in the above are fixed. 
For a specific choice of the Yukawa coupling matrices 
 in the minimal $SO(10)$ model with the type II seesaw, 
 the proton decay rate has been calculated in \cite{gmnn}. 
In our analysis, we make no such a specific choice, 
 and perform detailed analysis 
 in general situations of the minimal $SO(10)$ model
 by varying the above five free parameters. 
The result for $\tan \beta =2.5$ is presented in Figure \ref{f3}. 
Here the distributions of the proton lifetime (log years) 
 for arbitrary choices of the five free parameters (normalized by 1) 
 is depicted. 
We can see that some special sets of the free parameters 
 can result the proton lifetime consistent with SuperK results. 
In that region, cancellation in the second line 
 in Eq.~(\ref{rate1}) occurs 
 by tuning of the free parameters in the Higgsino mass matrix. 
Note that number of free parameters is not enough 
 to cancel both of the process 
 $p \to K^{+} \overline{\nu_\tau}$ (dominant mode) and 
 $p \to K^{+} \overline{\nu_\mu}$ (sub-dominant mode), 
 and thus the proton lifetime has an upper bound 
 in the $SO(10)$ model. 
For $\tan\beta = 10$, we obtain the same figure 
 depicted in Figure \ref{f4} 
 but the lifetime is scaled by roughly $(2.5/10)^2$, 
 which is consistent with the naive expectation 
 that the lifetime is proportional to $1/\tan^2 \beta$. 
Whole region is excluded in the case with $\tan \beta=10$. 

In the case of nondegenerate masses of $M_T$, 
the parameters increase from three to $5+1$ 
(the last 1 is the ratio of masses). 
However, the results only slide by the square of 
this mass ratio and do not show the special cancellation. 

We have found that for $\tan \beta=2.5$ 
 some special sets of the parameters 
 predict the proton decay rate consistent with the SuperK results, 
 where the cancellation for the dominant modes 
 of the proton decay amplitude occurs 
 by tuning of the parameters. 
Although there exists the allowed region, it is very narrow. 
Our result is consistent with the one in the previous work \cite{gmnn} 
 for only one specific choice of the Yukawa coupling matrices. 
It has been found that the resultant proton decay rate 
 is proportional to $\tan^2 \beta$ as expected 
 and the allowed region eventually disappears 
 as $\tan \beta$ becomes large, even for $\tan \beta=10$. 

There are some theoretically possible ways to 
 extend the proton lifetime. 
One way is to adopt a large mass hierarchy 
 between the sfermions and the Wino 
 as can be seen in Eq.~(\ref{rate1}). 
The proton lifetime is pushed up 
 according to the squared powers of the mass hierarchy, 
 and the allowed region becomes wide. 
How large the hierarchy can be depends on 
 the mechanism of the SUSY breaking and its mediation. When we assume the minimal supergravity scenario, 
 the cosmologically allowed region \cite{CMSSM} 
 consistent with the recent WMAP satellite data \cite{WMAP} 
 suggests that the masses of sfermion and Wino 
 are not so hierarchical and the value we have taken 
 in our analysis seems to be reasonable. 
Another way to evade fast proton decay is to abandon the assumption 
 of Higgsino degeneracy at the GUT scale, 
 and to make the mass eigenvalues of 
 the effective colored Higgsino mass matrix heavy. 
We can examine this possibility based 
 on a concrete Higgs sector. 
However, this seems to be a very difficult task 
 even if we introduce a minimal Higgs sector 
 in the minimal $SO(10)$ model discussed 
 in \cite{Fukuyama:2004xs} \cite{Bajc:2004xe} since there are lots of free parameters in the Higgs sector. 
Furthermore, even in the minimal Higgs sector, 
 there are lots of Higgs multiplets involved 
 and the beta function coefficients of the gauge couplings are huge. 
It seems to be very hard 
 to succeed the gauge coupling unification 
 before blowing up of the gauge couplings. 
Therefore, the assumption that all the Higgs multiplets 
 are degenerate at the GUT scale would be natural.  

Consequently our results show the typical properties of 
SO(10) GUT but are not exhaustive. Also there is possibility to 
vary GUT phases $\beta '$, $\gamma '$ of \bref{phases} generically.

\paragraph{General Higgs superpotential}
Also we may consider the more general superpotential for completeness \cite{fuku1}. 
\bea
W &=& \frac{1}{2} m_{1} \Phi^2 + m_{2} \overline{\Delta} \Delta + \frac{1}{2} m_{3} H^2 
\nonumber\\
&+& \frac{1}{2} m_{4} A^2 + \frac{1}{2} m_{5} E^2 + \frac{1}{2} m_{6} D^2 
\nonumber\\
&+& \lambda_{1} \Phi^3 + \lambda_{2} \Phi \overline{\Delta} \Delta
+ \left(\lambda_3 \Delta + \lambda_4 \overline{\Delta} \right) H \Phi
\nonumber\\
&+&
\lambda_{5} A^2 \Phi -i \lambda_{6} A \overline{\Delta} \Delta
+ \frac{\lambda_7}{120} \varepsilon A \Phi^2
\nonumber\\
&+& E \left( \lambda_{8} E^2 + \lambda_{9} A^2 + \lambda_{10} \Phi^2 
+ \lambda_{11} \Delta^2 + \lambda_{12} \overline{\Delta}^2 + \lambda_{13} H^2 
\right)
\nonumber\\
&+& D^2 \left( \lambda_{14} E + \lambda_{15} \Phi \right)
\nonumber\\
&+& D \left\{ \lambda_{16} H A + \lambda_{17} H \Phi + \left( 
\lambda_{18} \Delta + \lambda_{19} \overline{\Delta} \right) A 
+ \left( \lambda_{20} \Delta + \lambda_{21} \overline{\Delta} \right) \Phi  
\right\}.  
\label{potential}
\eea
Here $A = {\bf 45}$, $\Delta = {\bf 126}$, $\Phi = {\bf 210}$ 
and $E = {\bf 54}$ -plets.  
For general coupling 
constants $\lambda_1,\cdots,\lambda_{21}$, $m_1,\cdots,m_8$, the solutions 
with higher symmetries are specified by following relations.
Solutions with higher symmetries are characterized by:
\begin{enumerate}
\item
$SU(5) \times U(1)_{X}$ and $(SU(5) \times U(1))^{\mathrm{flipped}}$ symmetry solutions
\bea
\label{G51vac}
\left\{
\begin{array}{lll}
E  &=& v_{R} \ =\ 0,
\\
\Phi_1 & = & \frac{\varepsilon}{\sqrt{6}} \,\Phi_3, \quad
\Phi_2 \ =\ \frac{\varepsilon}{\sqrt{2}} \, \Phi_3,
A_{1} \ = \ \frac{2\varepsilon}{\sqrt{6}} A_{2},  
\end{array}
\right. 
\eea
where $\varepsilon = 1$ and $\varepsilon = -1$ correspond to the 
$SU(5)\times U(1)_{X}$ symmetric vacua and 
$(SU(5) \times U(1))^{\mathrm{flipped}}$ symmetric vacua, respectively.
For the concrete matter contents, see the nextsection.
\item
$SU(5)$ symmetry solutions
\bea
\label{G5vac}
\left\{
\begin{array}{lll}
E  &=& 0,
\\
\Phi_1 & = & \frac{1}{\sqrt{6}} \,\Phi_3, \quad
\Phi_2  \ =\ \frac{1}{\sqrt{2}}\, \Phi_3, \quad
A_{1}\ = \ \frac{2}{\sqrt{6}}\, A_{2}, \quad
v_R \ \neq\ 0.
\end{array}
\right.
\eea
\item
$G_{422} \equiv SU(4) \times SU(2)_{L} \times SU(2)_{R}$ symmetry solutions
\bea
\label{G422vac}
\left\{
\begin{array}{lll}
\Phi_2 & =& \Phi_3 \ =\ A_{1}\ =\ A_{2} \ =\ v_{R} \ =\ 0, 
\\
\Phi_1 & \neq& 0, \quad E \ \neq\ 0. 
\end{array}
\right.
\eea
\item
$G_{3221} \equiv SU(3)_{C} \times SU(2)_{L} \times SU(2)_{R} \times U(1)_{B-L}$ symmetry solutions
\bea
\label{G3221vac}
\left\{
\begin{array}{lll}
\Phi_3 &=& A_1 \ =\ v_R \ =\ 0,
\\
\Phi_1  &\neq& 0,  \quad \Phi_2 \ \neq\ 0,  \quad A_2 \ \neq\ 0,  \quad E \ \neq\ 0.
\end{array}
\right. 
\eea
\item
$G_{421} \equiv SU(4) \times SU(2)_{L} \times U(1)$ symmetry solutions
\bea
\label{G421vac}
\left\{
\begin{array}{lll}
\Phi_2 &=& \Phi_3\ =\ A_2\ =\ v_R\ =\ 0, 
\\
\Phi_1  &\neq& 0,  \quad A_1 \ \neq\ 0, \quad  E \ \neq\ 0.
\end{array}
\right.
\eea
\item
$G_{3211} \equiv SU(3)_{C} \times SU(2)_{L} \times U(1)_{R} \times U(1)_{B-L}$ 
symmetry solutions
\bea
\label{G3211vac}
\left\{
\begin{array}{lll}
v_{R} &=&0, 
\\
\Phi_i &\neq& 0\ (i=1,2,3), \quad A_i\ \neq\ (i=1,2), \quad E\ \neq\ 0. 
\end{array}
\right. 
\eea
\end{enumerate}
The higher symmetry solutions given in Eqs. (\ref{G51vac})-(\ref{G3211vac}) lead
to the crucial consistency checks for all results mentioned before.
In order to keep the successful gauge coupling unification as usual, it is desirable that all Higgs multiplets 
 have masses around the GUT scale, but some Higgs fields 
 develop VEVs at the intermediate scale. 
More Higgs multiplets and some parameter tuning in the Higgs sector 
 are necessary to realize such a situation. 

In addition to the issue of the gauge coupling unification, 
 the minimal SO(10) model potentially suffers from the problem 
 that the gauge coupling blows up around the GUT scale. 
This is because the model includes many Higgs multiplets of 
 higher dimensional representations. 

According to the line of thoughts from \bref{cutoff1} to \bref{cutoff2}, it was natural to consider
\be
L_{GUT}=L_{ren}''+\frac{L_3}{\Lambda_3}
\label{cutoff3}
\ee
up to $M_{P}$.
Here $\Lambda_3={\cal O}(M_{P})$ and gravitation (spacetime structure) appears as a subdominant term. However the blow-up before $M_{P}$ problem shows that such scheme does not exist in its naive sense. 

 The minimal SO(10) model also is faced on the fast proton decay \cite{Fukuyama3}.

These facts strongly (though not indipensablly) suggest the presence of extra dimensions, which not only solves the above problems but also gives new insights for SUSY breaking mechanism
\cite{Luty}.

\section{Solutions to the Problems of the Renormalizable SO(10) GUT-Why Extra Dimension ?}

When we accept the superpotential of \bref{lee}, is the breakdown of the gauge coupling unification inevitable ?

\subsection{Model modifications in 4D}
There are several approaches to this problem making leave the theory in 4D.

First let us try to remedy the pathologies mentioned in the previous section preserving the principles of the renormalizability but discarding minimality of  SO(10) GUT, which is to add ${\bf 120}$.

The great advantage of minimal SO(10) model was its high predictivity, implying that all quark-leptons mass matrices including Dirac and Majorana neutrinos, are completely determined.

The reason why the gauge coupling unification is broken is as follows.
The renormalizable SUSY GUT with Higgs fields of high dimensional representation has many Standard Model vacua. However such intermediate energy scale is fixed by only single parameter as was shown in \bref{vev1}-\bref{nuR} and also \cite{Mimurap}
\be
\frac{c_{10}}{c_{126}}=-\frac{3(v-1)(v+1)(2v-1)(v^3+5v-1)}{8v^6-27v^5+38v^4-70v^3+87v^2-31v+3},
\ee
where $v\equiv \frac{\phi_3}{{\cal M}_1}$.
So if we add another Higgs, if we retain renormalizability, ${\bf 120}$ by virtue of which
$\frac{c_{10}}{c_{126}}$ can be free.

Since ${\bf 120}$ has two SM doublets $({\bf 1,2,2})$ and $({\bf 15,2,2})$, mass matrices become \cite{Fukuyama4}
\bea
M_u&=&c_{10}M_{10}+c_{120}^{(1)}M_{120}+c_{126}M_{126}\nonumber\\
M_d&=&M_{10}+M_{120}+M_{126}\nonumber\\
M_D&=&c_{10}M_{10}+c_{120}^{(2)}M_{120}-3c_{126}M_{126}\\
M_e&=&M_{10}+c_{120}^{(3)}M_{120}-3M_{126}\nonumber\\
M_L&=&c_LM_{126}\nonumber\\
M_R&=&c_RM_{126}.\nonumber
\label{mass2}
\eea
Here
\be
c_{120}^{(1)}=\frac{\la \phi_+\ra+\la \phi'_+\ra}{-\la \phi_-\ra+\la \phi'_-\ra},~~c_{120}^{(2)}=\frac{\la \phi_+\ra-3\la \phi'_+\ra}{-\la \phi_-\ra+\la \phi'_-\ra},~~c_{120}^{(3)}=\frac{-\la \phi_-\ra-3\la \phi'_-\ra}{-\la \phi_-\ra+\la \phi'_-\ra},
\ee
where
$\la\phi_{\pm}\ra$ are expectation values of ${\bf (1,2,2)}$ of ${\bf 120}$, and$\la\phi'_{\pm}\ra$ are those of ${\bf (15,2,2)}$ of ${\bf 120}$.

In the original model, ${\bf 126}$ takes part of Majorana neutrinos, as well as Dirac Fermions \bref{massmatrix}. In other word, $Y_{126}$ is of $O(1)$ as $Y_{10}$ to recover the wrong SU(5) mass relation \bref{SU5mass}.

The mass of heavy right handed Majorana neutrino is surely several orders smaller than $M_{GUT}$ (we recognised that type II seesaw is subdominant), which means that we are forced to have the vev $v_R$ of intermediate energy scale.
However, we have additionally many parameters and can use ${\bf 126}$ for determining $M_R$ and $M_L$ independently on the determination of Dirac fermion mass matrices. That is $Y_{126}$ is free from order one unlike the minimal case and vevs are free from having the intermediate energy scales and we can recover the gauge coupling unifications. This seems to  be fine at least for data fittings of low energy. This model has been extensively discussed, especially on suppression of proton decay, by \cite{Mimura2}

In order that such theory becomes the SM of next generation, we must also study Doublet-Triplet (D-T) problem and SUSY breaking mechanism. We will see this point soon later.

One of the other approaches is to use Split Susy \cite{Bajcsplit} with light gauginos and higgsinos in 100 TeV range and superheavy squarks and sleptons in energy scale close to GUT. However, it is essentially non SUSY and seems to be unnatural.

The other, for instance, is to adopt non SUSY SO(10) GUT with ${\bf 45}+\overline{{\bf 126}}$ Higgs \cite{Bertolini4}. 

However, it seems to us these are too conservative to construct new model BSM.
\subsection{Flipped GUT model}

Before we consider flipped SO(10) model we go back to flipped SU(5) model.
In SU(5) GUT, the SM contents are embedded in different multiplets of ${\bf 10}$ and $\overline{\bf 5}$.
Then hypercharge assignment is not unique in these two sets.
This may easily understood if we consider two series of symmetry breaking (see the results of the previous section)
\be
SO(10) \supset SU(5)\times U(1)_V ,~~SU(5)\supset SU(3)_C\times SU(2)_L\times U(1)_Z
\label{SU(5)}
\ee
and
\be
SO(10)\supset SU(4)_{PS}\times SU(2)_L\times SU(2)_R\supset SU(3)_c\times SU(2)_L\times SU(2)_R\times U(1)_X.
\label{PS0}
\ee

In Eq.\bref{SU(5)}, there are two assignments of hypercharge:
\bea
\label{G-G}
&&\frac{Y}{2}=Z\\
&&\frac{Y}{2}=-\frac{1}{5}(Z+V). 
\label{flip}
\eea
Eq.\bref{G-G} corresponds to the usual Georgi-Glashow, and Eq.\bref{flip} does to the flipped SU(5) \cite{flip} given from the Georgi-Glashow model by the interchange of $SU(2)_R$ doublets
\be
u^c\leftrightarrow d^c,~~e^c\leftrightarrow \nu^c
\ee
and we obtain harmless relation
\be
M_u=M_\nu
\label{SU5mass2}
\ee
in place of \bref{SU5mass} of Georgi-Glashow SU(5) model.
Moreover, it is attractive from D-T splitting:
Higgs superpotential has the form
\be
W_H={\bf 10}\times {\bf 10}\times {\bf 5}+\overline{\bf 10}\times\overline{\bf 10}\times\overline{\bf 5}, 
\label{MP}
\ee
which gives rise to triplet mass
\be
\la (1,1,0)_{10}\ra(\overline{3},1;1/3)_{10}(3,1;-1/3)_5+\la (1,1;0)_{\overline{10}}\ra(3,1;-1/3)_{\overline{10}}(\overline{3},1;1/3)_{\overline{5}}
\ee
but has no doublet mass since $5+\overline{5}$ has no partner in $10+\overline{10}$
(Missing partner mechanism).

This is a solution to the D-T problem without additional adjoint Higgs.

However, in flipped SU(5) corresponding to the assigment \bref{assign5}, $\nu^c$ does not belong to SU(5) single but to {\bf 10}-plet, which drives us to
unrenormalizable heavy Majorana neutrino mass term,
\be
{\bf 10}_i{\bf 10}_j\overline{\bf 10}_H\overline{\bf10}_H/M_P,
\ee
responsible for seesaw mechanism.

For SO(10) case, SM matter contents are embedded in single ${\bf 10}_i$. So if we consider flipped SO(10), we are forced to
enlarge group like $SO(10)\times U(1)_{V'}\subset E_6$.

This implies the addition of matters other than SM matters since
\be
{\bf 27}={\bf 16}_1+{\bf 10}_{-2}+{\bf 1}_4.
\ee

There are three kinds of assignments in SO(10) \cite{Harada}:
\bea
{\bf 16}&=&d^c+e+\nu+u^c+u+d+e^c+\nu^c\nonumber\\
{\bf 10}&=&D+E^c+N^c+D^c+E+N\\
{\bf 1}&=&S,\nonumber
\label{assign1}
\eea
\bea
{\bf 16}&=&D^c+E^c+N^c+d^c+u+d+\nu^c+S\nonumber\\
{\bf 10}&=&D+E+N+u^c+e+\nu\\
{\bf 1}&=&e^c,\nonumber
\label{assign2}
\eea
\bea
{\bf 16}&=&D^c+E+N+u^c+u^c+u+d+e^c+S\nonumber\\
{\bf 10}&=&D+E^c+N^c+d^c+e+\nu\\
{\bf 1}&=&\nu^c.\nonumber
\label{assign3}
\eea
These sets are each classified into two different sets and therefore six kinds of classification in SU(5).
For the first set
\bea
{\bf 16}&=&(d^c+e+\nu)+(u^c+u+d+e^c)+\nu^c\nonumber\\
{\bf 10}&=&(D+E^c+N^c)+(D^c+E+N)\\
{\bf 1}&=&S\nonumber
\label{assign4}
\eea
and
\bea
{\bf 16}&=&(u^c+e+\nu)+(d^c+u+d+\nu^c)+e^c\nonumber\\
{\bf 10}&=&(D+E+N)+(D^c+E^c+N^c)\\
{\bf 1}&=&S.\nonumber
\label{assign5}
\eea
Flipped SO(10) is given from one of the second set
\bea
{\bf 16}&=&(d^c+E^c+N^c)+(D^c+u+d+S)+\nu^c\nonumber\\
{\bf10}&=&(D+e+\nu)+(u^c+E+N)\\
{\bf1}&=&e^c.\nonumber
\eea

Unfortunately, there is no renormalizable terms making work of Missing Partner mechanism unlike flipped SU(5). In this case the counter example is \cite{Yamashita}
\be
W_H\supset \overline{{\bf 16}}_1{\bf 16}_2{\bf 16}_2\overline{{\bf 16}}_1/M_P+
{\bf 16}_1\overline{{\bf 16}}_2\overline{{\bf 16}}_2{\bf 16}_1/M_P
\ee
\be
W_H\supset \la{\bf 1}\ra\la {\bf 10}\ra{\bf 10}~{\bf 5}/M_P+\la {\bf 1}\ra\la\overline{{\bf 10}}\ra\overline{{\bf 10}}~\overline{{\bf 5}}/M_P
\label{flippedSO(10)}
\ee
which corresponds to \bref{MP}, leading to massive triplets and massless doublets.
However, we can not incorporate this interaction naively in $E_6$ model since
\be
{\bf 27}_1{\bf 27}_1\overline{\bf 27}_2\overline{\bf 27}_2
\ee
includes \bref{flippedSO(10)} with vev of ${\bf 16}_1$ but simultaneously it also includes
\be
{\bf 5}_1{\bf 5}_{\overline{2}}\overline{\bf 10}_{\overline{2}}~~~\mbox{and}~~~
\overline{\bf 5}_{\overline{1}}\overline{\bf 5}_{2}{\bf 10}_2,
\ee
which give mass to doublet with vev of ${\bf16}_2$.

Another approach is to use Dimopoulos-Wilczek mechanism \cite{Dimopoulos}. 
Bertolini et al \cite{Bertolini3} considered flipped $SO(10)\otimes U(1)$ model with ${\bf 45}\oplus 2\times ({\bf 16}\oplus \overline{\bf 16}$) Higgs fields whose Higgs superpotetial is given by
\be
W_H=\frac{\mu}{2}Tr{\bf 45}^2+\rho_{ij}{\bf 16}_i\overline{\bf 16}_j+\tau_{ij}{\bf 16}_i{\bf 45}\overline{\bf 16}_j~~(i,j=1,2),
\ee
leading directly from $SO(10)\otimes U(1)$ to the SM.
This sounds good, however, they are forced to introduce unrenormalizable Yukawa coupling like
\be
W_Y=Y_U{\bf 16}_F{\bf 10}_F{\bf 10}_H+\frac{1}{M_P}[Y_E{\bf 10}_F{\bf 1}_F\overline{\bf 16}_H\overline{\bf 16}_H+Y_D{\bf 16}_F{\bf 16}_F\overline{\bf 16}_H\overline{\bf 16}_H]
\ee
to achiev a realistic texture. It seems to spoil the very merits of SO(10) GUT.

We have discussed D-T splitting in both missing partner mechanism and Dimopoulos-Wilczek mechanism. However, there exists no arguments for why this is necessarily the case.
To assert that D-T splitting is completely solved, we must also explain why $\mu$ term is so small, which requires a symmetry, most probably R-symmetry in SUSY \cite{Nelson}. Before we address this problem, we argue briefly in the next three subsections on GUT from different angles not mentioned so far.
\subsection{Perturbative SO(10) GUT \label{perturbative}}
Our renormalizable model use high dimensional Higgs like {\bf 126} and {\bf 210}, which makes the unified gauge coupling blow up after GUT scale, probably before the Planck scale.
However, GUT scale is O($10^{16}$)GeV rather near to the reduced Planck scale O($10^{18}$)GeV, around which the renormalization may lose its conventional meaning.
Anyhow many physicists prefer to adopt low dimensional Higgs fields, which assures the validity of perturbation to the Planck scale and is called perturbative SO(10) GUT. Such perturbative GUT leads us inevitably to unrenormalizable Yukawa coupling and lose a definite criterion to construct Yukawa couplings.
As a typical example let us consider the case of Raby's model \cite{Raby3},
\be
W=W_f+W_\nu,
\ee
where
\bea
W_f&=&{\bf 16}_3{\bf 10}_H{\bf 16}_3+{\bf 16}_a{\bf 10}_H\chi_a\nonumber\\
&+&\overline{\chi}_a\left(M_\chi\chi_a+{\bf 45}_H\frac{\phi_a}{\hat{M}}{\bf 16}_3+{\bf 45}\frac{\tilde{\phi}_a}{\hat{M}}{\bf 16}_a+A{\bf 16}_a\right),\\
W_\nu&=&\overline{{\bf 16}}\left(\lambda_2N_a{\bf 16}_a+\lambda_3N_3{\bf 16}_3\right)+\frac{1}{2}\left(S_aN_aN_a+S_3N_3N_3\right)
\eea
with $a=1,2$. Yukawa coupling unification at $M_{GUT}$ is realized only for the third generation.
Their number of parameters is 24,  whereas ours is 17 (see subsection \bref{parameter}). Also their number counting is quite differnt from ours. Their mass matrices are approximated as real but ours are generic.
As we pointed out in \cite{Fukuyama1}, complex phases are very important even in matching up mixing angles as well as in CP violating processes.
Also it is very difficult to construct Higgs superpotential generically in the case of perturbative SO(10) GUT case.

\subsection{Linear and nonlinear realization}
In the previous subsection we have discussed the larger group $E_6$ and might discuss larger group $E_7$ or $E_8$.
The fields transform linearly under the corresponding gauge group, which are called linear realization.
Let me consider a symmetry breaking from G to H, there appear NG bosons.
If we consider their superpartners as the SM matters and if we consider Lagrangian in terms of only NG bosons, Lagrangian is automatically invariant.
Let me explain it \cite{Kugo}.
In this case, the numbers of NG bosons are equal to dim(G/H)=dim G-dim H and transforms linearly under $h\in H$ transformation but nonlinearly under $g\in G$.
Let us describe the representatives of G/H as $\xi(\pi)$, parametrizing in terms of $\pi^a$ as
\be
\xi(\pi)=e^{i\pi(x)/f_\pi},~~\pi (x)\equiv \sum_{a\in {\it G-H}}\pi^a(x)X^a,
\ee
where $\pi$ are NG bosons.
\be
g\xi(\pi)=\xi(\pi')h(\pi,g),~~h(\pi,g)\in H
\ee
NG field $\pi(x)$ is transformed as
\be
\xi(\pi)\rightarrow \xi(\pi')=g\xi(\pi)h^{-1}(\pi,g),
\ee
under the global $g\in G$ transformation.
Since this transformation from $\pi$ to $\pi'$ is nonlinear w.r.t. $\pi$ and the representation of G in terms of $\pi$ is called nonlinear realization.
The low energy effective Lagrangian on G/H is constructed as
\be
{\cal L}=\frac{f_\pi^2}{4}Tr(\pa_\mu U^\dagger\pa^\mu U),
\ee
where
\be
U\equiv \xi\xi^T.
\ee
Thus even if we consider the same higher gauge group G, we have different low energy contents depending on whether we adopt linear or nonlinear realization.
In the case we see quite different pattern of symmetry breaking $E_8$ or $E_7\rightarrow E_4\times U(1)^4 $ from the former case. Indeed, Kugo-Yanagida showed that
three generations $3\times (10+5^*+1)+5$. Unfortunately it includes an additional one {\bf 5}-plet \cite{Kugo-Yanagida}.
This is quite interesting because it gives new insight on the relation between gauge symmetry breking patterns and family symmetry.

Thus even if we fix gauge group, we have different aspects depending on linear or nonlinear representation.

\subsection{Constraints from the string theory}
We have discussed GUT so far from the bottom-up approach.
It is needless to say that top-down approach is also important.

There are arguments that the string theory does not allow high dimensional Higgs \cite{Dienes}.
This is concerned with Heterotic string model and due to perturbation.
Non perturbative F theory is out of this constraints, and {\bf 126} does not necessarily denied. However, it may be useful to consider such counter example.

\begin{table}[t]
\centerline{
\begin{tabular}{l|l|l|l}
{}~~~$k=1$ & ~~~$k=2$ & ~~~$k=3$ & ~~~$k=4$ \\
\hline
\hline
              (\rep{10},~1/2) &
              (\rep{10},~9/20) &
             (\rep{10},~9/22) &
             (\rep{10},~3/8) \\
             (\rep{16},~5/8) &
             (\rep{16},~9/16) &
             (\rep{16},~45/88) &
             (\rep{16},~15/32) \\
            ~ &
             (\rep{45},~4/5) &
             (\rep{45},~8/11) &
             (\rep{45},~2/3) \\
            ~ &
             (\rep{54},~1) &
             (\rep{54},~10/11) &
             (\rep{54},~5/6) \\
           ~ &
           ~ &
             (\rep{120},~21/22) &
             (\rep{120},~7/8) \\
           ~ &
           ~ &
             (\rep{144},~85/88) &
             (\rep{144},~85/96) \\
           ~ &
           ~ &
           ~ &
             (\rep{210},~1) \\
\hline
\end{tabular}
  }
\caption{Unitary, potentially massless representations of $SO(10)$
realized at affine levels $k=1,2,3,4$.
Each representation $R$ is listed as $({\bf dim\,R},h_{R})$
where $h_R$ is its conformal dimension.
Singlets and conjugate representations are not explicitly
written, but understood \cite{Dienes}.}
\label{tableone}
\end{table}

As we have shown $\rep{126}$ takes very important roles in renormalizable minimal SO(10)
GUT model. However, one would require SO(10) at Kac-Moody levels $k\geq 1$ (see Table \ref{tableone}.
Figure caption is described in terms of string theory terminologies. Affine levels are rank of SO(10) in our language.
${\bf 10},~ {\bf 45},~ {\bf 120},~ {\bf 210}$, and ${\bf 126}$ are rank $1,~ 2,~ 3,~ 4$, and $5$ antisymmetric tensor, respectively.
${\bf 54}$ is symmetric rank 2 tensor \cite{fuku1}.

The central charge of SO(10) at level k is given by 

\be
c=\frac{kdim(G)}{k+\tilde{h}_G}=\frac{45k}{(k+8)}
\label{string1}
\ee
for SO(10). Here $\tilde{h}_G$ is coxeter.
The perturbative heterotic string central charge must be
\be
c\leq 22.
\label{string2}
\ee
It goes from \bref{string1} and \bref{string2} that conformal anomaly free condition leads to $k\leq 4$.
Thus if we assume GUT is top-downed from heterotic string \cite{Gross} perturbatively, we can not produce ${\bf 126}$.

F-theory GUT tries to predict masses of quark-leptons and magnitude of mixing matrices in close accord with experiments \cite{Heckman}.

There are few papers which have tried to unify bottom-up and top-down approaches rather closely \cite{Maekawa}, and these approaches will become more important hereafter.  

Then we will go back to the main flow in the next subsection.

\subsection{SUSY breaking in 4D}
The tree level potential is given by
\be
V=F_aF_a^\dagger+\frac{1}{2}D_AD^A.
\ee
Here 
\be
F^a=W^{\dagger a}=\frac{\pa W^\dagger}{\pa Q_a^\dagger},~~F_a^\dagger=\frac{\pa W}{\pa Q^a},~~D_A=g_AQ_a^\dagger(T_A)^a_bQ^b
\ee
So SUSY is spontaneously broken unless all $F_a=0,~D_A=0$.

$F_a\neq 0$ (called O'Raifeartaigh mechanism) is dominant in many cases and we will consider mainly this case.
$D_A\neq 0$ (called Fayet-Iliopoulos mechanism) is considered only in anomalously mediated symmetry breaking in this review.
If SUSY is spontaneously broken at tree level it must satisfy
\be
str(M^2)=tr(M_0^2)-2tr(M_{1/2}^\dagger M_{1/2})+3tr(M_1)^2)=0.
\ee
This directly leads us to the result that the scalar superpartners must be lighter than the heaviest observed fermion,
which contradicts with observation \cite{Martin}.
So we must consider some non-renormalizable process or loop corrections.

We must separate visible sector from invisible sector where SUSY is broken.
There are mainly two SUSY breaking mechanisms, gravity mediation and gauge mediation.
In either case, gravity mediates both sectors.
For F-term breaking, gravitino mass is
\be
m_{3/2}=\frac{\la F\ra}{\sqrt{3}M_P}.
\ee
However, the value $\la F\ra$ is quite different in these two mechanisms.
\paragraph{Gravity mediation mechanism}
The gravity mediation mechanism \cite{Lahanas} is to consider gravitational interaction which communicates SUSY breaking in breaking sector to visible sector \cite{GM1} in 4D.
We assume here that SUSY is broken in the hidden sector by the F component of a field $X$.
Taking the fact that ultra-violet gravity plays the role into consideration, we may relax in this case the renormalizability so far assumed.
The most general interactions X and the visible sector's fields are \cite{Luty}
\bea
\Delta {\cal L}&=&\int d^4\theta \left[\frac{(8z_Q)^i_j}{M_P^2}X^\dagger XQ^\dagger _iQ_j+...\right.\nonumber\\
&+&\left.\frac{b}{M_P}XH_uH_d+\frac{b'}{M_P}X^\dagger XH_uH_d+h.c.\right]\\
&+&\int d^2\theta\left[\frac{s_1}{M_P}XW_1^aW_{1a}+...\right]+h.c.\nonumber\\
&+&\int d^2\theta\left[\frac{a_{ij}}{M_P}XQ^iH_u(u^c)^j+...\right],\nonumber
\eea
and we find SUSY breking soft mass,
\be
m_{soft}=\frac{\la F\ra}{M_P}.
\ee
$m_{soft}$ is of ${\cal O}(1)$ TeV and $\sqrt{\la F\ra}\approx {\cal O}(10)^{11}$ GeV.
This value is very impressive since it seems to be common with the breaking energy scales of B-L and Peccei-Quinn symmetries.It should be remarked that $b$ and $b'$ terms give $\mu$ and $B\mu$ terms, and that $s_i$ does gaugino mass. 
This model can explain why $\mu$ and $B\mu$ are suppressed compared with $\la F\ra$. However, it simultaneously accompanies flavour problem unless the off diagonal part of $|a_{ij}|$ is highly suppressed.
To assume $a_{ij}$ diagonal is unnatural unless there are flavour symmetry at Planck scale. Such symmetry at $M_P$ is unlikely from Black Hole evaporation \cite{Luty}. In this sense minimal SUGRA ansatz
\be
(z_Q)^i_j=(z_L)^i_j=...=z_0\delta_{il}, ~~z_{H_u}=z_{H_d}=z_o
\ee
is unnatural.

\paragraph{Gauge mediation mechanism}
Another mechanism is to assume that SUSY breaking in invisible sector is communicated to the visible sector by heavy chiral fields, called messengers, through ordinary $SU(3)_c\times SU(2)_L\times U(1)_Y$ gauge interactions (GMSB).
Contrast to gravity mediation, it is natural that GMSB is flavour blind at $M_P$ since gauge theory does not discriminate flavour.

Gauge mediation \cite{GM1}, \cite{Giudice}, \cite{GM2} suffers from the anomalously small gaugino masses compared to the scalar masses.
\be
m_\lambda (\mu)=\frac{N_mg^2(\mu)}{16\pi^2}\frac{F}{M},
\ee
where $N_m=b-b'$.
Here $b$ and $b'$ are the beta functions of MSSM and MSSM+messengers, respectively.
Also it has $\mu-B\mu$ problem.
\bea
\frac{m_Z^2}{2}&=&-|\mu|^2-\frac{m_{H_u}^2\mbox{tan}^2\beta-m_{H_d}^2}{\mbox{tan}^2\beta-1} \label{bmu1}\\
\mbox{sin}2\beta&=&\frac{2B\mu}{2|\mu|^2+m_{Hu}^2+m_{H_d}^2}\label{bmu2}.
\eea
If we consider $\mu$ is generated by the SUSY breaking, 
\be
W=\lambda X H_uH_d.
\ee
Here $X$ is the backgroud chiral superfield, and $X=M+\theta^2 F$.
Then we obtain
\be
\mu=\lambda M\approx \frac{1}{16\pi^2}\frac{F}{M_{mess}},~~B\mu=\lambda F\approx 16\pi^2\lambda M\mu=16\pi^2 \mu^2\gg \mu^2
\ee
\be
m_{soft}\approx \frac{\alpha_a}{4\pi}\frac{\la F\ra}{M_{mess}}= \frac{\alpha_a}{4\pi}\Lambda.
\ee
MSSM scalar masses are
\be
m_{\phi_i}^2=2\Lambda^2\left[\left(\frac{\alpha_3}{4\pi}\right)^2C_3(i)+\left(\frac{\alpha_2}{4\pi}\right)^2C_2(i)+\left(\frac{\alpha_1}{4\pi}\right)^2C_1(i)\right].
\ee
If we go beyond 4D and go to extra dimensions, we have two new SUSY breaking mechanism, gaugino mediation \cite{gMSB} and
anomaly mediation (AMSB) \cite{RS2} \cite{GLMR}. We will discuss on these mechanisms in the next chapters.

Above the SUSY breaking mechanism, the initial condition of SUSY breaking is also very important, which will be discuused in the last subsection of the LHC results.

\subsection{No-go theorem in 4D \label{no-go}}
There is arguments that it is impossible to construct a GUT in 4D with a finite number of multiplets that leads to the MSSM with a residual R symmetry \cite{Ratz}, whose no-go theorem is not applicable to extra dimensions.
This is very important, and let me explain this:
SUSY invariant action is assumed to be invariant under global $U(1)_R$ transformation (for N=1 supersymmetry as we consider in this review),
\be
\theta\rightarrow e^{i\alpha}\theta,~~\theta^\dagger\rightarrow e^{-i\alpha}\theta^\dagger,
\label{U1R}
\ee
impling that R-charge of $\theta$ and $\theta^\dagger$ are 1 and -1, respectively.  Chiral superfield is expressed as
\be
\Phi=\phi(y)+\sqrt{2}\theta\psi(y)+\theta\theta F(y)
\ee
with
\be
y^\mu\equiv x^\mu+i\theta^\dagger\overline{\sigma}^\mu\theta.
\ee
Vector superfield is real and its R-charge =0. Vector superfield in Wess-Zumino gauge is
\be
V=\theta^\dagger\overline{\sigma}^\mu\theta A_\mu+\theta^\dagger\theta^\dagger\theta\lambda+\theta\theta\theta^\dagger\lambda^\dagger+\frac{1}{2}\theta\theta\theta^\dagger\theta^\dagger D
\ee
and $A_\mu,~\lambda,~D$ have R-charge 0,1,0, repectively.

Nelson and Seiberg discussed the relation between R symmetry and SUSY breaking \cite{Nelson}.
They showed under the condition 
\begin{itemize}
\item[i)]
Superpotential is generic, and
\item[ii)] low energy theory can be described by a supersymmetric Wess-Zumino model
\end{itemize}
that
\begin{itemize}
\item[a)]
R symmetry is necessary for SUSY breaking, and
\item[b)]
spontaneous R symmetry breaking is sufficient for SUSY breaking.
\end{itemize}
Concretely speaking, this will be explained as follows \cite{Nelson}.
Let us consider N superfields $\Phi_i$ whose $U(1)_R$ charges are $R(\Phi)=\phi_i$.
The R charge of superpotential is 2 and at least one superfield must have nonzero R charge, which we take it as N'th field $\Phi_N$, $\phi_N\neq0$,  without losing generality.
If $W$ is R symmetric, then $W$ can be described as
\be
W(\Phi_i)=\Phi_N^{2/\phi_N}\widetilde{W}(\widetilde{\Phi}_a),
\ee
where
\be
\widetilde{\Phi}_a=\frac{\Phi_a}{\Phi_N^{\phi_a/\phi_N}},~~a=1,...,N-1,
\ee
\be
\frac{\partial W}{\partial \widetilde{\Phi}_i}=\left\{\begin{array}{l}
\frac{2}{\phi_N}\Phi_N^{\frac{2}{\phi_N}-1}\widetilde{W}\\ \\
\Phi_N^{2/\phi_N}\frac{\partial \widetilde{W}}{\partial \widetilde{\Phi}_a}
\end{array}\right.\;.
\label{Fflat}
\ee
Now if $U(1)_R$ are spontaneously broken, $\langle \Phi_N\rangle \neq 0$, \bref{Fflat} leads to
\be
\widetilde{W}=0~~ \mbox{and}~~\frac{\partial \widetilde{W}}{\partial \widetilde{\Phi}_a}=0.
\ee
Generically we can not satisfy the above $N$ constraints by $N-1$ fields.

Thus if we have no U(1) symmetry we have appropriate SUSY vacuum, that is,
U(1) symmetry is necessary for SUSY breaking (condition (a)).

If there is U(1) symmetry and it is spontaneously broken, SUSY is automatically broken (condition (b)).

So the problem is how to impose $U(1)_R$ symmetry in superpotential of GUT.

Reflecting these situations, Fallbacher et al. \cite{Ratz} concluded that no MSSM model with either a ${\bf Z}_{M\geq 3}^R$ or U(1)$_{R}$ symmetry can be completed by a four dimensional GUT in the ultraviolet.
The essential point is explained for SU(5) GUT as follows.
$SU(5)\times Z_M^R$ is broken to the SM$\times  Z_M^R$ by the vev of the SM singlet of ${\bf 24}$.
${\bf 24}$ has zero R-charge since $Z_M^R$ is unbroken, and 
\be
{\bf 24}={\bf (8,1)}_0\oplus {\bf (1,3)}_0\oplus {\bf (1,1)}_0\oplus {\bf (3,2)}_{-5/6}\oplus {\bf (\overline{3},2)}_{5/6}.
\ee
Here $\la {\bf (1,1)}_0\ra\neq 0$, and ${\bf (3,2)}_{-5/6}$ and ${\bf (\overline{3},2)}_{5/6}$ get absorved to the longitudinal part of gauge bosons. The remaining ${\bf (8,1)}_0$ and ${\bf (1,3)}_0$ must be massive and therefore require mass term
$m{\bf 24}\times {\bf 24}$. However, it is prohibited because ${\bf 24}$ have 0 R-charge but superpotential must be 2 R-charge.
this is the case for more general mutiplet and more general gauge group including SO(10), The details should be referred to \cite{Ratz}.
On the other hand in the case of Pati-Salam case,
PS group to the SM need to reduce rank by one, which is done by (4,1,2) and break B-L
quantum number and there give rise to no problem.
Therefore, the minimum group subject to no-go theorem is SU(5).

Of course there are a loophole of this no-go theorem.
For instance it is for meta-stable supersymmetry breaking vacuum, where $U(1)_R$ is broken explicitly \cite{Intrigator}.
That is, let us consider 
\be
W=-k\Phi_1+m\Phi_2\Phi_3+\frac{y}{2}\Phi_1\Phi_3^2,
\ee
which is $U(1)_R$ symmetric with R-charge, $R_{\Phi_1}=R_{\Phi_2}=2, R_{\Phi_3}=0$. They introduced a broken term,
\be
\Delta W=\frac{1}{2}\epsilon m\Phi_2^2,
\ee
where $\epsilon$ is a small dimensionless parameter. However, this time we must explain why $\epsilon$ is so highly tuned to satisfy longevity of metastable state $\Phi_1=\Phi_2=\Phi_3=0$ and we do not adopt this scenario.

On the other hand, no-go theorem can not be applied in an extra dimensions, where new ways of GUT symmetry breaking mechanisms appear \cite{Witten} \cite{Breit} \cite{Kawamura}. This is one of very strong motivations for us to proceed to extra dimension. \footnote{The roles of more general no-go theorems in GUT were discussed in \cite{Fukuopen}}.

We may consider \bref{U1R} from string theory.
In string theory \cite{Polchinski}, it has originally global space-time SO(10) symmetry and is broken to $SO(4)\times SO(6)$ in 4D.
This SO(6) is isomorphic to SU(4).
The spinor in ten space-time dimensions has $16_L+16_R$ components. (Do not confuse with flabour group so far discussed.)  
In the splitting from 10 to (4+6) dimensions, this spinor is divided into four 4-component spinor, $\theta_a^{(i)}, \theta_{\dot{a}}^{(i)}~~(a=1,2),~~i=1,2,3,4$
So there is $SU(4)_R$ transformation
\be
\theta'^{(i)}=U^i_j\theta^{(j)}.
\ee

Taking all the arguments on the problems in the minimal SO(10) into considerations, we will mainly study the possibility of extending GUT to extra dimensions in the susequent sections.

Going beyond 4D to 5D or more extra dimensions, we achieve much more merits than demerits as we will explain. All previous correct predictions are remained valid, whereas the small mismatch comes from the tight mass relation of 
\bref{massmatrix} will be improved. 
Gauge coupling unification is recovered. Fast proton decay is evaded.

Though we have discussed bottom-up approach,  it is useful or indispensable to consider top-down approach also \cite{Raby2}\cite{Dienes}. In this case, as the scale of the bottom-up approach is close to Planck scale, it is expected to coincide with top-down approaches from string world \cite{Raby}.
The mutual coincidence and compatibility give also some informations to both approaches.

\paragraph{Comment against the Eingorn-Zhuk's arguments}

Before discussing SUSY GUT in 5D, we discuss the Eingorn-Zhuk's argument \cite{Eingorn} since many physicists consider it as the no-go theorem against the presence of the extra dimension.

They considered the reduced action in D+1 dimensions under the assumption that
\be
g_{ij}\rightarrow \eta_{ij} ~~(i,j=0,1,...,D) 
\ee
at spatial infinity,
\be
S=-Et+M\psi+S_3(r_3)+S_4(x^4)+....+S_D(x^D).
\ee
Here $E$ and $M$ are conserved energy and angular momentum of the system, respectively.
However, this action is consistent only if S is independent on the coordinates of extra dimensions, $x^4,...,x^D$.
In that case only, $M\psi$ term can be consistent and conserved quantity since
angular momentum in general defined by
\be
M^{\mu\nu}=\sum (p^\mu x^\nu-p^\nu x^\mu).
\ee
In their model, spatial isotropy is assumed only in usual three dimensions and therefore test particles have zero momentum
\be
p^\mu =0
\ee
in extra dimensions, which means
\be
\partial_\mu S=0 ~\mbox{for} ~\mu=4,..,D.
\ee
Under these assumptions they obtained
\be
S_3(r_3)=\int\left[\left(2Em+\frac{E^2}{c^2}\right)+\frac{1}{r_3}\left(m^2c^2r_g+2\frac{2(D-1)}{D-2}mEr_g\right)-\frac{1}{r_3}\left(M^2-\frac{Dm^2c^2r_g^2}{2(D-2)}\right)\right]^{1/2}dr_3,
\ee
where $r_g=\frac{2GM}{c^2}$ with the usual Newton constant $G$ and source mass $M$. 
For the Mercury perihelion precession is given by \cite{L-L}
\be
\delta \psi=\frac{D\pi m^2c^2r_g}{2(D-2)M^2},
\ee
which is consistent with observation if $D=3$, irrelevantly to the size of the extra dimensions.
It follows from these arguments that the extra dimensions are excluded if
\begin{itemize}
\item
Space-time is asymptotically flat.

and
\item
Fields other than gravitation does not enter into extra dimensions.
\end{itemize}
Of course, GUT models in extra dimensions which will be discussed in the next chapter are free from these constraints.
 
We consider SO(10) is realized in 5D and the symmetry in 4D is replaced by the Pati-Salam invariance. As the result of this change, the mass matrix of heavy right-handed neutrino which was represented by $M_{126}$ \bref{massmatrix}, one of the partners of mass matrices of charged fermions and Dirac neutrinos \bref{massmatrix}, is replaced by the independent mass matrix \bref{massmatrix2}. For that reason, mismatch of the data is resolved.
Also we can eliminate the Higgs component which was involved in ${\bf 126}$ and induced fast proton decay.
However, this is not the all reasons why we must go beyond 4D. We can recover the gauge coupling unification spoiled by the intermediate energy scales. 
Furthermore, it is free from the internal inconsistency like no-go theorem on the SUSY breaking mechanism.

\chapter{Orbifold SO(10) GUT in Five Dimensions}
In this chapter we exploit the orbifold SO(10) GUT.
By the compactification on the orbifold \cite{Kawamura},
 $N=1$ SUSY of the five dimensional theory, 
 corresponding to $N=2$ SUSY in the four dimensional point of view, 
 is broken down to four dimensional $N=1$ SUSY. 
Supersymmetric Lagrangian of this system can be described 
 in terms of the superfield formalism 
 of four dimensional $N=1$ SUSY theories \cite{SUSYL1, SUSYL2, SUSYL3}. 
There are lots of possibilities to construct such a model, 
 where some fields reside in the bulk and some reside 
 on branes. 
Bulk itself has two possibility that it is flat or warped.
We first discuss the case of warped model and, subsequently main flat case, orbifold GUT.
\section{Warped SO(10) GUT}

We embed the minimal SO(10) model \cite{F-K-O2}
 into a warped extra dimension model \cite{RS}. 
In this scenario, the warped metric gives rise to 
 an effective cutoff in 4-dimensional effective theory, 
 which is warped down to a low scale from the fundamental mass scale 
 of the original model (a higher dimensional Planck scale). 
We choose appropriate model parameters so as to realize 
 the effective cutoff scale as GUT scale. 
Furthermore, in the context of a warped extra dimension 
 we can propose a simple setup that naturally generates 
 right-handed neutrino masses at intermediate scale 
 even with Higgs field VEVs at the GUT scale. 
Thus, the gauge coupling unification remains 
 as usual in the MSSM. 
The fifth dimension is compactified on the orbifold $S^1/Z_2$ 
 with two branes, ultraviolet (UV) and infrared (IR) branes, 
 sitting on each orbifold fixed point. 
With an appropriate tuning for cosmological constants 
 in the bulk and on the branes, 
 we obtain the warped metric \cite{RS}, 
\begin{eqnarray}
 d s^2 = e^{-2 k r_c |y|} \eta_{\mu \nu} d x^{\mu} d x^{\nu} 
 - r_c^2 d y^2 \; , 
\end{eqnarray}
 for $-\pi\leq y\leq\pi$, where $k$ is the AdS curvature, and 
 $r_c$ and $y$ are the radius and the angle of $S^1$, respectively.

The most important feature of the warped extra dimension model 
 is that the mass scale of the IR brane is warped down to 
 a low scale by the warp factor $ \omega = e^{-k r_c \pi}$ in four dimensional effective theory. 
For simplicity, we take the cutoff of the original five dimensional theory 
 and the AdS curvature as 
 $M_5 \simeq k \simeq M_P=2.4 \times 10^{18}$ GeV, 
 the four dimensional (reduced) Planck mass,
 and so we obtain the effective cutoff scale 
 as $\Lambda_{IR}= \omega M_P$ in effective four dimensional theory. 
Now let us take the warp factor so as for the GUT scale 
 to be the effective cutoff scale 
 $ M_{\rm GUT}= \Lambda_{IR}=\omega M_P$, 
 namely $\omega \simeq 0.01$ \cite{Nomura:2006pn}. 
As a result, we can realize, as four dimensional effective theory, 
 the minimal SUSY SO(10) model 
 with the effective cutoff at the GUT scale.

Before going to a concrete setup of the minimal SO(10) model 
 in the warped extra dimension, 
 let us see Lagrangian for the hypermultiplet in the bulk, 
\begin{eqnarray}
{\cal L} &=& \int dy \left\{ 
\int d^4 \theta \; r_c \; e^{- 2 k r_c |y|} 
 \left( 
 H^{\dagger} e^{- V} H + H^{c} e^{ V}H^{c \dagger} 
 \right) \right. \nonumber \\
&+& 
\left. 
\int d^2 \theta e^{-3 k r_c |y|}
 H^{c} \left[ 
  \partial_{y} - \left( \frac{3}{2}-c \right) k r_c \epsilon(y) 
 - \frac{\chi}{\sqrt{2}}  \right]  
 H  +h.c. \right \} \; , 
\label{bulkL}
\end{eqnarray}
where $c$ is a dimensionless parameter, 
$\epsilon(y)=y/|y|$ is the step function, 
 $H, ~H^c$ is the hypermultiplet charged under some gauge group, 
 and 
\bea 
V &=& - \theta \sigma^\mu \bar{\theta} A_\mu
    -i \bar{\theta}^2 \theta \lambda_1  +i \theta^2 \bar{\theta} \bar{\lambda}_1
    + \frac{1}{2} \theta^2 \bar{\theta}^2 D \; , 
\nonumber \\
\chi &=&  \frac{1}{\sqrt{2}}(\Sigma + i A_5) +\sqrt{2} \theta\lambda_2
     + \theta^2 F 
\eea
 are the vector multiplet and the adjoint chiral multiplets, 
 which form an $N=2$ SUSY gauge multiplet. 
 $Z_2$ parity for $H$ and $V$ is assigned as even, 
 while odd for $H^c$ and $\chi$.

When the gauge symmetry is broken down, 
 it is generally possible that the adjoint chiral multiplet 
 develops its VEV \cite{Kitano:2003cn}. 
Since its $Z_2$ parity is odd, 
 the VEV has to take the form, 
\bea
\left<\Sigma \right> = 2 \alpha k r_c  \epsilon(y) \; , 
\eea
where the VEV has been parameterized by a parameter $\alpha$. 
In this case, the zero mode wave function of $H$ 
 satisfies the following equation of motion:
\bea
\left[\partial_y -
 \left(\frac{3}{2}-c + \alpha \right) k r_c \epsilon(y) \right]H =0 \; ,
\eea
which yields 
\bea
H = \frac{1}{\sqrt{N}} 
 e^{ (3/2-c + \alpha) kr_c |y|} \; h(x^\mu) \; , 
\eea
where $h(x^\mu)$ is the chiral multiplet in four dimensions. 
Here, $N$ is a normalization constant 
 by which the kinetic term is canonically normalized, 
\be
\frac{1}{N} 
 =  \frac{(1-2 c+2 \alpha) k }
{e^{(1-2 c+2 \alpha) k r_c \pi}-1} \; . 
\ee
Hence, at $y=\pi$, the wave function becomes 
\bea
H(y=\pi) \simeq 
\sqrt{ (1-2 c + 2 \alpha ) k } \;  \omega^{-1} \; h(x^\mu) 
\eea
 if $e^{ (1/2 - c + \alpha ) k r_c \pi}  \gg 1$, while 
\bea 
H(y=\pi) \simeq 
 \sqrt{- (1-2 c + 2 \alpha ) k } \;  \omega^{-1} 
 e^{ (1/2 - c +  \alpha ) k r_c \pi}  \; h(x^\mu) 
\eea 
 for $ e^{ (1/2 -  c +  \alpha ) k r_c \pi}  \ll 1$. 

Lagrangian for a chiral multiplets on the IR brane is given by 
\bea 
 {\cal L}_{IR}=  
 \int d^4 \theta \;  \omega^\dagger \omega \;  \Phi^\dagger \Phi 
 +\left[   \int d^2 \theta \;  \omega^3 \; W(\Phi) + h.c.   
 \right] \; ,
\eea 
where we have omitted the gauge interaction part 
 for simplicity. 
If it is allowed by the gauge invariance, 
 we can write the interaction term 
 between fields in the bulk and on the IR brane, 
\bea 
{\cal L}_{int}= \int d^2 \theta \omega^3 
 \frac{Y}{\sqrt{M_5}} \Phi^2 H(y=\pi) +h.c. \; ,  
\label{IR-Yukawa} 
\eea  
where $Y$ is a Yukawa coupling constant, 
 and $M_5$ is the five dimensional Planck mass 
 (we take $M_5\sim M_P$ as mentioned above, for simplicity). 
Rescaling the brane field $\Phi \rightarrow  \Phi/\omega$ 
 to get the canonically normalized kinetic term 
 and substituting the zero-mode wave function of the bulk fields, 
 we obtain Yukawa coupling constant 
 in effective four dimensional theory as 
\bea 
  Y_{4D} \sim Y 
\eea 
 for $e^{ (1/2 - c +  \alpha ) k r_c \pi}  \gg 1$,  
 while 
\bea 
  Y_{4D} \sim Y 
  \times e^{ (1/2 -  c +  \alpha ) k r_c \pi}  \ll Y \; , 
\label{suppression}
\eea  
 for $e^{ (1/2 - c + \alpha ) k r_c \pi }  \ll 1$. 
In the latter case, we obtain a suppression factor 
 since $H$ is localized around the UV brane.

Now we give a simple setup of the minimal SO(10) model 
 in the warped extra dimension. 
We put all ${\bf 16}$ matter multiplets on the IR ($y=\pi$) brane, 
 while the Higgs multiplets ${\bf 10}$ and $\overline{\bf 126}$ 
 are assumed to live in the bulk. 
In Eq.~(\ref{IR-Yukawa}), replacing the brane field into the matter 
 multiplets and the bulk field into the Higgs multiplets, 
 we obtain Yukawa couplings in the minimal SO(10) model. 
The Lagrangian for the bulk Higgs multiplets are given 
 in the same form as Eq.~(\ref{bulkL}), 
 where $\chi$ is the SO(10) adjoint chiral multiplet, ${\bf 45}$. 
As discussed above, since the SO(10) gauge group is broken 
 down to the SM one, 
 some components in $\chi$ which is singlet under the SM gauge group 
 can in general develop VEVs. 
Here we consider a possibility that 
 the ${\rm U}(1)_X$ component 
 in the adjoint $\chi ={\bf 45}$ under the decomposition 
 SO(10) $\supset {\rm SU}(5) \times {\rm U}(1)_X$ has 
 a non-zero VEV\footnote{
Since $\chi$ has an odd $Z_2$ parity, its non-zero VEV leads to 
 the Fayet-Iliopoulos D-terms localized 
 on both the UV and IR branes \cite{D-term}, 
 which should be canceled to preserve SUSY. 
For this purpose, we have to introduce new fields on both branes 
 by which the D-terms are compensated. 
If such fields are in the same representations 
 as matters or Higgs fields like ${\bf 16}$ or $\overline{\bf 126}$, 
 we would need to impose some global symmetry to distinguish them. 
}, 
\bea
 {\bf 45} = {\bf 1}_0
 \oplus {\bf 10}_{+4} \oplus \overline{\bf 10}_{-4}
 \oplus {\bf 24}_0 \; .   \nonumber 
\eea
The ${\bf 10}$ Higgs multiplet and 
 the $\overline{\bf 126}$ Higgs multiplet 
 are decomposed under ${\rm SU}(5) \times {\rm U}(1)_X$ as
\bea
{\bf 10} &=& {\bf 5}_{+2} \oplus \overline{\bf 5}_{-2} \; , 
 \nonumber \\
\overline{\bf 126} &=& 
 {\bf 1}_{+10} 
 \oplus {\bf 5}_{+2} \oplus \overline{\bf 10}_{+6}
 \oplus {\bf 15}_{-6} 
 \oplus \overline{\bf 45}_{-2} \oplus {\bf 50}_{+2} \; . 
 \nonumber  
\eea
In this decomposition, 
 the coupling between a bulk Higgs multiplet and 
 the ${\rm U}(1)_X$ component in $\chi$ is proportional 
 to  ${\rm U}(1)_X$ charge, 
\be
{\cal L}_{int} \supset \frac{1}{2} \int d^2 \theta \omega^3
Q_X \langle \Sigma_X \rangle H^c H + h.c. \;,
\ee
 and thus each component effectively obtains 
 the different bulk mass term,  
\bea 
  \left( \frac{3}{2} - c \right) k r_c  
   + \frac{1}{2}Q_X \langle \Sigma_X \rangle, 
\label{bulkmass}  
\eea
 where $Q_X$ is the ${\rm U}(1)_X$ charge of corresponding Higgs multiplet, 
 and $\Sigma_X$ is the scalar component of 
 the ${\rm U}(1)_X$ gauge multiplet (${\bf 1}_0$). 
Now we obtain different configurations of the wave functions 
 for these Higgs multiplets. 
Since the ${\bf 1}_{+10}$ Higgs has a large ${\rm U}(1)_X$ charge 
 relative to other Higgs multiplets, 
 we can choose parameters $c$ and $\langle \Sigma_X \rangle$ 
 so that Higgs doublets are mostly localized around the IR brane 
 while the ${\bf 1}_{+10}$ Higgs is localized around the UV brane. 
For example, the parameter choice, 
 $c=-7/2$ for both ${\bf 10}$ and $\overline{\bf 126}$ Higgs multiplets 
 and $ \langle \Sigma_X \rangle = - k r_c$, 
 can realize this situation.

Using the decomposition of matter multiplets, 
\bea
 {\bf 16}^i = {\bf 1}_{-5}^i 
 \oplus \overline{\bf 5}_{+3}^i  \oplus {\bf 10}_{-1}^i  \; ,
 \nonumber  
\eea
the Yukawa couplings between matters and 
 the $\overline{\bf 126}$ Higgs multiplet 
 on the IR brane are decomposed into  
\bea
 W_{Y_{126}} &=& 
  Y_u^{ij} {\bf 5}_{+2} {\bf 10}_{-1}^i {\bf 10}_{-1}^j 
+ Y_d^{ij} \overline{\bf 45}_{-2} 
  \overline{\bf 5}_{+3}^i {\bf 10}_{-1}^j \nonumber \\
&+& Y_D^{ij} {\bf 5}_{+2} {\bf 1}_{-5}^i \overline{\bf 5}_{+3}^j 
 + Y_e^{ij} \overline{\bf 45}_{-2} \overline{\bf 5}_{+3}^i {\bf 10}_{-1}^j
\nonumber \\ 
&+& 
  Y_{\nu_L}^{ij} {\bf 15}_{-6}
   \overline{\bf 5}_{+3}^i  \overline{\bf 5}_{+3}^j 
+ Y_{\nu_R}^{ij} {\bf 1}_{+10} {\bf 1}_{-5}^i {\bf 1}_{-5}^j \; . 
\eea
Here, all the Yukawa couplings coincide with 
 the original Yukawa coupling $Y_{126}$ 
 up to appropriate CG coefficients. 
As discussed above, the ${\bf 1}_{+10}$ Higgs multiplet 
 giving masses for right-handed neutrinos
 is localized around the UV brane and, therefore, 
 we obtain a suppression factor 
 as in Eq.~(\ref{suppression}) 
 for the effective Yukawa coupling between 
 the Higgs and right-handed neutrinos. 
In effective four dimensional description, 
 the GUT mass matrix relation is partly broken down, 
 and the last term in Eq.~(\ref{Yukawa3}) is replaced into 
\bea 
 Y_{126}^{ij} v_R \rightarrow Y_{126}^{ij} (\epsilon v_R) \; ,  
\eea
where $\epsilon$ denotes the suppression factor. 
By choosing appropriate parameters 
 so as to give $\epsilon \simeq 10^{-2}$, 
 we can take $v_R \simeq M_{\rm GUT}$ 
 and keep the successful gauge coupling unification in the MSSM. 
In fact, the above parameter set, 
 $c=-7/2$ and $ \langle \Sigma_X \rangle = - k r_c$, 
 leads to $\epsilon = \omega = M_{\rm GUT}/M_P \simeq 10^{-2}$. 
The other Higgs multiplets are localized around the IR brane, 
 so that there is no suppression factor for other effective 
 Yukawa couplings.

In our setup, all the matters reside on the brane 
 while the Higgs multiplets reside in the bulk. 
This setup shares the same advantage as 
 the so-called orbifold GUT \cite{Kawamura,
Altarelli:2001qj, Hall:2001pg}. 
We can assign even $Z_2$ parity 
 for MSSM doublet Higgs superfields 
 while odd for triplet Higgs superfields, 
 as a result, the proton decay process through 
 dimension five operators are forbidden. 

In order to solve these problems, we have considered 
 the minimal SO(10) model in the warped extra dimension. 
As a simple setup, we have assumed that matter multiplets 
 reside on the IR brane 
 while the Higgs multiplets reside in the bulk. 
The warped geometry leads to a low scale effective cutoff
 in effective four dimensional theory, 
 and we fix it at the GUT scale. 
Therefore, the four dimensional minimal SO(10) model 
 is realized as the effective theory with the GUT scale cutoff.

After the GUT symmetry breaking, the adjoint scalar in  
 the gauge multiplet in five dimensional SUSY 
 can generally develop a VEV, 
 which plays a role of bulk mass for the bulk Higgs multiplets. 
This bulk mass is proportional to the charge 
 of each Higgs multiplets and cause the difference 
 between wave functions of each Higgs multiplet. 
We have found the possibility that 
 the singlet Higgs which provides right-handed neutrino with masses 
 is localized around the UV brane and the geometrical suppression factor 
 emerges in Yukawa couplings of the right-handed neutrinos. 
As a result, we can set the mass scale of the right-handed neutrinos 
 at the intermediate scale, nevertheless the singlet Higgs VEV is around the GUT scale. 
All Higgs multiplets naturally have masses around GUT scale 
 and the gauge coupling unification in the MSSM remains unchanged.

We give some additional comments. 
One can easily extend our setup to put some 
 of matter multiplets in the bulk 
 \cite{Grossman:1999ra, Chang:1999nh, Gherghetta:2000qt}. 
In this case, we may explain the fermion mass hierarchy 
 in terms of the different overlapping of 
 fermion wave functions between different generations. 
In this section, we have assumed that GUT gauge symmetry 
 is successfully broken down to the SM one. 
There are several possibilities for GUT symmetry breaking. 
It is easy to introduce appropriate Higgs multiplets 
 and superpotential so as to break the GUT symmetry 
 on a brane as in four dimensional SO(10) models. 
We also be able to introduce an appropriate boundary conditions 
 for bulk gauge multiplets to (explicitly) break 
 the GUT symmetry to a subgroup with rank five in total, 
 as in the orbifold GUTs. 

\section{Model Setup of SO(10) GUT in 5D}
From this section we will realize the new model compatible with no-go theorem discussed in the last part of 
previous chapter.
That is, we will discuss SO(10) orbifold GUT proposed by Ref.~\cite{F-O1}. 
From the view of string theory, we should consider ten dimensional world.
However, we will discuss essentially bottom up approach and incorporate the indispensable effect of extra dimension as minimally as possible. 

The model is described in 5D and 
 the fifth dimension is compactified 
 on the orbifold $S^1/{Z_2 \times Z_2^\prime}$. \footnote{Such orbifold condition is also geometrically derived from the singularity free condition \cite{Eguchi}.}
A circle $S^1$ with radius $R$ is divided by 
 a $Z_2$ orbifold transformation $y \to -y$ 
 ($y$ is the fifth dimensional coordinate $ 0 \leq y < 2 \pi R$)
 and this segment is further divided by a $Z_2^\prime$ transformation 
 $y^\prime \to -y^\prime $ with $y^\prime = y + \pi R/2$. 
There are two inequivalent orbifold fixed points at $y=0$ and $y=\pi R/2$. 
Under this orbifold compactification, a general bulk wave function 
 is classified with respect to its parities,  
 $P=\pm$ and $P^\prime=\pm$, under $Z_2$ and $Z_2^\prime$, respectively.

Assigning the parity ($P,P^\prime $) 
 the bulk SO(10) gauge multiplet as listed in Table~I, 
 only the PS gauge multiplet has zero-mode 
 and the bulk 5D N=1 SUSY SO(10) gauge symmetry is broken 
 to 4D N=1 SUSY PS gauge symmetry. 
Since all vector multiplets has wave functions  
 on the brane at $y=0$, SO(10) gauge symmetry is respected there, 
 while only the PS symmetry is on the brane at $y=\pi R/2$ (PS brane). 

\begin{table}[h]
\begin{center}
\begin{tabular}{|c|c|c|}
\hline
$(P,P')$ & bulk field & mass\\
\hline 
& & \\
$(+,+)$ & $V(15,1,1)$, $V(1,3,1)$, $V(1,1,3)$ & $\frac{2n}{R}$\\
& & \\
\hline
& & \\
$(+,-)$ &  $V(6,2,2)$ & $\frac{(2n+1)}{R}$ \\
& & \\
\hline
& & \\
$(-,+)$ &  $\Phi (6,2,2)$
& $\frac{(2n+1)}{R}$\\
& & \\
\hline
& & \\
$(-,-)$ & $\Phi (15,1,1)$, $\Phi (1,3,1)$, $\Phi (1,1,3)$ & $\frac{(2n+2)}{R}$ \\
& & \\
\hline
\end{tabular}
\end{center}
\caption{
 ($P,~P^\prime$) assignment and masses ($n \geq 0$) of fields 
 in the bulk SO(10) gauge multiplet $(V,~\Phi)$ 
 under the PS gauge group. 
$V$ and $\Phi$ are the vector multiplet and adjoint chiral 
 multiplet in terms of 4D N=1 SUSY theory. 
}
\label{t1}
\end{table}

Boundary condition in brains are very elegant introduction to
GUT model, whose original idea was introduced to circumvent the
mirror particles. These particles appear when we incorporate all matters of
three generations preserving chiral symmetry.

We place the all matter and Higgs multiplets on the PS brane, 
 where only the PS symmetry is manifest 
 so that the particle contents are in the representation 
 under the PS gauge symmetry, not necessary to be 
 in SO(10) representation.   
For a different setup, see \cite{Raby}. 
The matter and Higgs in our model is listed in Table \ref{t1}. 
For later conveniences, let us introduce the following notations: 
\bea
H_1&=&({\bf 1},{\bf 2},{\bf 2})_H, ~H_1^{\prime}=({\bf 1},{\bf 2},{\bf 2})'_H,
\nonumber \\
H_6&=&({\bf 6},{\bf 1},{\bf 1})_H, ~H_{15}=({\bf 15},{\bf 1},{\bf 1})_H,
\nonumber \\ 
H_L&=&({\bf 4},{\bf 2},{\bf 1})_H,
~\overline{H_L} =(\overline{{\bf 4}},{\bf 2},{\bf 1})_H,  
\nonumber \\
\phi&=&({\bf 4},{\bf 1},{\bf 2})_H,
~\bar{\phi}=(\overline{{\bf 4}},{\bf 1},{\bf 2})_H.
\label{PSHiggs}
\eea
under the PS group $SU(4)_c\times SU(2)_L\times SU(2)_R$.

Superpotential relevant to fermion masses is given by%
\footnote{
For simplicity, we have introduced only minimal terms 
 necessary for reproducing observed fermion mass matrices. 
}
\bea
W_Y&=& Y_{1}^{ij} F_{Li} F_{Rj}^c H_1 
+\frac{Y_{15}^{ij}}{M_5} F_{Li} F_{Rj}^c 
 \left(H_1^{\prime} H_{15} \right) \nonumber\\ 
&+&\frac{Y_R^{ij}}{M_5} F_{Ri}^c F_{Rj}^c 
 \left(\phi \phi \right)  
 +\frac{Y_L^{ij}}{M_5} F_{Li}F_{Lj} 
 \left(\overline{H_L} \overline{H_L} \right).
\label{Yukawa}
\eea 
Here the notations are as follows: $M_5$ is the 5D Planck scale. 
$F_{Li}$ and $F_{Ri}^c$ are matter multiplets 
 of i-th generation in $({\bf 4, 2, 1})$ and $({\bf \bar{4}, 1, 2})$ 
 representations, respectively. $H_1=({\bf 1},{\bf 2},{\bf 2})$,  
$H_1'=({\bf 1},{\bf 2},{\bf 2})'$, $H_{15}=({\bf 15},{\bf 1},{\bf 1})_H$,~
$H_6=({\bf 6},{\bf 1},{\bf 1})_H$ (See \bref{HiggsW}), $\phi=({\bf 4},{\bf 1} ,{\bf 2})_H$, $\overline{\phi}=(\overline{{\bf 4}},{\bf 1},{\bf 2})_H$, $H_L=({\bf 4},{\bf 2},{\bf 1})_H$, $\overline{H_L}=(\overline{{\bf 4}},{\bf 2},{\bf 1})_H$ are Higgs multiplets.

The product, $H_1^{\prime} H_{15}$, effectively works 
 as $({\bf 15},{\bf 2},{\bf 2})_H$, 
 while $\phi \phi$ and $\overline{H_L}\overline{H_L}$ 
 effectively work as $({\bf 10},{\bf 1},{\bf 3})$ and 
 $(\overline{{\bf 10}},{\bf 3},{\bf 1})$, respectively, 
 and are responsible for the left- and the right-handed 
 Majorana neutrino masses. Thus $W_Y$ inherits the essential part of the minimal SO(10) GUT model. Providing VEVs for appropriate Higgs multiplets, 
 fermion mass matrices are obtained. 

\begin{table}[h]
{\begin{center}
\begin{tabular}{|c|c|}
\hline
& brane at $y=\pi R/2$ \\ 
\hline
& \\
Matter Multiplets & $\psi_i=F_{Li} \oplus F_{Ri}^c \quad (i=1,2,3)$ \\
 & \\
\hline
 & \\
Higgs Multiplets & 
$({\bf 1},{\bf 2},{\bf 2})_H$,  
$({\bf 1},{\bf 2},{\bf 2})'_H$,
$({\bf 15},{\bf 1},{\bf 1})_H$,
$({\bf 6},{\bf 1},{\bf 1})_H$ \\  & 
$({\bf 4},{\bf 1},{\bf 2})_H$, 
$(\overline{{\bf 4}},{\bf 1},{\bf 2})_H$, 
$({\bf 4},{\bf 2},{\bf 1})_H$, 
$(\overline{{\bf 4}},{\bf 2},{\bf 1})_H$  \\
& \\
\hline
\end{tabular}
\end{center}}
\caption{
Particle contents on the PS brane. 
$F_{Li}$ and $F_{Ri}^c$ are matter multiplets 
 of i-th generation in $({\bf 4, 2, 1})$ and $({\bf \bar{4}, 1, 2})$ 
 representations, respectively. 
}
\end{table}

\begin{eqnarray}
\label{massmatrix2}
 M_u &=& c_{10} M_{1,2,2}+ c_{15} M_{15,2,2} \; , 
 \nonumber \\
 M_d &=& M_{1,2,2} + M_{15,2,2} \; ,   
 \nonumber \\
 M_D &=& c_{10} M_{1,2,2} - 3 c_{15} M_{15,2,2} \; ,  \\
 M_e &=& M_{1,2,2} - 3 M_{15.2,2} \; , 
 \nonumber \\
 M_L &=& c_L M_{10,3,1} \; ,
 \nonumber \\ 
 M_R &=& c_R M_{10,1,3} \; . \nonumber
\end{eqnarray}  

Two important remarks are in order. \\
1. $M_{15,2,2}$ is, in general, not symmetric unlike $M_{126}$. However, we imposed the
L-R symmetry ${\bf 4,1,2}\leftrightarrow {\bf \bar{4},2,1}$, which implies that bothy  $M_{1,2,2}$ and $M_{15,2,2}$ matrices are symmetric and mass structure of charged fermions and Dirac neutrino is same as that in SO(10).\\
2. $M_L$ and $M_R$ are independent on those of the charged fermions and the Dirac neutrino unlike the SO(10) case \bref{massmatrix}.
So the precise data fitting becomes possible without changing $Y_\nu$.
This is very important especially for LFV and leptogenesis as will be discussed
in the subsequent sections.

$H_6$ is necessary to make the color triplet heavy. However, there arises no D-T problem since they are not involved in the same multiplet. There are sufficient numbers of free parameters 
 to fit all the observed fermion masses and mixing angles. 

We introduced Higgs superpotential invariant under the PS symmetry \cite{F-O1}
 such as \footnote{It is possible to consider a different superpotential by introducing a singlet chiral superfield, which will be discussed in section \ref{smoothhb}.}
\bea
W_{PS} &=& 
 \frac{m_1}{2} H_1^2 + \frac{m_1^\prime}{2} H_1^{\prime 2} 
 + m_{15}~{\rm tr}\left[H_{15}^2 \right] 
  +m_4 \left(\overline{H_L}H_L+\overline{\phi}\phi\right) \nonumber\\
&+& 
\left(H_L \overline{\phi}+ \overline{H_L} \phi \right) 
\left( \lambda_1 H_1 + \lambda_1^\prime H_1^\prime \right) 
+\lambda_{15} \left(\overline{\phi} \phi + \overline{H_L} H_L\right) 
H_{15} \nonumber\\
&+&
\lambda~{\rm tr}\left[H_{15}^3 \right]+
\lambda_6 
 \left( H_L^2+ \overline{H_L}^2 + \phi^2 + \overline{\phi}^2 \right) 
 H_6 .
\label{HiggsW}
\eea

Parameterizing 
 $ \langle H_{15} \rangle =\frac{v_{15}}{2 \sqrt{6}} 
 {\rm diag}(1,1,1,-3)$, 
 SUSY vacuum conditions from Eq.~(\ref{HiggsW}) and 
 the D-terms are satisfied by solutions,  
\bea
v_{15} =\frac{2 \sqrt{6}}{3 \lambda_{15}} m_4,~~~
\langle           \phi  \rangle = 
\langle \overline{\phi} \rangle = 
\sqrt{
 \frac{8 m_4}{3 \lambda_{15}^2} 
 \left( m_{15} -\frac{\lambda}{\lambda_{15}} m_4 \right) }
 \equiv v_{PS} 
\eea 
and others are zero, by which the PS gauge symmetry is broken 
 down to the SM gauge symmetry.  
We choose the parameters so as to be 
 $ v_{15} \simeq \langle \phi \rangle = \langle \overline{\phi} \rangle$.  
Note that the last term in Eq.~(\ref{HiggsW}) is necessary 
 to make all color triplets in $\phi$ and $\overline{\phi}$ heavy.

Weak Higgs doublet mass matrix is given by
\begin{equation}
\left(
       \begin{array}{ccc}
        H_1, & H_1^\prime, & H_L \end{array}
\right)\left(
        \begin{array}{ccc}
        m_1 &  0         & \lambda_1        \langle \phi \rangle \\
        0   & m_1^\prime & \lambda_1^\prime \langle \phi \rangle \\ 
        \lambda_1 \langle \overline{\phi} \rangle &  
        \lambda_1^\prime \langle \overline{\phi} \rangle & m_4 
        \end{array}
\right)\left(
        \begin{array}{c}
        H_1\\
        H_1^\prime \\
        \overline{H_L}
        \end{array}
\right).
\end{equation} 
In order to realize the MSSM at low energy, 
 only one pair of Higgs doublets out of the above tree pairs 
 should be light, while others have mass of the PS symmetry breaking scale. 
This doublet-doublet Higgs mass splitting requires 
 the fine tuning of parameters to satisfy 
\bea
\det M=m_1 m_1^\prime m_4 - 
 (m_1 \lambda_1^{\prime 2} + m_1^\prime \lambda_1^2) v_{PS}^2=0.
\label{vev}
\eea

Suppose a Higgs superpotential which provides 
 the same VEVs ($v_{\rm PS}$) for Higgs multiplets to break the PS symmetry 
 and leaves only the particle contents of the minimal supersymmetric 
 Standard Model (MSSM) at low energies. 
In Ref.~\cite{F-O1}, assuming $M_c=v_{\rm PS}$ and imposing 
 the left-right symmetry, the gauge coupling unification was examined. 
Analyzing the gauge coupling running in the MSSM, 
 $M_c=v_{\rm PS}$ is fixed as the scale 
 where the SU(2)$_L$ and SU(2)$_R$ gauge couplings coincide 
 with each other, which is found to be  
 $M_c=v_{\rm PS}=1.2 \times 10^{16}$ GeV. 
For the scale $\mu \geq M_c=v_{\rm PS}$,  
 we have only two independent gauge couplings, 
 SU(4)$_c$ and SU(2)$_L$ (or SU(2)$_R$) gauge couplings. 
After taking KK mode contributions into account, 
 it was shown that the gauge coupling is successfully unified at  
 $M_{\rm GUT}=4.6 \times 10^{17}$ GeV 
 (see Figure~1 from Ref.~\cite{F-O1}). 
We assume that a more fundamental SO(10) GUT theory 
 takes place at $M_{GUT}$, 
 and it would be natural to assume $M_{GUT} \sim M_5$. 
In fact, the relation between 4D and 5D Planck scales, 
 $M_5^3/M_c \simeq M_P^2$ 
 ($M_P=2.44 \times 10^{18}$ GeV is the reduced Planck scale), 
 supports this assumption with $M_c=1.2 \times 10^{16}$ GeV. 
When we abandon the left-right symmetry, 
 there is more freedom for the gauge coupling unification 
 with two independent parameters $v_{PS}$ and $M_c$.

Our model gives a large $v_{PS}$ relative to 
 other 5D orbifold SO(10) models \cite{Raby}. 
The high value of $v_{PS}$ or $M_c$ is advantageous 
 for dangerous proton decay due to dimension six operators. 
 From Eq.~(\ref{Yukawa}), the right-handed neutrino mass scale 
 is given by $M_R \sim Y_R v_{PS}^2/M_5 \sim Y_R v_{PS}^2/M_{\rm GUT}$. 
The scale $M_R = {\cal O}(10^{14}~\mbox{GeV})$ 
 preferable for the seesaw mechanism can be obtained 
 by a mild tuning of the Yukawa coupling $Y_R \sim 0.1$.

In the next section, we show that this model is well fitted 
 to the smooth hybrid inflation model and the model predictions 
 are compatible with the cosmological observations 
 like the power spectrum of the curvature perturbations, 
 the scalar spectral index, the ratio of scalar-to-tensor 
 fluctuations and so on. 
\paragraph{Proton decay}

Dimension five operator in nucleon decay does not appear since $H_6$ does not couple with matters
as in \bref{Yukawa}. Dimension six operator is also suppressed because $M_X$ is increased in 5D.
\section{Gauge Coupling Unification in 5D}
In the orbifold GUT model, we assume that 
 the GUT model takes place at some high energy 
 beyond the compactification scale.  
For the theoretical consistency of the model, 
 the gauge coupling unification should be realized 
 at some scale after taking into account 
 the contributions of Kaluza-Klein modes 
 to the gauge coupling running \cite{Hall} \cite{GHU}. 

In our setup, the evolution of gauge coupling 
 has three stages, $G_{321}$ (SM+MSSM), $G_{422}$ (the PS) 
 and $M_c =1/R$. 
For simplicity, we assume $v_{PS}=M_c$ in our analysis. 
Furthermore, since we have imposed the left-right symmetry, 
 SU(2)$_L$ and SU(2)$_R$ gauge couplings 
 must coincide with each other at the scale $\mu=v_{PS}$. 
As a consequence, the PS scale is fixed 
 from the gauge coupling running in the MSSM stage. 

In the $G_{321}$ stage, we have 
\bea
\frac{1}{\alpha_i (\mu)}=\frac{1}{\alpha_i(M)}
 +\frac{1}{2\pi}b_i\mbox{ln}\left(
\frac{M}{\mu}\right); ~~(i=3,2.1), 
\eea
were $b_i$s are
\bea
b_3=-7,~b_2=-19/6,~b_1=41/10
\eea
for $M_Z <\mu < M_{SUSY}$ and
\bea
b_3=-3,~b_2=1,~b_1=33/5
\eea
for $M_{SUSY} < \mu < M_c=v_{PS}$. 
At the PS scale, the matching condition holds
\bea
\alpha_3^{-1}(M_c)&=&\alpha_4^{-1}(M_c)\nonumber\\
\alpha_2^{-1}(M_c)&=&\alpha_{2L}^{-1}(M_c)\nonumber\\
\alpha_1^{-1}(M_c)&=&[2\alpha_4^{-1}(M_c)+3\alpha_{2R}^{-1}(M_c)]/5
\label{maching}
\eea
In the PS stage $\mu > M_c$, the threshold corrections 
 $\Delta_i$ due to KK mode in the bulk are added, 
\bea
\frac{1}{\alpha_i (\mu)}=\frac{1}{\alpha_i(M_c)}
+\frac{1}{2\pi}b_i\mbox{ln}\left(
\frac{M_c}{\mu}\right)+\Delta_i.~~(i=4,2_L,2_R)
\eea
The beta functions from the matter and Higgs multiplets 
 on the PS brane are 
\be
 b_4=3, ~b_{2L}=b_{2R}=6 .
\label{lrs}
\ee
KK mode contributions are given by 
\bea
\Delta_i&=& \frac{1}{2\pi}b_i^{even}\sum_{n=0}^{N_l}
\theta(\mu-(2n+2)M_c)\mbox{ln}\frac{(2n+2)M_c}{\mu} \nonumber\\  
&+&
\frac{1}{2\pi}b_i^{odd}\sum_{n=0}^{N_l}
\theta(\mu-(2n+1)M_c)\mbox{ln}\frac{(2n+1)M_c}{\mu}
\eea
with 
\bea
b_i^{even}&=&(-8,-4,-4) , \nonumber\\ 
b_i^{odd}&=&(-8,-12,-12)  
\eea
under $G_{422}$. 
\begin{figure}[t]
\begin{center}
\includegraphics[scale=1.2]{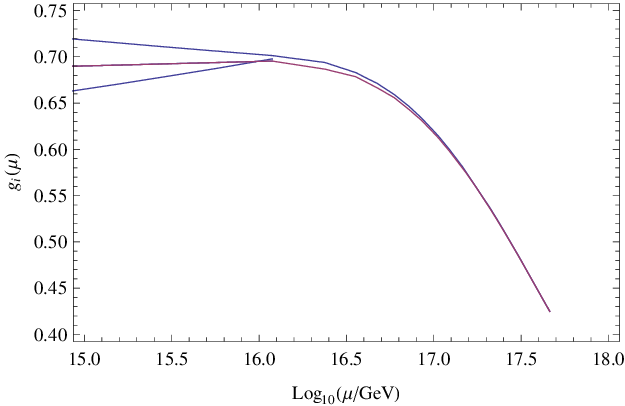}
\caption{
Gauge coupling unification in the left-right symmetric case, 
 taken from Ref.~\cite{F-O3}. 
Each line from top to bottom corresponds to 
 $g_3$, $g_2$ and $g_1$ for $ \mu < M_c=v_{\rm PS}$, 
 while  $g_3=g_4$ and $g_2=g_{2R}$ for $ \mu > M_c=v_{\rm PS}$. 
}
\label{GCU2}
\end{center}
\end{figure}
Fig.\ref{GCU2} shows the gauge coupling unification 
 for the left-right symmetric case. 
As the characteristic property of gaugino mediation, KK modes
affect on the asymptotic free direction.
The PS (compactification) scale, $M_c$, is determined 
 from the gauge coupling running 
 in the MSSM stage by imposing the matching condition, 
 $\alpha_2^{-1}(M_c)= \alpha_{2R}^{-1}(M_c) 
 =(5\alpha_1^{-1}(M_c)-2\alpha_3^{-1}(M_c))/3$, 
 and we find 
\bea 
 v_{PS}=M_c = 1.2\times 10^{16} \mbox{GeV} . 
\eea 
 for the inputs, 
 $(\alpha_1(M_Z), \alpha_2(M_Z), \alpha_3(M_Z))
 = (0.01695, 0.03382, 0.1176)$ and $M_{SUSY}=1$ TeV. 
For the scale $\mu > M_c$, there are only two independent 
 gauge couplings $\alpha_4$ and $\alpha_2=\alpha_{2R}$, 
 and so the gauge coupling unification is easily realized. 
We find the unification scale as 
\bea 
  M_{GUT} = 4.6 \times 10^{17} \mbox{GeV} . 
\label{GUT2}
\eea  
As mentioned before, we assume that a more fundamental 
 SO(10) GUT theory takes place at $M_{GUT}$, 
 and it would be natural to assume $M_{GUT} \sim M_5$. 
In fact, the relation between 4D and 5D Planck scales, 
 $M_5^3/M_c \simeq M_P^2$ 
 ($M_P=2.4 \times 10^{18}$ GeV is the reduced Planck scale), 
 supports this assumption 
 with $M_c=1.2 \times 10^{16}$ GeV. 
When we abandon the left-right symmetry, 
 there is more freedom for the gauge coupling unification 
 with two independent parameters $v_{PS}$ and $M_c$. 
We will consider the gauge coupling unification for
\be
M_c>v_{PS}
\label{Mc}
\ee
in section \ref{SUSYBM}. \bref{Mc} is necessary for neutralino LSP in the gaugino mediation.
\section{Confrontation with Cosmology--Smooth Hybrid Inflation\label{smoothhb}}
Inflation model is already accepted as the new Standard Cosmology
since it solves the shortcomings of old Standard Big-Bang Model \cite{InflationRev}. Inflation is supposed to have occurred at GUT scale and must have strong relationships with GUT. After old inflation \cite{Guth}, there appeared many inflation models: slow roll inflation \cite{chaotic} and K-inflation \cite{K} are driven by the potential and kinetic enery of scalar fields, respectively. Also there are models due to modified gravity \cite{Starobinsky} and due to Bose-Einstein condensatation \cite{Morikawa}. Inflation model has the observational supports from WMAP.
Inflation is considered to have occurred at GUT scale and its scenario crucially depends on GUT model. Also WMAP data give them the severe constraints mainly due to baryon acaustic oscillation (BAO) \cite{Ellis} \cite{Ellis2}.

The inflation have three main observables; power spectrum $n_s$, tensor-to-scalar ratio, and (non) Gaussianity \cite{WMAP}. Let me introduce them.
The spatial elements of the metric tensor $g_{\mu\nu}$ are divided into the trace part ${\cal R}$ and the traceless parts $\left(e^h\right)_{ij}$,
\be
g_{ij}=a^2(t)e^{2{\cal R}}\left(e^h\right)_{ij}
\ee
with det$\left(e^h\right)_{ij}=1$.
The power spectrum of ${\cal R}$ is defined by
\be
\la {\cal R}_{{\bf k}}{\cal R}_{{\bf k}'}\ra =(2\pi)^3\delta({\bf k}+{\bf k}')P_{\cal R}(k)
\ee
and the spectral index $n_s$ is defined as
\be
P_{\cal R}(k)\propto k^{n_s-4}.
\ee
The current data is $n_s=0.96\pm 0.01$ (68\% CL).
The tensor-to-scalar ratio $r$ is defined by
\be
r\equiv \frac{P_h(k_0)}{P_{\cal R}(k_0)}.
\ee
Here $P_h(k)$ is
\be
\la h_{ij,{\bf k}}h^{ij,{\bf k}'}\ra=(2\pi)^3\delta({\bf k}+{\bf k}')P_h(k)
\ee
and $k_0=0.002$Mpc${}^{-1}$ is some reference wavenumber at which r is defined.
The current upper bound is $r<0.24$(95\% CL).
The three-point function
\be
\la{\cal R}_{{\bf k}_1}{\cal R}_{{\bf k}_2}{\cal R}_{{\bf k}_3}\ra=(2\pi)^3\delta_D({\bf k}_1+{\bf k}_2+{\bf k}_3)B_{\cal R}(k_1,k_2,k_3)
\ee
vanishes for Gaussian fluctuations. Single field inflation is nearly Gaussian, that is,
\be
\frac{6}{5}f_{NL}\equiv \frac{B_{\cal R}(k_1,k_2,k_3)}{P_{\cal R}(k_1)P_{\cal R}(k_2)+(2\mbox{perm})}
\ee
almost vanishes, and non-Gaussian signal indicates the exclusion of it.
The current bound is $f_{NL}=32\pm 21$(68\% CL) and $n_s=0.96\pm 0.01$ (68\% CL) \cite{WMAP}. The Planck data are expected to be open in 2013 \cite{Planck}.

There are some deficits for a single-field inflation model. It gives rise to fine tuning problem. For single field inflation $f_{NL}$ is related with $n_s$ as
\be
f_{NL}=\frac{5}{12}(1-n_s).
\ee
Hence if $f_{NL}\geq 1$ will be observed, it rules out single field inflation.
Here we restrict our arguments on hybrid inflation \cite{HIRev} since single field inflation suffers from several tensions from observation.
The most naive superpotential may be 
\be
W=\kappa S(-M^2+\overline{\phi}{\phi}),
\ee
where $\overline{\phi}$ and $\phi$ is a pair of the SM singlet.
However, this model is flat along the inflation valley ($\overline{\phi}=\phi=0$) and must include one-loop correction of the Coleman-Weinber type. We want to corelate $\phi$ with high dimensional GUT field. However, $\phi$ and $\overline{\phi}$ develop vevs during the waterfall phase and lead to copious production of monopole, monopole problem.
One solution to evade the monopole problem is the shifted inflation model \cite{ShiftedHI}
\be
W=\kappa S(-M^2+\overline{\phi}{\phi})-\beta\frac{S(\overline{\phi}\phi)^2}{M_{String}^2},
\ee
where $\phi$ and $\overline{\phi}$ are defined in \bref{PSHiggs} and $M_{String}$ is a string scale.
In this and next smooth cases, $\phi$ and $\overline{\phi}$ have vevs during the inflation phase and monopoles are diluted away.
However, in this section, we consider a simpler model, smooth hybrid inflation model \cite{SmoothHI} 
 in the context of the orbifold GUT model 
 discussed in the previous section. 
For this purpose, we introduce a singlet chiral superfield $S$ as before.  
Needless to say, this singlet field causes 
 no change for the gauge coupling unification. 
Let us consider the superpotential 
 for the smooth hybrid inflation \cite{SmoothHI}%
\footnote{
The renormalizable term, $S (\bar{\phi} \phi)$, 
 can be forbidden by introducing a discrete symmetry \cite{SmoothHI}, 
 for example, $\phi \to -\phi$ and $\bar{\phi} \to \bar{\phi}$. 
}, 
\bea
 W_S= S \left( -\mu^2+\frac{(\bar{\phi} \phi)^2}{M^2} \right).
\label{smooth}  
\eea
Here $\phi$ and $\overline{\phi}$ are defined in \bref{PSHiggs} and we have omitted possible ${\cal O}(1)$ coefficients. 
\bref{smooth} has $U(1)_R$ symmetry
\bea
&&\overline{\phi}\phi\rightarrow \overline{\phi}\phi\nonumber \\
&&S\rightarrow e^{i\alpha}S,~~W\rightarrow e^{i\alpha}W
\eea
under \bref{U1R}.
SUSY vacuum conditions lead to non-zero VEVs for 
 $\langle \phi \rangle = \langle \bar{\phi} \rangle = \sqrt{\mu M}$, 
 by which the PS symmetry is broken down to the SM one, and thus  
\bea 
  v_{\rm PS} = \sqrt{\mu M}.   
\eea 
$U(1)_R$ is conserved and no problem mentioned in subsection \ref{no-go} occurs. Total Higgs superpotential including terms irrelevant to inflation is
\be
W=W_{PS}+W_S,
\label{SP2}
\ee
where $W_{PS}$ is one given by \bref{HiggsW} eliminating $\lambda_{15}$ term.
\footnote{If we accept more PS singlets other than $S$, we can construct an alternative PS invariant superpotential \cite{Antoniadis}. However it is less predictive for quark-lepton mass marices.}
In the following analysis of inflation, we treat $M$ as a
 free parameter with $v_{\rm PS} = \sqrt{\mu M}=1.2 \times 10^{16}$ GeV    
 fixed by the analysis of gauge coupling unification. 
Since $M$ is involved in the non-renormalizable term, 
 it is theoretically natural that 
 $M \sim M_5 \sim M_{\rm GUT}=4.6 \times 10^{17}$ GeV \bref{GUT2}. 
Notwithstanding this condition is independent of the inflation scenario, 
 we will find from the following analysis that 
 $M \sim M_{\rm GUT}$ is in fact consistent with the cosmological
 observations.
Therefore, our GUT model is suitable for the inflation models 
 with the parameters $v_{\rm PS}$ and $M_{\rm GUT}$ 
 fixed by the analysis for the gauge coupling unification. 

It would be worth mentioning that the Higgs superpotential 
 includes a term $W \supset H_6 (\phi^2+\bar{\phi}^2)$ 
 through which all color triplets in $\phi$ and $\bar{\phi}$ become heavy (\bref{HiggsW}). 
Only the superpotential of Eq.~(\ref{smooth}) is relevant to inflation.

\begin{figure}
\begin{center}
\includegraphics[scale=0.3]{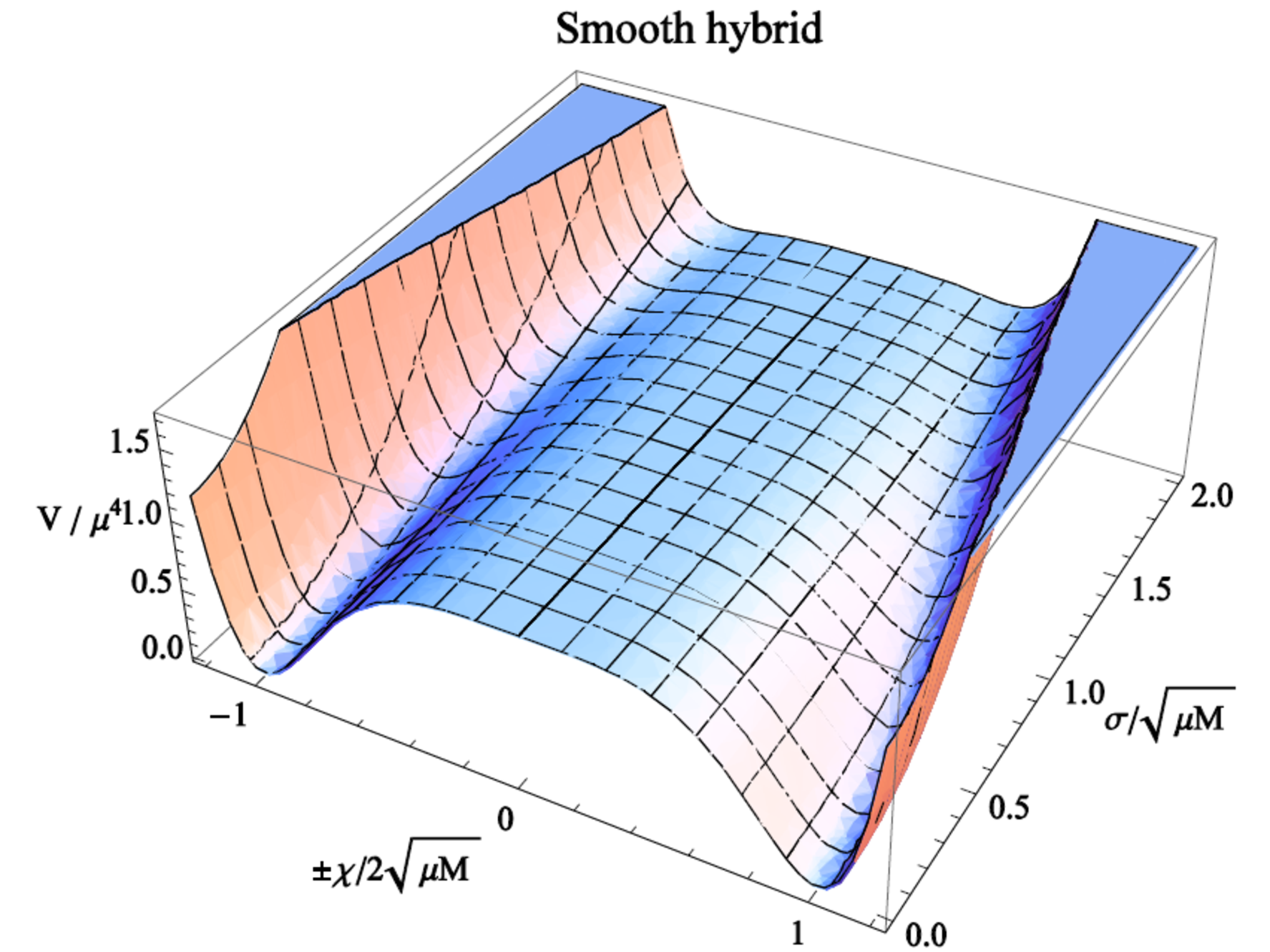}
\caption{ The potential diagram of the smooth hybrid inflation. Inflation occurs along the valley and smoothly tranmitted to the global minimum. 
}
\label{fig:hybrid}
\end{center}
\end{figure}

Now we examine the smooth hybrid inflation scenario \cite{F-O2}. 
The scalar potential from  Eq.~(\ref{smooth}) is given by%
\footnote{
Although supergravity effects from Kahler potential 
 can play an important role \cite{sugra}, 
 we do not consider effects form supergravity in this paper, 
 assuming a special Kahler potential for the inflaton field. 
}
\bea
 V= \left| -\mu^2+\frac{(\bar{\phi} \phi)^2}{M^2}\right|^2
 +4S^2 \frac{|\phi|^2 |\bar{\phi}|^2}{M^4} 
 \left(|\phi|^2 + |\bar{\phi}|^2\right). 
\label{vsmooth}
\eea
Considering the D-flatness condition, we normalize
\bea
 |\phi|=|\bar{\phi}|=\frac{\chi}{2},~~|S|=\frac{\sigma}{\sqrt{2}} 
\eea
and then V becomes
\bea
 V=\left( \mu^2-\frac{\chi^4}{16M^2} \right)^2 
 +\frac{\chi^6\sigma^2}{16M^4}. 
\eea
For a fixed $\sigma$, $V$ has a minimum at 
\bea
 \chi^2 = -6 \sigma^2 + \sqrt{36 \sigma^4 + 16 v_{\rm PS}^4}
        \simeq \frac{4}{3} \frac{v_{\rm PS}^4}{\sigma^2}, 
\eea
 where we have used an approximation for 
 $\sigma^2 \gg v_{PS}^2$ (satisfied during inflation) 
 in the last expression. 
The inflation trajectory is along this minimum and 
 we obtain the potential along this path 
\bea
  V \simeq \mu^4 
  \left( 1 - \frac{2 v_{\rm PS}^4}{27 \sigma^4} \right) .  
\eea
The potential diagram of the smooth hybrid inflation is depicted in figure \ref{fig:hybrid}
Note that the PS symmetry is spontaneously broken 
 anywhere on the inflation trajectory  and thus, 
 no topological defects such as strings, monopoles, 
 or domain walls are produced at the end of inflation
\cite{SmoothHI}.

The slow-roll parameters ($\epsilon$, $\eta$) 
 and the parameter ($\xi^2$), which enters the running of the 
 spectral index, are defined as \cite{InflationRev} 
\bea
 \epsilon &=& \frac{M_P^2}{2} \left( \frac{V'}{V} \right)^2 
 \simeq \frac{32 M_P^2 v_{\rm PS}^8}{729 \sigma^{10}} 
 \simeq - \frac{4 v_{\rm PS}^4}{135 \sigma^4} \eta, \nonumber \\
 \eta &=& M_P^2 \left( \frac{V''}{V} \right) = 
  - \frac{40 M_P^2 v_{\rm PS}^4}{27 \sigma^6},  \nonumber \\
 \xi^2 &=& M_P^4 \left( \frac{V' V'''}{V^2} \right) 
   \simeq 
   \frac{640 M_P^4 v_{\rm PS}^8}{243 \sigma^{12}},  
 \label{slow-roll}
\eea
where the prime denotes derivative with respect to $\sigma$. 
The slow-roll approximation is valid if the conditions,  
 $\epsilon$ and $|\eta| \ll 1$, hold. 
In this case, 
 the spectral index ($n_{\rm s}$), 
 the ratio of tensor-to-scalar fluctuations ($r$) 
 and the running of the spectral index ($\alpha_{\rm s}$) 
 are given by 
\bea 
 n_{\rm s} &\simeq &1-6 \epsilon + 2 \eta,  \nonumber \\
 r &\simeq &16 \epsilon,  \nonumber \\
 \alpha_{\rm s} &=& \frac{d n_{\rm s}}{d \ln k}
  \simeq 16 \epsilon \eta -24 \epsilon^2 -2 \xi^2. 
\label{spectral}
\eea

The number of e-folds $N_k$ after the comoving scale $\ell_0=2 \pi/k_0$ 
 has crosses the horizon is given by 
\bea
N_k = \frac{1}{M_P^2} \int^{\sigma_k}_{\sigma_f} 
  d\sigma \frac{V}{V'} 
  \simeq \frac{9}{16 M_P^2 v_{\rm PS}^4} 
  \left( \sigma_k^6 - \sigma_f^6  \right),  
 \label{e-folds1}
\eea
where $\sigma_k$ is the value of the inflaton field 
 when the scale corresponding to $k_0$ 
 exits the horizon, and $\sigma_f$ is the value of the inflaton field 
 when the inflation ends, which is determined by $|\eta|=1$ 
 so that 
\bea
\sigma_f^6 &=& \frac{40}{27} M_P^2 v_{\rm PS}^4.  
\eea
For $\sigma_k^6 \gg \sigma_f^6$, several formulas given above 
 are reduced into simpler forms, for example, 
\bea 
 N_k \simeq - \frac{5}{6 \eta_k}, ~~
 n_s \simeq 1- \frac{5}{3 N_k}, ~~ 
 \alpha_{\rm s} \simeq -\frac{5}{3 N_k^2} .  
\eea 
The number of e-folds $N_k$ required for solving the horizon  
 and flatness problems of the standard big bang cosmology, 
 for $k_0=0.002$ Mpc$^{-1}$, is given by \cite{InflationRev}
\bea 
 N_k \simeq 51.4 
 + \frac{2}{3} \ln \left( \frac{V(\sigma_k)^{1/4}}{10^{15}~{\rm GeV}} \right)
 + \frac{1}{3} \ln \left( \frac{T_{\rm rh}}{10^{7}~{\rm GeV}} \right), 
\label{e-folds2}
\eea 
 where we have assumed a standard thermal history, 
 and $T_{\rm rh}$ is the reheating temperature after inflation. 
If gravitino has mass around 100 GeV as in the gravity mediated 
 SUSY breaking, the reheating temperature is severely constrained
 (gravitino problem) from the fact that the gravitino decay products do not to destroy 
 the light elements successfully synthesized during 
 big bang nucleosynthesis \cite{gravitino}, 
\bea
 T_{\rm rh} \leq  10^6-10^7 \textrm{GeV}.  
\eea
The power spectrum of the primordial curvature perturbation 
 at the scale $k_0$ is given by 
\bea 
 {\cal{P_R}}^{1/2} \simeq 
\frac{1}{2 \sqrt{3} \pi M_P^3} \frac{V^{3/2}}{|V'|} 
\simeq 
 \frac{27}{16 \sqrt{3} \pi } \frac{\sigma_k^5}{M_P^3 M^2}.  
 \label{power}
\eea
This should satisfy the observed value by the WMAP \cite{WMAP}, 
 ${\cal{P_R}} \simeq  2.457 \times 10^{-9}$. 

\begin{figure}[th]
\begin{center}
\includegraphics[scale=0.6]{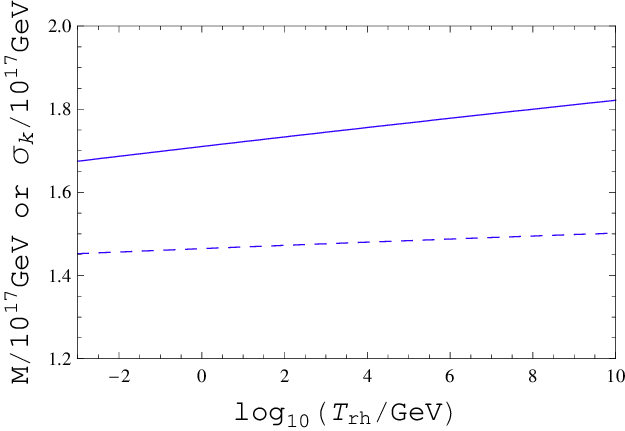}
\caption{
$M$ and $\sigma_k$ as a function of $T_{\rm rh}$. 
The solid and dashed lines correspond to 
 $M$ and $\sigma_k$, respectively. 
}
\label{sigma-T}
\end{center}
\end{figure}
\begin{figure}[t]
\begin{center}
\includegraphics[scale=0.6]{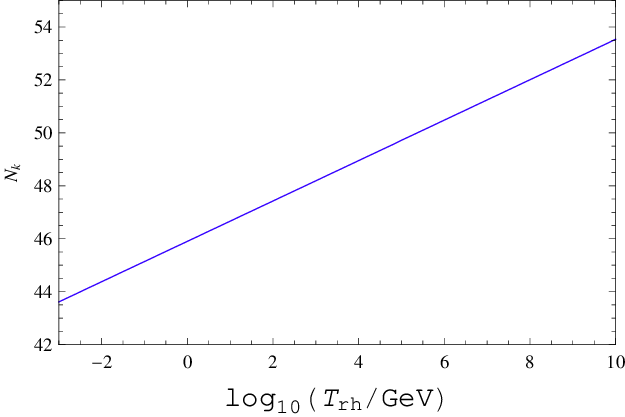}
\caption{ 
The number of e-folds versus $T_{\rm rh}$. 
}
\label{efold-T}
\end{center}
\end{figure}
\begin{figure}
\begin{center}
\includegraphics[scale=0.6]{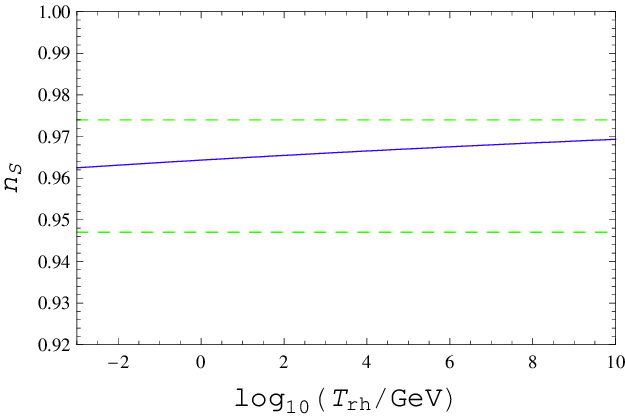}
\caption{ 
The spectral index as a function of $T_{\rm rh}$. 
The dashed lines correspond to the WMAP 5-year data, 
 $n_{\rm s} = 0.960^{+0.014}_{-0.013}$. 
}
\label{ns}
\end{center}
\end{figure}

Now we solve Eqs.~(\ref{e-folds1}), (\ref{e-folds2}) 
 and (\ref{power}) and fix the model parameters of the inflation scenario. 
In our analysis, we have three independent free parameters,  
 $M$, $\sigma_k$ and $T_{\rm rh}$, 
 with the fixed $v_{\rm PS}=1.2 \times 10^{16}$ GeV. 
We solve the equations for a given $T_{\rm rh}$ 
 in the range 1 MeV $\leq T_{\rm rh} \leq 10^{10}$ GeV 
 and find $M$ and $\sigma_k$. 
Fig.~\ref{sigma-T} shows the results for $M$ and $\sigma_k$ 
 as a function of $T_{\rm rh}$. 
We can check that $\sigma_k^6 \gg \sigma_f^6$ 
 and the slow-roll conditions are satisfied. 
The number of e-folds is depicted in Fig.~\ref{efold-T}. 
Using these outputs and Eq.~(\ref{slow-roll}), 
 we evaluate the spectral index (Fig. \ref{ns}), 
 the tensor-to-scalar ratio and the running of the spectral index: 
\bea 
   0.963 \leq  & n_{\rm s} & \leq 0.968,   \nonumber \\ 
  4.0 \times 10^{-7} \geq  & r &   \geq  3.1 \times 10^{-7} ,
  \nonumber \\ 
  -8.4 \times 10^{-4} \leq & \alpha_{\rm s} & \leq -6.1 \times 10^{-4} 
\eea  
 for 1 MeV $\leq T_{\rm rh} \leq 10^7$ GeV. 
The tensor-to-scalar ratio and the running of the spectral index 
 are negligibly small. 
These results are consistent with the WMAP 5-year data \cite{WMAP}: 
 $n_{\rm s} = 0.960^{+0.014}_{-0.013}$, 
 $ r <  0.24$ (95\% CL) 
 and $ \alpha_{\rm s} = -0.032^{+0.021}_{-0.020}$ (68\% CL) 
 (consistent with zero in 95\% CL). The tiny value of $r$ is in contrast with that of single field inflation, $\lambda_\alpha S^\alpha~\alpha=4,3,2,1,2/3$, which gives rather large $r$ around $0.1$ ($\alpha=2$).
\begin{figure}
\begin{center}
\includegraphics[scale=0.6]{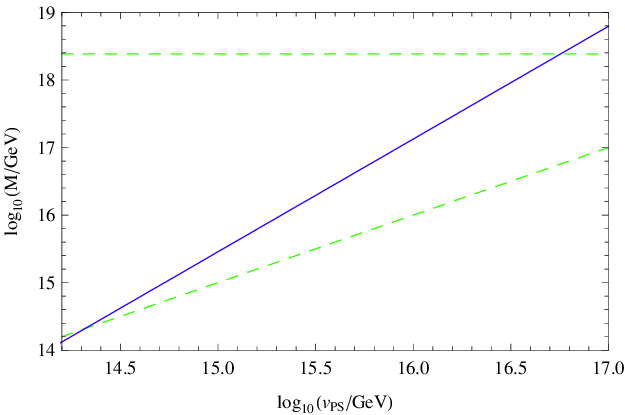}
\caption{ 
$M$ as a function of $v_{\rm PS}$ (solid line). 
The lower and upper dashed lines specified 
 the theoretical consist region for $M$, 
 $ v_{\rm PS} \leq M \leq M_P$. 
}
\label{M-vps}
\end{center}
\end{figure}
\begin{figure}
\begin{center}
\includegraphics[scale=0.6]{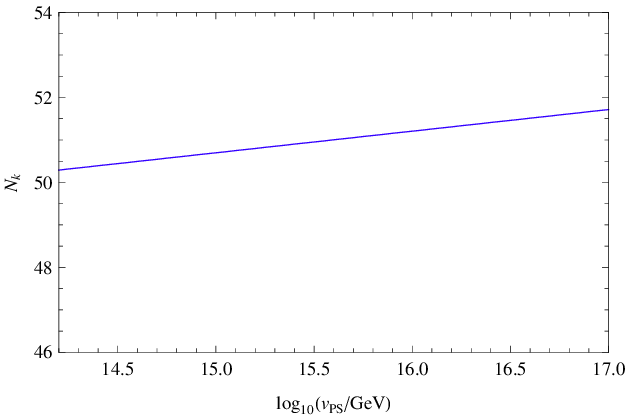}
\caption{ 
The number of e-folds versus $v_{\rm PS}$. 
}
\label{efold-vps}
\end{center}
\end{figure}
\begin{figure}
\begin{center}
\includegraphics[scale=0.6]{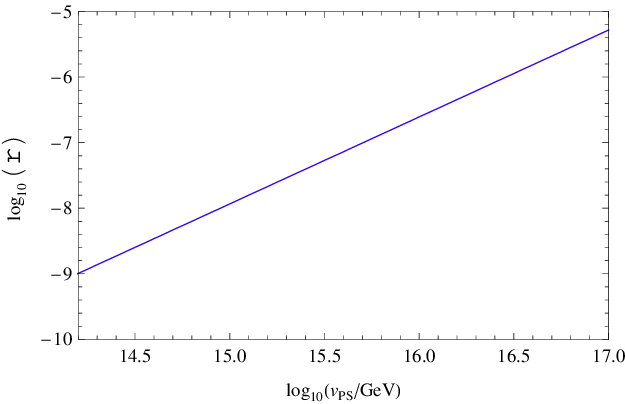}
\caption{ 
The tensor-to-scalar ratio versus $v_{\rm PS}$. 
}
\label{tensor-scalar}
\end{center}
\end{figure}
In general, we do not need to impose either the left-right symmetry 
 or $M_c=v_{\rm PS}$ on the model. 
In this case, $v_{\rm PS}$ can be varied \cite{Raby}, 
 and we repeat the same analysis for this general case 
 with $v_{\rm PS}$ as a free parameter. 
Fixing, for example, $T_{\rm rh}=10^7$ GeV, 
 $M$ can be obtained as a function of $v_{\rm PS}$ 
 as shown in Figure~\ref{M-vps}. 
Since $M$ appears in the non-renormalizable term, 
 it would be natural to identify $M$ as an effective cutoff. 
Thus, the theoretical consistency leads to the condition, 
 $v_{\rm PS} \leq M \leq M_P$, from which we obtain 
 $1.9 \times 10^{14}$ GeV $\leq v_{\rm PS} \leq 5.6 \times 10^{16}$ GeV. 
The number of e-folds and the tensor-to-scalar ratio versus $v_{\rm PS}$ are shown in Figs.~\ref{efold-vps} and \ref{tensor-scalar}, respectively. 
 
The other outputs, the spectral index and its running, 
 are found as 
\bea  
  0.967 \leq &n_{\rm s}& \leq 0.968,  \nonumber \\   
  -6.4 \times 10^{-4} \leq  &\alpha_{\rm s}&  
  \leq -6.1 \times 10^{-4}.  
\eea 
These are consistent with the WMAP data. 
\newpage
\section{SUSY Breaking Mechanism and Dark Matter\label{SUSYBM}}

\begin{figure}[ht]
\begin{center}
\includegraphics*[width=.6\linewidth]{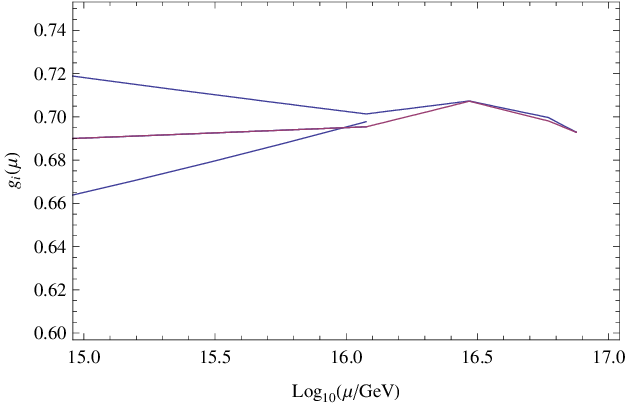}
\caption{
Gauge coupling unification for $M_c>v_{PS}$. 
Each line from top to bottom corresponds to 
 $g_3$, $g_2$ and $g_1$ for $ \mu < v_{PS}$,
 while  $g_3=g_4$ and $g_2=g_{2R}$ for $ \mu > v_{PS}$. 
Here, we have taken $M_c= 2.47 \times v_{PS}$.
}
\label{Figgauge}
\end{center}
\end{figure}

For the scale $\mu \geq v_{PS}$, there are only two independent 
 gauge couplings $\alpha_4$ and $\alpha_2=\alpha_{2R}$, 
 so that the gauge coupling unification is easily realized 
 with a suitable $M_c$. 
Fig. \ref{Figgauge} shows the gauge coupling evolutions for $\mu>v_{PS}$.
For the scale $\mu\geq v_{PS}$, there are only two independent gauge couplings $\alpha_4$ and $\alpha_2=\alpha_R=\alpha_L$, so that the gauge coupling unification is easily realized with a suitable $M_c$. In this Fig. \ref{Figgauge}, we have taken (corresponding to the result in the following neutral LSP arguments)
\bea 
 M_c = 2.47 \times v_{PS} = 2.95 \times 10^{16} \;  \mbox{GeV}  
\eea 
 and after including Kaluza-Klein threshold contributions 
 into the gauge coupling evolutions, 
 the gauge coupling unification is realized at 
\bea 
  M_{\rm GUT} = 7.54 \times 10^{16} \; \mbox{GeV} . 
\eea  
As $M_c$ is raised, $M_{\rm GUT}$ becomes smaller. 
As mentioned before, we assume that a more fundamental 
 SO(10) GUT theory takes place at $M_{\rm GUT}$. 
The origin of SUSY breaking and its mediation to the MSSM sector 
 is still a prime question in any phenomenological SUSY models. 
Since flavor-dependent soft SUSY breaking masses are 
 severely constrained by the current experiments, 
 a mechanism which naturally transmits SUSY breaking 
 in a flavor-blind way is the most favorable one.

In higher dimensional models, the sequestering \cite{RS-SUSY} 
 is the easiest way to suppress flavor-dependent SUSY breaking 
 effects to the MSSM matter sector. 
Since all matters reside on the PS brane in our model, 
 the sequestering scenario is automatically realized 
 when we simply assume a SUSY breaking sector 
 on the brane at $y=0$. SUSY breaking is propagated to PS brane by the SM gauge multiplet in the bulk (gaugino mediation) \cite{gMSB}. 
There is another SUSY breaking scenario.
That is, gauge multiplets are also confined on MSSM brane and SUSY breaking is transmitted only by supergravity effect, the soft terms on the visible brane can be understood by the anomalous violation of a local superconformal invariance and called anomalous SUSY breaking (AMSB) \cite{RS-SUSY}, \cite{GLMR}. 
AMSB suffers from the problem that the slepton masses become tachyonic.
There are positive contributions from modulus \cite{KKLT} and D-term.
It should be remarked that SO(10) GUT including high dimensional representation discussed here have $U(1)_{B-L}$ and D term.
So in general we can combine modulus and D-term contributions \cite{Dterm}.
Anyhow we adopt the gauge multiplet in bulk and do not discuss AMSB.
\paragraph{Gaugino mediation} 
The SO(10) gauge multiplet is in the bulk and 
 can directly communicate with the SUSY breaking sector  
 through the higher dimensional operator of the form,  
\bea 
 {\cal L} \sim \delta (y) 
 \int d^2 \theta \; 
  \frac{X}{M_5^2} 
 {\rm tr} \left[ {\cal W}^\alpha {\cal W}_\alpha \right].  
\label{W}
\eea 
Here $X$ is a singlet chiral superfield 
 which breaks SUSY by its F-component VEV, $X= \theta^2 F_X$. Also
${\cal W}$ is the field strength chiral superfield constructed from the vector superfield $V$ and chiral covariant $D_\alpha$
,
\bea
{\cal W}_\alpha (x,\theta, \bar{\theta})
&=&-\frac{1}{4}D^\dagger D^\dagger \left(e^{-V}D_\alpha e^V\right)\nonumber\\
&=&- i \lambda_\alpha(x + i \theta \sigma_\mu \bar{\theta})
+\left[\delta_\alpha^\beta D(x + i \theta \sigma_\mu \bar{\theta})
- i (\sigma^{\mu \nu})_{\alpha \dot{\beta}} 
F_{\mu \nu}(x + i \theta \sigma_\mu \bar{\theta}) \right]\theta_\beta
\nonumber\\
&+& \theta \theta (\sigma_\mu)_{\alpha \dot{\beta}} \partial^\mu 
\bar{\lambda}_{\dot{\beta}}(x + i \theta \sigma_\mu \bar{\theta}) 
\eea
and
\be
D_\alpha=\frac{\pa}{\pa \theta^\alpha}-i(\sigma^\mu\theta^\dagger )_\alpha\pa_\mu.
\ee
Therefore, the bulk gaugino obtains the SUSY breaking soft mass, 
\bea 
 M_\lambda \sim \frac{F_X M_c}{M_5^2}
 \simeq \frac{F_X}{M_P} \left(\frac{M_5}{M_P} \right) ,  
\label{gaugino}
\eea
where $M_c$ comes from the wave function normalization 
 of the bulk gaugino, 
 and we have used the relation between the 4D and 5D Planck scales,  
 $M_5^3/M_c \simeq M_P^2$ in the last equality. 
As usual, we take $M_\lambda =$100 GeV-1 TeV.  
Once the gaugino obtains non-zero mass, 
 SUSY breaking terms for sfermions are automatically 
 generated through the RGE from the compactification scale 
 to the electroweak scale. 
This scenario is nothing but the gaugino mediation \cite{gMSB} 
 and flavor-blind sfermion masses are generated 
 through the gauge interactions. 
In this setup, a typical gaugino mass in Eq.~(\ref{gaugino}) 
 is smaller than the gravitino mass $m_{3/2} \simeq F_X/M_P$ 
 by a factor $M_5/M_P < 1 $.

As discussed in Ref.~\cite{F-O1}, for $M_c=v_{PS}$, 
 we find that the right-handed slepton (normally, stau) 
 is the LSP, because the sfermion mass spectrum is 
 obtained from the boundary condition with vanishing soft masses 
 at $M_c=v_{PS}=1.19 \times 10^{16}$ GeV. 
This result is problematic for cosmology. 
As pointed out in Ref.~\cite{gMSB2}, 
 this stau LSP problem is cured by the soft mass RGE running 
 from the compactification scale to the GUT scale in a GUT model. 
In the following, we apply this idea to our model 
 with $M_c > v_{PS}$.

For the scale, $v_{PS} \leq \mu \leq M_c$,  
 we are in the PS stage and the RGEs of gaugino and sfermion masses 
 are given by 
\bea 
&&  \frac{d}{d t} \left( \frac{M_4}{\alpha_4} \right) = 
 \frac{d}{d t} \left( \frac{M_{2L}}{\alpha_{2L}} \right) = 
 \frac{d}{d t} \left( \frac{M_{2R}}{\alpha_{2R}} \right) =0,  \nonumber \\ 
&& \frac{d m^2_{\tilde{F}}}{d t} = 
 - \frac{15}{4 \pi} \alpha_4 M_4^2 
 - \frac{3}{2 \pi} \alpha_{2L} M_{2L}^2,  \nonumber \\ 
&& \frac{d m^2_{\tilde{F}^c}}{d t} =  
 - \frac{15}{4 \pi} \alpha_4 M_4^2 
 - \frac{3}{2 \pi} \alpha_{2R} M_{2R}^2 ,
\label{RGEs}
\eea
where $t=\ln(\mu/M_c)$, 
 $\alpha_4$ and $\alpha_{2L}=\alpha_{2R}$ are 
 the PS gauge coupling of the corresponding gauge groups 
 (whose RGE solutions are obtained in the previous section), 
 and $M_4$, $M_{2L}$ and $M_{2R}$ are 
 the corresponding gaugino masses. 
Sfermion mass spectrum is obtained by solving the RGEs 
 with the boundary conditions, 
 $m_{\tilde{F}}(M_c)=m_{\tilde{F}^c}(M_c)=0$. 
Analytic solutions of Eq.~(\ref{RGEs}) at $\mu =v_{PS}$ 
 are easily found: 
\bea 
  m^2_{\tilde{F}}(v_{PS}) &=& 
    \frac{5}{4} M_4^2(v_{PS}) \left[ 
     \left( \frac{\alpha_4(M_c)}{\alpha_4(v_{PS})} \right)^2 -1 \right] 
   +  \frac{1}{4} M_{2L}^2(v_{PS}) \left[ 
    \left( \frac{\alpha_{2L}(M_c)}{\alpha_{2L}(v_{PS})} \right)^2 -1 
 \right], \nonumber \\ 
  m^2_{\tilde{F}^c}(v_{PS}) &=& 
   \frac{5}{4} M_4^2(M_c) \left[ 
    \left( \frac{\alpha_4(v_{PS})}{\alpha_4(v_{PS})} \right)^2 -1 \right] 
   + \frac{1}{4} M_{2R}^2(v_{PS}) \left[ 
    \left( \frac{\alpha_{2R}(M_c)}{\alpha_{2R}(v_{PS})} \right)^2-1
  \right]. 
\eea 
Note that the PS model is unified into a more fundamental SO(10) model 
 and this unification leads to the well-known relation, 
\bea 
 \frac{M_4}{\alpha_4}  = 
 \frac{M_{2L}}{\alpha_{2L}}  = 
 \frac{M_{2R}}{\alpha_{2R}}  = 
 \frac{M_{1/2}}{\alpha_{\rm GUT}}  , 
\eea
where $M_{1/2}$ is the universal gaugino mass 
 at the unification scale. 
Thus, the formula
as for sfermion masses are simplified as 
\bea 
 && m^2_{\tilde{F}}(v_{PS}) = m^2_{\tilde{F}^c}(v_{PS}) \nonumber \\
 &=&  
  \left( \frac{M_{1/2}}{\alpha_{\rm GUT}} \right)^2 
  \left[ \frac{5}{4} 
     \left( \alpha_4^2(M_c) - \alpha_4^2 (v_{PS}) \right)
  +  \frac{1}{4} 
     \left( \alpha^2_{2L}(M_c) - \alpha^2_{2L} (v_{PS}) \right) 
  \right]. 
\eea

Solving the RGEs in the MSSM with the universal boundary condition 
 $ m^2_{\tilde{F}}(v_{PS}) = m^2_{\tilde{F}^c}(v_{PS}) 
 = m_0^2$ at $\mu = v_{PS}$, we obtain the sfermion masses 
 at the electroweak scale. 
In our model, the PS scale is almost the same as the usual SUSY GUT 
 scale in the MSSM ($M_{\rm GUT} \simeq 2 \times 10^{16}$ GeV) 
 and the gauge couplings are roughly unified at the PS scale, 
 $ \alpha_4(v_{PS}) \simeq \alpha_{2}(v_{PS})= \alpha_{2R}$ 
 (see Fig.~\ref{Figgauge}). 

\begin{figure}[h]
\begin{center}
{\includegraphics*[width=.4\linewidth]{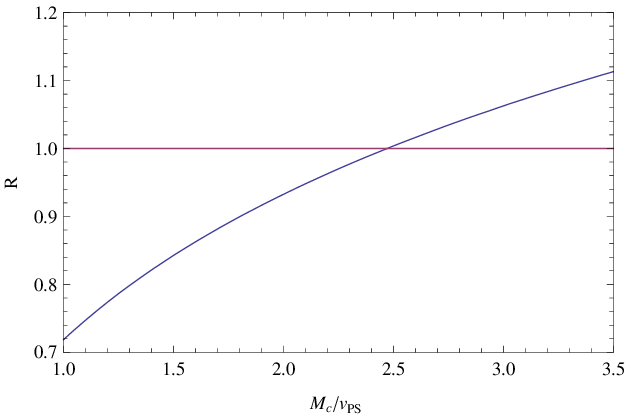}}
\caption{
The ratio, $R \equiv m_{\tilde{e^c}}^2/M_{126}^2$, 
 as a function of $M_c/v_{PS}$. 
Here, soft masses have been evaluated at $\mu=M_{\rm SUSY}$. 
$R=1$ when $M_c/v_{PS} = 2.47$. 
}
\label{GCU3}
\end{center}
\end{figure}
Therefore, our study on the sfermion masses are 
 almost the same as the one usual in the constrained MSSM. 
For a small $\tan \beta$ (say, $\tan \beta=10$), 
 we neglect Yukawa coupling contributions to the soft masses 
 of right-handed sleptons, and the analytic solutions 
 of the MSSM RGEs are found to be 
\bea 
 M_{126} (\mu) &=& \alpha_1(\mu) 
  \left(  \frac{M_{1/2}}{\alpha_{\rm GUT}} \right) , \nonumber \\ 
 m_{\tilde{e^c}}^2 (\mu) &=& 
  \left(  \frac{M_{1/2}}{\alpha_{\rm GUT}} \right)^2 
   \frac{2}{11} 
   \left[ \alpha_1^2(v_{PS})- \alpha_1^2(\mu) \right] 
   + m_0^2.  
\eea 
If $m_0$ is large enough, the slepton (stau) mass is bigger 
 than the bino mass ($M_{126}$). 
In our model, $m_0$ is given as a function of $M_c$. 
Fig.~\ref{GCU3} shows the ratio, 
\bea 
  R \equiv \frac{m^2_{\tilde{e^c}}}{M_{126}^2},  
\eea
 as a function of $M_c/v_{PS}$. 
We can obtain $ R \geq 1$ for $M_c/v_{PS} \geq 2.47$.

Now, for $M_c/v_{PS} \geq 2.47$, the bino-like neutralino 
 will be the LSP and a good candidate for the cold dark matter 
 in cosmology \cite{DM}. 
For a small $ \tan \beta$, 
 the annihilation processes of the bino-like neutralino 
 are dominated by p-wave and are not so efficient. 
As a result, the neutralino relic density tends to exceed 
 the upper bound of the observed dark matter density. 
This problem can be avoided if the neutralino  is quasi-degenerate 
 with the next LSP slepton (stau), and the co-annihilation process 
 with the next LSP can lead to the right dark matter relic density. 
In our result, such a situation appears for 
 $M_c \simeq 2.47 \times v_{PS} \simeq 2.95 \times 10^{16}$ GeV. 
It would be interesting to note that 
 the discrepancy of the abundance of ${}^7$Li 
 between the observed values in WMAP and in metal poor halo stars 
 may be explained the degeneracy between 
 the LSP neutralino and stau \cite{Jittoh}. 
\section{Leptogenesis in 5D \label{leptogenesis5D}}
In section \ref{leptogenesis}, we discussed non-therml leptogenesis in 4D, where $\lambda=10^{-8}$. 
This gave rise to a naturaliness problem.
In the following, we discuss how this problem is modified in the smooth hybrid inflation model \footnote{In ths section we discuss on the sneutrino case (See section \ref{leptogenesis} for neutrino.).  Also there is a contribution from soft SUSY breaking terms (soft Leptogenesis) \cite{Fong}.}.

It goes from \bref{Yukawa} and \bref{smooth} that the interaction terms relevant to inflaton decay are
\be
 W= S \left( -\mu^2+\frac{(\bar{\phi} \phi)^2}{M^2} \right)+\frac{Y_R^{ij}}{M_5} \tilde{F}_{Ri}^c \tilde{F}_{Rj}^c 
 \left(\phi \phi \right)  
\label{nu_R}
\ee
The potential V is
\be
V=\left|S\frac{2\bar{\phi}(\bar{\phi} \phi)}{M^2}+\frac{Y_R^{ij}}{M_5} \tilde{F}_{Ri}^c \tilde{F}_{Rj}^c  \phi \right|^2
\label{potential}
\ee
So  $S\tilde{F}_{Ri}^c \tilde{F}_{Rj}^c$ interaction term is
\be
H_{S\tilde{F}\tilde{F}}=8\frac{M_{PS}^4}{M^2M_5}Y_RS\tilde{F}_{Ri}^c \tilde{F}_{Rj}^c, 
\ee
and the decay ratio at tree level is
\be
\Gamma(S\rightarrow \tilde{F}_{Ri}^c \tilde{F}_{Rj}^c)=\frac{64}{8\pi}Y_R^2\frac{M_{PS}^8}{M_5^2M^4}\frac{1}{M_S}.
\ee
Thus $\lambda$ corresponds in this paper to 
\be
\lambda ^2\approx Y_R^2\frac{M_{PS}^8}{M_G^6M_S^2},
\label{BR}
\ee
where we assumed $M=M_5$. $M_S$ is obtained from \bref{potential}, $M_S^2=\frac{8M_{PS}^6}{M_5^4}$. Substituting the values obtained, we get $Y_R=1.0\times 10^{-6}$ if $BR(S\rightarrow \tilde{F}_{R1}^c \tilde{F}_{R1}^c)=1$.

In the SO(10) in 4D model,
\be
M_{R1}=1.64\times 10^{11},~M_{R2}=2.50\times 10^{12},~M_{R3}=8.22\times 10^{12}
 \mbox{GeV}
\ee
In this model $Y_{Rij}$ was related with quark-lepton masses via ${\bf \overline{126}}$ Higgs field. On the other hand, as is well known from \bref{Yukawa} $Y_{Rij}$ in the present model are independent on the quark-lepton mass matrices.
However, Dirac neutrino mass matrices $Y_{\nu ij}$ are same in both models 
and therfore eigen values of $M_R$ may not be so different to each other.
So
$M_S>~ M_{R3},~M_{\phi}$ and all decay channels mentioned above are open with the same decay ratio from SUSY \cite{plumacher}.

Single inflaton model suffered from the fine tuning problem $\lambda=O(10^{-14})$,
which was one of the motivation to have introduced hybrid model.
As we mentioned above, this problem has reappeared as $Y_R=1.0\times 10^{-6}$.
This constraint comes from $T_R\leq 10^6$ GeV.
However, we are free from this condition if gravitino mass is of order $100$ TeV. Let us consider \bref{W} in more details, and we put
\be
 {\cal L} =c_g \delta (y) 
 \int d^2 \theta \; 
  \frac{X}{M_5^2} 
 {\rm tr} \left[ {\cal W}^\alpha {\cal W}_\alpha \right].
\ee
From which we obtain the gaugino mass
\bea
M_\lambda&=&c_g\frac{F_X}{M_5^2}M_c=c_g\frac{F_X}{M_P}\left(\frac{M_5}{M_P}\right)\nonumber\\
&=&c_gm_{3/2}\left(\frac{M_5}{M_P}\right)
\eea
If we adopt $c_g=0.1$, then we obtain $m_{3/2}\approx 100$TeV heavy enough to be free from gravitino problem and from the constraint $T_R\leq 10^6$ GeV.
One gravitino decays to one LSP and we obtain
\be
n_{LSP} / s = n_{3/2} / s.
\ee
LSP as a DM must satisfy the observational constraint,
\be
\Omega_{LSP} h^2 \lesssim 0.1,
\ee
which leads us to
\be
n_{3/2} / s =
n_{LSP} / s = 3 \times 10^{-12}
             \times (m_{LSP} / 100 GeV)^{-1}
             \times (\Omega_{LSP} h^2 / 0.1 ).
\ee
If gravitino is generated in the reheating heat bath after the inflation, $n_{3/2}$ and $T_R$ are related with
\be
n_{3/2} / s \sim 1.5 \times 10^{-12}
                 \times (T_R / 10^{10} GeV)
\ee
and therefore
\be
T_R \lesssim 2 \times 10^{10} GeV
             \times (m_{LSP} / 100 GeV)^{-1}
             \times (\Omega_{LSP} h^2 / 0.1 )
\ee
So $M_R$ must be non-thermally generated.

\section{LFV in 5D}
Recently MEG collaboration \cite{MEG2} reported new results of 
a search for the $\mu \to e \gamma$ decay and a maximally 
 likelihood analysis sets an upper limit at 90\% C.L. on 
the branching ratio, 
\be
\BR (\mu \to e \gamma) < 2.4 \times 10^{-12}, 
\label{MEGsignal2}
\ee

Before the data of \bref{MEGsignal2}, we heard that they found  3 events as the best value for the number of signals 
in the maximally likelihood fit, which corresponds to \cite{MEG}
\bea 
 \BR (\mu \to e \gamma) = 3 \times 10^{-12}   
 \label{MEGsignal}
\eea
 for the center value.  This result was denied and reduced to \bref{MEGsignal2}.
However, our model gives rather large branching ratio marginal to the experimental upper bound (see Fig. \ref{Fig2a}). 
So it is very interesting to consider what happens if the observation gives non-null result around this value \cite{F-O-MEG}.

The relic abundance of the cold dark matter (CDM) in 2$\sigma$ range 
 has been measured as
\begin{eqnarray}
 \Omega_{CDM} h^2 = 0.1131 + \pm 0.0034 . 
\label{WMAP} 
\end{eqnarray}
As is well-known, in SUSY models with the $R$-parity conservation, 
 a neutralino, if it is the lightest sparticle (LSP), 
 is the promising candidate for the CDM in the present universe. 

For $\tan \beta=45$, $\mu>0$ and $A_0=0$, for example, 
 we can find more precise relation between $m_0$ and $M_{1/2}$ than \bref{relation} such as 
\begin{eqnarray}
 m_0(\mbox{GeV}) = 
  125.3 
+ 0.329 M_{1/2}(\mbox{GeV}) 
+ 5 \times 10^{-5} \left( M_{1/2}(\mbox{GeV}) \right)^2,  
 \label{relation2} 
\end{eqnarray} 
along which the observed relic abundance, 
 $\Omega_{CDM} h^2 = 0.113$, is realized. 
In our analysis, we have employed 
 the {\tt SoftSUSY3.1.4} package \cite{softsusy} 
 to solve the MSSM RGEs and produce mass spectrum. 
Then, the relic abundance of the neutralino dark matter 
 is calculated by using the {\tt micrOMEGAs 2.4} \cite{micromega} 
 with the output of SoftSUSY in the SLHA format \cite{slha}. 

Anyhow let us continue our arguments without adopting \bref{MEGsignal}.
Through the seesaw mechanism \cite{seesaw}, 
 the light neutrino mass matrix is given by
\bea
 m_\nu=Y_\nu^T M_R^{-1} Y_\nu v_u^2 = U_{MNS} D_\nu U_{MNS}^T 
\label{seesaw}
\eea
 in the basis where the mass matrix of charged lepton is diagonal 
 with real and positive eigenvalues.
Here $v_u$ is the VEV of the up-type Higgs doublet in the MSSM, 
 $D_\nu$ is the diagonal mass matrix of light neutrinos, 
 and $U_{MNS}$ is neutrino mixing matrix. 
This is equivalent to the expression of the right-handed neutrino 
 mass matrix as 
\bea 
 M_R = v_u^2 \left( 
  Y_\nu U_{TBM}^*D_\nu^{-1}U_{TBM}^\dagger Y_\nu^T  \right). 
\label{MRformula}
\eea 
Once the information of the Dirac Yukawa coupling, 
 the mass spectrum of the light neutrinos, 
 and the neutrino mixing matrix is obtained, 
 we can fix the right-handed neutrino mass matrix. 
In order to determine $M_R$, we follow the manner in \cite{F-O4}.

We consider the normal hierarchical case 
 for the light neutrino mass spectrum, for simplicity, 
 and describe $D_\nu$ in terms of the lightest mass eigenvalue 
 $m_1$ and the mass squared differences: 
\bea
 D_\nu=\mbox{diag}\left(m_1,~\sqrt{\Delta m_{12}^2 + m_1^2},
 ~\sqrt{\Delta m_{13}^2 + m_1^2}\right).  
\label{Dnu}
\eea 
Here we adopted the neutrino oscillation data \cite{NuData}: 
\bea 
  \Delta m_{12}^2=7.59\times 10^{-5}~\mbox{eV}^2,
~~\Delta m_{13}^2=2.43\times 10^{-3}~\mbox{eV}^2 
\label{nudata}
\eea
Contrary to the case of the minimal SO(10) GUT in 4D, right-handed heavy neutrino mass matrix is independent on the charged fermion mass matrices, and we assume the neutrino mixing matrix of 
 the so-called tri-bimaximal form \cite{TBM} 
\bea
 U_{TBM} = 
\left(
 \begin{array}{ccc}
 \sqrt{\frac{2}{3}}  & \sqrt{\frac{1}{3}} &  0 \\
-\sqrt{\frac{1}{6}}  & \sqrt{\frac{1}{3}} & \sqrt{\frac{1}{2}} \\ 
-\sqrt{\frac{1}{6}}  & \sqrt{\frac{1}{3}} & -\sqrt{\frac{1}{2}}
      \end{array} \right),
\label{TBM}
\eea
 which is in very good agreement with the current best fit 
 values of the neutrino oscillation data \cite{NuData}. \footnote{This is found now correct upto a zero-th order approximation after the non zero discovery of $\theta_{13}$ \bref{Daya-Bay}.}
As has been discussed above,  
 the data fit for the realistic charged fermion mass matrices  
 is same as in the minimal SO(10) model, and we here use 
 the numerical value of $Y_\nu$ in Eq.~(\ref{Ynu}) 
 at the GUT scale for $\tan \beta=45$, 
 in the basis where the charged lepton mass matrix is diagonal. 
Substituting Eqs.~(\ref{Ynu}), (\ref{Dnu}), 
 (\ref{nudata}) and (\ref{TBM}) into Eq.~(\ref{MRformula}), 
 we obtain the right-handed neutrino mass 
 matrix as a function of only $m_1$. 
Since it has been shown \cite{F-O4} that the simple 5D SO(10) model 
 can reproduce the observed baryon asymmetry of the present universe 
 for $m_1 \simeq 1.8 \times 10^{-3}$ eV through non-thermal leptogenesis, 
 we take this for a reference value of $m_1$. 
In this way, $M_R$ is now completely determined, but not 
 yet diagonalized. 
Changing the basis where $M_R$ is also diagonal, 
 we find the neutrino Dirac Yukawa matrix, 
\begin{eqnarray}
 Y_\nu = 
\left( 
 \begin{array}{ccc}
-0.00119 + 0.0000268i & -0.00108 - 0.000485 i & -0.000392 + 0.000421 i \\
 0.00135 + 0.00167 i  & -0.0253 + 0.00154 i & 0.0237 + 0.000851 i \\
-0.0265 - 0.0173 i    & 0.0609 + 0.0275 i & 0.790 - 0.0436i 
 \end{array}   \right) ,   
\label{Ynu2}
\end{eqnarray}   
 and the diagonal right-handed neutrino mass matrix 
 with eigenvalues (in GeV) 
\begin{eqnarray}
 M_{R_1}=1.03 \times 10^{10},  \; 
 M_{R_2}=7.55 \times 10^{11},  \;  
 M_{R_3}=3.22 \times 10^{15}.   
\label{MR2}
\end{eqnarray}

Reflecting the independence of $M_R$ from the charged leptons, $Y_\nu$ of \bref{Ynu2} is different from \bref{Ynu}.
Now we are ready to analyze the $\mu \to e \gamma$ decay rate 
 by using the completely determined neutrino Dirac Yukawa 
 coupling matrix and the right-handed neutrino mass eigenvalues%
\footnote{
Previous analysis for the minimal SO(10) model 
 with general parameter sets, see \cite{F-K-O}.}, 
 Eqs.~(\ref{M_R}) and  (\ref{Ynu}) for the minimal SO(10) model 
 while Eqs.~(\ref{Ynu2}) and  (\ref{MR2}) 
 for the simple 5D SO(10) model. 
For $\tan \beta =45$, $A_0=0$ and $\mu > 0$, 
 the branching ratio of $\mu \to e \gamma$ 
 for the minimal SO(10) model is shown in Fig.~3.11 as a function of the universal gaugino mass 
 at the GUT scale, along the relation of Eq.~(\ref{relation2})
 to satisfy the observed relic density for the neutralino dark matter. 
The short dashed line corresponds to the branching ratio 
 of Eq.~(\ref{MEGsignal2}). 
 The universal gaugino mass becomes $M_{1/2} \geq 840$ GeV. 
\begin{figure}
\begin{center}
\epsfig{file=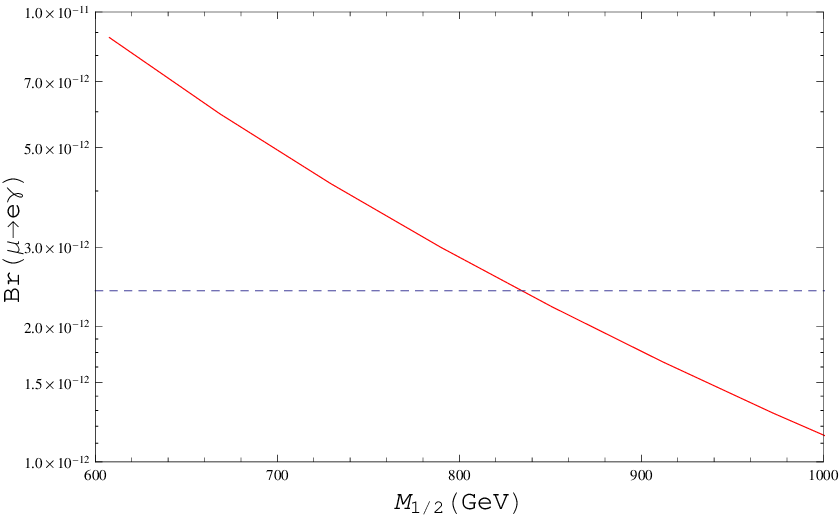, width=12cm}
\caption{ 
The branching ratio $\BR(\mu \to e \gamma)$ for the minimal SO(10) model 
 as a function of $M_{1/2}$ (GeV) 
 along the cosmological constraint of Eq.~(\ref{relation2}). 
The short dashed line corresponds to the MEG result, 
 $\BR(\mu \to e \gamma)<2.4 \times 10^{-12}$. This diagram is same as the left-upper panel of Fig.2.3 except for the use of Eq.~(\ref{relation} for the latter case.}
\end{center}
\end{figure}
\begin{figure}
\begin{center}
\epsfig{file=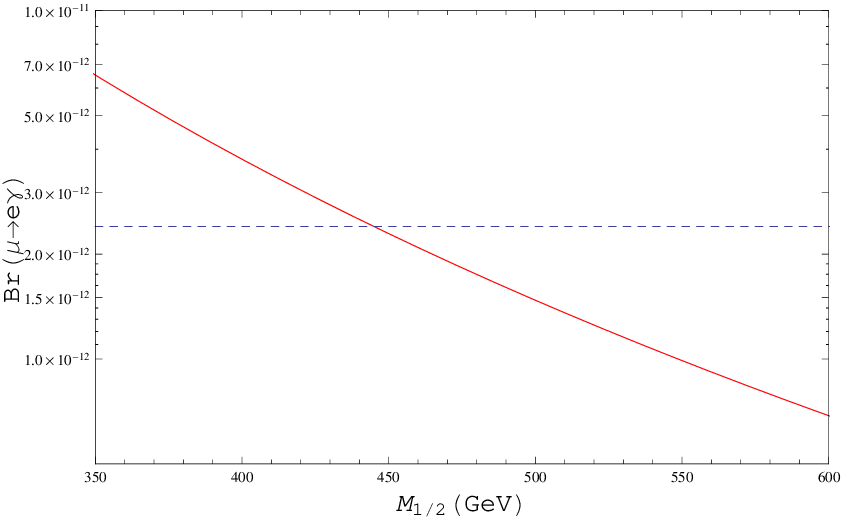, width=12cm}
\caption{ 
The same figure as Fig.~3.11 but for the simple 5D SO(10) model. 
}
\end{center}
\end{figure}
\begin{table}
\caption{
Mass spectra (in GeV) and phenomenological constraints 
for the two SUSY SO(10) models 
with the universal boundary conditions in the CMSSM cited from \cite{F-O-MEG}. $A_0$s in the first and second frames are the trilinear term and CP-odd neutral scalar, repectively. See also Table \ref{MSSM2} reflecting the discovery of Higgs-like boson around 125 GeV by the LHC.
} \label{table}
\begin{center}
\begin{math}
\begin{array}{|c|c|c|}
\hline  & $minimal SO(10) model$ & $simple 5D SO(10) model$ \\
\hline m_0     & 415   & 272  \\ 
       M_{1/2} & 790   & 420  \\ 
       A_0     &   0   &   0  \\ 
    \tan \beta &  45   &   45 \\ 
\hline 
 h_0   & 119 & 115  \\ 
 H_0   & 786 & 449  \\ 
 A_0   & 787 & 449  \\
 H^\pm & 791 & 457  \\  
\hline 
\tilde{g} & 1756 & 981 \\ 
{\tilde{\chi}^0}_{1,2,3,4} & 
  333, 631, 928, 938 & 171, 324, 535, 548 \\ 
{\tilde{\chi}^{\pm}}_{1,2} & 
  631, 938 & 324, 548 \\
\hline 
\tilde{d},\tilde{s}_{R,L} & 1576, 1645 & 898, 934 \\
\tilde{u},\tilde{c}_{R,L} & 1582, 1643 & 901, 931 \\
\tilde{b}_{1,2} & 1409, 1473 & 784, 849 \\
\tilde{t}_{1,2} & 1266, 1475 & 698, 864 \\
\hline 
\tilde{\nu}_{e,\mu,\tau} 
 & 667, 667, 619 & 386, 386, 355 \\
\tilde{e},\tilde{\mu}_{R,L} 
 & 511, 672 & 317, 395  \\
\tilde{\tau}_{1,2} 
 & 342, 642 & 186, 392 \\
\hline 
\BR(b \rightarrow s \gamma) & 
 3.27 \times 10^{-4} & 2.36 \times 10^{-4} \\
\BR(B_s \rightarrow \mu^+ \mu^-) & 
 1.04 \times 10^{-8} & 4.95 \times 10^{-9} \\ 
\Delta a_{\mu} & 
 12.0 \times 10^{-10} & 37.7 \times 10^{-10} \\
\hline \Omega h^2 & \multicolumn{2}{|c|}{0.113}  \\
\hline 
\end{array}
\end{math}
\end{center}
\end{table}



Fig.~3.12 depicts the same result as in Fig.~3.11, but 
 for the simple 5D SO(10) model. 
In this case, the universal gaugino mass is relatively low
 and becomes $M_{1/2}\geq 440$ GeV.
This is because the components in the neutrino Dirac Yukawa matrix 
 relevant to the LFV between 1st and 2nd generations 
 are smaller than those in the minimal SO(10) model 
 and a lighter sparticle mass spectrum is necessary 
 in order to achieve the same branching ratio 
 (see Eq.~(\ref{LFVrough})).  

The set of the CMSSM parameters proposes a good benchmark point 
 for the SUSY search at the LHC and we presented 
 (sparticle) mass spectra for each model in Table~3.3, 
 along with other observables which can be compared 
 with the experimental bounds \cite{F-O-MEG} before the dicovery of the Higgs-like particle. (We remained the old data before the anouncement at July 2012 by the LHC since they are useful for exhibiting what should be correcyted. We will discuss on the impact of the recent LHC results at the end of this section.)
In both results, the lower bound on the Higgs boson mass 
 $m_h \geq 114.4$ GeV \cite{Higgs} is satisfied. 
However, squark masses in 5D SO(10) model are disfavored.
Other phenomenological constraints we consider here are 
\begin{eqnarray}
& 2.85 \times 10^{-4} \leq \BR(b \rightarrow s  \gamma) \leq 4.24
\times 10^{-4} \; (2 \sigma ) & \hspace{1cm} \cite{bsgamma} 
\label{bsg}, \\
&  \BR(B_{s} \rightarrow \mu^{+} \mu^{-} ) < 5.8 \times 10^{-8}  & \hspace{1cm}
\cite{bsmumu} \label{bsmm}.
\end{eqnarray}
\footnote{\bref{bsmm} is updated to
$\BR(B_{s} \rightarrow \mu^{+} \mu^{-} ) <4.5 (3.8) \times 10^{-9}$ at 95\% (90\%) C.L. \cite{Aaij}.}
The results for the minimal SO(10) model satisfy these constraints, 
 while $\BR(b \rightarrow s  \gamma)$ for the simple 5D SO(10) model 
 is marginal (about 3.4$\sigma$ away from the center value). 

We can accommodate our results with muon anomalous magnetic moment, another important signal of New Physics BSM.

Table~3.3 also includes the sparticle contributions to $\Delta a_\mu$. 
We can see that the result for the minimal 5D SO(10) model favors 
 the deviation obtained by using the data of the hadronic
 $\tau$ decay, while the deviation from the $e^+ e^-$ data 
 is favored by the result for the simple SO(10) model.  These are the results before the anouncement of the LHC at 7 Tev and 8 TeV run.


\section*{GUT and the LHC results}
We study the impact of the recent LHC discoveries of Higgs-like object around $m_h=125$ GeV, $h\rightarrow \gamma\gamma$, SUSY search etc. in connection with SO(10) GUT models.
Here the consistencies of the SUSY have been analyzed without considering the details of model but with the pattern of soft SUSY breaking. 

One loop correction to the lightest Higgs mass in CMSSM is \cite{Harber2}\cite{Carena}
\be
m_h^2\approx M_Z^2\mbox{cos}^22\beta+\frac{3}{4\pi^2}\frac{m_t^4}{v^2}\left[\mbox{ln}\frac{M_S^2}{m_t^2}+\frac{X_t^2}{M_S^2}\left(1-\frac{X_t^2}{12M_S^2}\right)\right],
\label{higgs1}
\ee
where
\be
M_S=\sqrt{m_{\tilde{t}_1}m_{\tilde{t}_2}},~~X_t=A_t-\mu\mbox{cot}\beta,~~v=174\mbox{GeV}
\ee
with the trilinear Higgs-stop coupling constant $A_t$. 
For large tan$\beta$, some negative corrections appear
\be
\Delta m_h^2\approx -\frac{h_b^4v^2}{16\pi^2}\frac{\mu^4}{M_S^4}
\ee
from sbottom, and
\be
\Delta m_h^2\approx -\frac{h_\tau^4v^2}{48\pi^2}\frac{\mu^4}{m_{\tilde{\tau}}^4}\ee
from stau.
Here the bottom Yukawa and the tau Yukawa are
\be
h_b\approx \frac{m_b}{v\mbox{cos}\beta(1+\mbox{tan}\beta \Delta h_b)},~~h_\tau\approx \frac{m_\tau}{v\mbox{cos}\beta(1+\mbox{tan}\beta \Delta h_\tau)}
\ee
with one-loop corrections of $\Delta h_b$ and $\Delta h_\tau$.
Anyhow these corrections may be subdominant.

The recent review of SUSY search at the LHC gives very severe constraint on CMSSM \cite{LHC2}.

Here we set Higgs-like object around 125 GeV as lightest Higgs $h$, and \bref{higgs1} indicates (a) rather large stop mass or (b) large $A_t$.
(a) implies a large discrepancy between $m_t$ and $m_{\tilde{t}}$ and is in the inverse direction of that of original SUSY motivation, the loop cancellation in Higgs mass hierarchy. This large stop mass need large $m_0$ or large $M_{1/2}$ for gaugino mediation, which is very severe from the LHC \cite{LHC2}. We did not adopt the second choice since we set $A_0=0$. We will be back on this point later.

As a result, someone assert that CMSSM is strogly disfavored \cite{Cao}. If it is indeed the case, it should be considered very worrisome since our analyses of LFV, leptogenesis, sparticles mass spectra have been based on the universal boundary condition of CMSSM from \bref{USB1} to \bref{USB3}.
As we mentioned, these universal boundary conditions are natural except for \bref{NUHM} in the framework of SO(10) group.
It is more natural that we relax \bref{NUHM} and take $m_{H_u}$ and $m_{H_d}$ as free parameters (NUHM1 or NUHM2).  It is not obvious to us that it improves the problems \cite{Buchmueller}. Here we only point out that this may suppress sfermion masses since
\be
16\pi^2\frac{d}{dt}m_{\phi_i}^2=-\sum_{a=1,2,3}8C_a(i)g_a^2|M_a|^2+\frac{6}{5}Y_ig_i^2S,
\ee
where 
\be
S\equiv Tr[Y_jm_{\phi_j}^2]=m_{H_u}^2-m_{H_d}^2+Tr[{\bf m}_Q^2-{\bf m}_L^2-2{\bf m}_{\overline{u}}^2+{\bf m}_{\overline{d}}^2+{\bf m}_{\overline{e}}^2].
\ee
Gaugino mediation which we adopted in 5D model makes the situation worse.
This is because all matters get their masses from gaugino by RGE and gaugino mass must be unacceptably large.
Also we may generalize \bref{USB2}: if the hidden sector field is not the singlet unlike \bref{W} but, for instance, {\bf 54}-plet
\bea 
 {\cal L} =  
 \int d^2 \theta \; 
  \frac{\Phi_{\alpha\beta}}{M_5} 
  {\cal W}^\alpha {\cal W}^\beta,
\label{W2}
\eea 
we have non-universal gaugino masses at GUT scale\cite{Martin2}
\be
M_3:M_2:M_1=2:-3:-1
\ee
in place of \bref{USB2}.
So, even if we adopt CMSSM or minimally extend it, we must first construct the most suitable soft mass conditions ($m_0,~M_{1/2},~A_0,~\mbox{tan}\beta$ for CMSSM or $m_0,~M_{1/2},~A_0,~\mbox{tan}\beta,~\mu,~ m_A$ for NUHM2 ) without singlet and with singlet (the next-to-minimal supersymmetric model (NMSSM) \cite{Ellwanger} and the nearly minimal supersymmetric model (nMSSM) \cite{Fayet} etc.) due to several SUSY breaking mechanism. 

The NMSSM and nMSSM are primarily motivated by $\mu$ problem and introduce gauge singlet superfield $S$.

The superpotential of NMSSM is
\be
W=W_{MSSM}+\lambda SH_uH_d+\frac{1}{3}\kappa S^3,
\label{NMSSM}
\ee
where $W_{MSSM}$ is the superpotential of the MSSM without $\mu$ term.
The F term is
\be
V_F=|\lambda|^2|S|^2\left(H_u^\dagger H_u+H_d^\dagger H_d\right)+|\kappa S^2|^2
\ee
S couples with Higgs only through W and has no gauge coupling. Therefore, D-term is same as MSSM
\be
V_D=\frac{1}{2}g_2^2|H_u^\dagger H_d|^2+\frac{1}{8}(g_1^2+g_2^2)\left(H_u^\dagger H_u-H_d^\dagger H_d\right)^2.
\ee
\be
V_{soft}=m_{H_u}^2H_u^\dagger H_u+m_{H_d}^2H_d^\dagger H_d+m_S^2|S|^2+\left(\lambda A_\lambda H_uH_dS+\frac{1}{3}\kappa A_\kappa S^3+c.c.\right).
\ee
The mass of the lightest Higgs boson becomes at tree level
\be
\left(m_h^{NMSSM}\right)^2~<~m_Z^2\left(\mbox{cos}^2(2\beta)+\frac{2|\lambda|^2\mbox{sin}^2(2\beta)}{g_1^2+g_2^2}\right),
\ee
though it works well for small tan$\beta$ around 1. 

The nMSSM is a model whose superpotential is
\be
W=W_{MSSM}+\xi_FM_n^2 S.
\ee
In this model DM is light singlino dominated neutralino, which greatly enhance the decay of $h$ into light neutralino. This greatly suppresses Br($h\rightarrow \gamma\gamma$) in comparion with the SM and is incompatible with the LHC observation.

One of the peculiar properties of our data fiiting among others are large tan$\beta$. 

It goes from \bref{leading} that this $A_t\approx A_\tau\approx A_0$ value enhances LFV in general.
\be
m_{\tilde{\tau}}^2=
\left(
    \begin{array}{cc}
     m_{L_3}^2+\Delta_{\overline{e}L} & v(A_\tau^*\mbox{cos}\beta-\mu y_\tau\mbox{sin}\beta) \\
     v(A_\tau\mbox{cos}\beta-\mu^* y_\tau\mbox{sin}\beta) & m_{\overline{e}_3}+\Delta_{\tilde{e}_R}
    \end{array}
   \right),
\ee
where $\Delta$ is the contribution from D term and
\be
\Delta_\phi=(T_{3\phi}-Q_\phi\mbox{sin}^2\theta_W)\mbox{cs}(2\beta)m_Z^2.
\ee
For large tan$\beta$ value, the following value is a check point \cite{Buras}\cite{Mahmoudi}
\bea
Br(B_s^0\rightarrow \mu^+\mu^-)&=&3.5\times 10^{-5}\left[\frac{\mbox{tan}\beta}{50}\right]^6\left[\frac{\tau_{B_s}}{1.5ps}\right]\left[\frac{F_{B_s}}{230MeV}\right]^2\left[\frac{|V_{ts}^{eff}|}{0.040}\right]^2\nonumber\\
&\times &\frac{\overline{m}_t^4}{M_A^4}\frac{(16\pi^2)^2\epsilon_Y^2}{(1+\tilde{\epsilon}_3\mbox{tan}\beta)^2(1+\epsilon_0\mbox{tan}\beta)^2}.
\eea
Here 
\be
\epsilon_Y=\frac{1}{16\pi^2}\frac{A_t-\mu\mbox{cot}\beta}{\mu}H_2(x^{Q/\mu}, x^{U/\mu})~~\mbox{etc.}
\ee
with
\be
x^{Q/\mu}=\frac{m_Q^2}{\mu^2},~~H_2(x,y)=\frac{x\mbox{ln}x}{(1-x)(x-y)}+\frac{y\mbox{ln}y}{(1-y)(y-x)}.
\ee
There may be $3 \sigma$ deviation from the SM in $h\rightarrow \gamma\gamma$ \cite{ATLAS}.
Its decay ratio is given by \cite{Gunion} \cite{Spira}
\bea
\Gamma(h\rightarrow \gamma\gamma)&=&\frac{G_F\alpha^2m_h^3}{128\sqrt{2}\pi^3}\mid\sum_fN_{cf}e_f^2g_f^hA_f^h(\tau_f)+g_W^hA_W^h(\tau_W)\nonumber\\
&+&g_{H^\pm}^hA_{H^\pm}^h(\tau_{H^\pm})
+\sum_{\tilde{\chi}^\pm} g_{\tilde{\chi}^\pm}^hA_{\tilde{\chi}^\pm}^h(\tau_{\tilde{\chi}^\pm})+\sum N_{cf}e_{\tilde{f}}^2g_{\tilde{f}}^hA_{\tilde{f}}^h(\tau_{\tilde{f}})\mid
\eea
where $\tau_a\equiv (m_h/2m_a)^2$. $A_i$ ($i$ specifies spin here) are defined by
\bea
A_1(\tau)&=&-[2\tau^2+3\tau+3(2\tau-1)g(\tau)]/\tau^2\\
A_{1/2}(\tau)&=&2[\tau+(\tau-1)g(\tau)]/\tau^2\\
A_0(\tau)&=&-[\tau-g(\tau)]/\tau^2,
\eea
where 
\bea
g(\tau)\equiv 
\left\{
\begin{array}{lll}
 \mbox{arcsin}^2\sqrt{\tau} & \mbox{for} & \tau\leq 1 \\
-\frac{1}{4}\left(\mbox{log}\frac{1+\sqrt{1-1/\tau}}{1-\sqrt{1-1/\tau}}-i\pi\right)^2 &\mbox{for}& \tau > 1
\end{array}
\right. 
\eea
$A_1$ is negative definite, and $A_0$ and $A_{1/2}$ are monotonicly increasing positive definite functions.
Thus W boson loop dominates and top quark suppreeses
the decay rate in the SM. However, if the contributions of sfermions dominate over the W boson loop, then $\Gamma(h\rightarrow \gamma\gamma)$ may be enhanced in comparison with the SM.
Taking all these situations into consideration, we will consider how these data constrain the model.
These properties are fit to the LHC data if we select parameter regions, for instance, as \cite{Carena}
\be
A_\tau~(\mbox{trilinear})~\approx 500 \mbox{[GeV]},~~\mu\approx 1 \mbox{TeV}~~\mbox{tan}\beta\approx 60,
\ee
which gives
\be
Br(h\rightarrow \gamma\gamma)\approx 1.5Br(h\rightarrow \gamma\gamma)^{SM}.
\ee
In this case a lighter stau mass $m_{\tilde{\tau}}=135$ GeV is very near to $m_h$. 
They give another example
\be
A_\tau\approx 1500 \mbox{[GeV]},~~\mu\approx 1030 \mbox{GeV}~~\mbox{tan}\beta\approx 60,
\ee
which gives a lighter stau mass $m_{\tilde{\tau}}=106 \mbox{GeV}~ <~m_h$.

Using {\tt SoftSUSY3.1.6}, we give an example in the table \ref{MSSM2} for tan$\beta=45$ realized in the minimal SO(10) GUT in 4D \footnote{We are grateful to H.Ishida for the help of numerical calculations.}.
\begin{table}
\caption{Mass spectra (in GeV) in various universal boundary conditions which give $m_{h_0}$ around 125 GeV. In all columns except for the first one we used \bref{relation2}. 
}
\label{MSSM2}
\begin{center}
\begin{math}
\begin{array}{|l|l|l|l|l|}
\hline  & \mbox{mass in units of GeV} \\
\hline
m_0 &2000 &1000 &1300 &1600\\
M_{1/2} &2000 &2000 &2500 &3000\\
A_0 &-3500 &-1800 &-2700 &-3500\\
\tan \beta &45 &45 &45 &45\\
\hline
m_{h_0} &125.1 &123.7 &124.7 &125.5\\
H_0 &2266 &2011 &2504 &2982 \\
A_0 &2266 &2011 &2504 &2982 \\
H^\pm &2267 &2013 &2505 &2984 \\
\hline
m_{\tilde{g}} &4268 &4218 &5204 &6179\\
\tilde{\chi}_{1,2}^0 &8702\,, 1656 &8615\,, 1639 &1085\,, 2059 &1309\,, 2478 \\
\tilde{\chi}_{3,4}^0 &-2754\,, 2757 &-2422\,, 2426 &-3057\,, 3060 &-3663\,, 3666 \\
\tilde{\chi}^\pm_{1,2} &1656\,, 2757 &1640\,, 2426 &2059\,, 3060 &2478\,, 3666 \\
\hline
\mu &2757 &2422 &3058 &3665 \\
\hline
\tilde{u}_{1,2} &4280\,, 4134 &3926\,, 3762 &4849\,, 4643 &5762\,, 5513 \\
\tilde{c}_{1,2} &4280\,, 4134 &3925\,, 3762 &4849\,, 4643 &5761\,, 5513 \\
\tilde{t}_{1,2} &2895\,, 3458 &2917\,, 3368 &3548\,, 4113 &4179\,, 4856\\
\tilde{d}_{1,2} &4281\,, 4118 &3926\,, 3743 &4849\,, 4619 &5762\,, 5484 \\
\tilde{s}_{1,2} &4281\,, 4117 &3926\,, 3743 &4849\,, 4619 &5762\,, 5483 \\
\tilde{b}_{1,2} &3428\,, 3560 &3315\,, 3383 &4064\,, 4134 &4811\,, 4881\\
\hline
\tilde{e}_{1,2} &2390\,, 2133 &1655\,, 1248 &2095\,, 1599 &2534\,, 1951 \\
\tilde{\mu}_{1,2} &2389\,, 2130 &1655\,, 1247 &2094\,, 1597 &2533\,, 1948 \\
\tilde{\tau}_{1,2} &1132\,, 2027 &6061\,, 1476 &6609\,, 1834 &7410\,, 2197 \\
\tilde{\nu_e}_{1,2} &2388 &1653 &2093 &2532 \\
\tilde{\nu_\mu}_{1,2} &2387 &1653 &2092 &2531 \\
\tilde{\nu_\tau}_{1,2} &2021 &1466 &1825 &2190 \\
\hline
{\rm Br} ( B_s \to \mu^+ \mu^-) &3.9 \times 10^{-9} &7.6 \times 10^{-9} &2.9 \times 10^{-9} &6.2 \times 10^{-10}\\
\hline
\end{array}
\end{math}
\end{center}
\end{table}
More elaborate sample calculations have been made in the CMSSM and NUHM1 \cite{Buchmueller} and in the pMSSM \cite{Sekmen}.
However, SUSY GUT has its advantages over MSSM and pMSSM etc. since it can fix the Yukawa coupling and tan$\beta$.
LFV and leptogenesis processes need the detailed structure of the Dirac Yukawa coupling $Y_\nu$.
As you can see from Figure \ref{Fig2a}, $M_{1/2}$ must satisfy $350<M_{1/2}<800$ [GeV] if $10<\delta a_\mu\times 10^{10} <40$ in the case of trilinear $A_0=0$. This is beacause $\delta a_\mu$ monotonically decreases as $M_{1/2}$ increases.

However, it can be relaxed for $A_0\neq 0$ and larger $M_{1/2}$ may be accepted.

After almost completing this review, LHCb Collaboration has anounced that they have found the first evidence of $Br(B_s\rightarrow \mu^+\mu^-)=(3.2^{+1.5}_{-1.2})\times 10^{-9}$ \cite{LHCb2}, which is in agreement with the SM prediction, $Br(B_s\rightarrow \mu^+\mu^-)=3.23\pm0.27)\times 10^{-9}$.
\section*{Discussion} 
We have reviewed SO(10) GUT mainly based on our works. The most peculiar property of minimal SO(10) GUT is its high predictivity and its predictions cover over all particle phenomena and over a wide range of energies, from $10^{-5}$ eV to $10^{16}$ GeV.
This is a rather peculiar property of the minimal SO(10) model.
This high predictivity, in other words, implies that this model suffers from continuous hard checks from many observations. Furthermore, not only data fitting over a wide range but also the internal consistencies like the compatibilities with a gauge coupling unifications and with the detailed SUSY breaking mechanism etc. have been considered on the minimal SO(10) GUT.

We first discussed the minimal SO(10) GUT in 4D. The data fitting with quark masses (6), CKM mixing angles and CP phase (3+1), charged lepton masses (3) were fitted with 14 parameters (see section 2.4.2 for a more generic form). 
Adjusting remaining only one parameter and one scaling factor $c_R$, we could fix full neutrino oscillation data; mass square differences (2), MNS mixing matrix (angles (3), CP phases (3)) as well as the Dirac neutrino ($Y_\nu$) and heavy right-handed neutrino ($M_R$) mass matrices. $Y_\nu$ and $M_R$ are very essential for LFV and Leptogenesis, respectively.
Unfortunately our predictions were found to deviate after Kamland in $\theta_{13}$ and the mass ratio $\Delta m_{23}^2/\Delta m_{12}^2$. However, this does not mean that minimal SO(10) is excluded. Our parameter search is natural but not exhaustive for the following reasons.

1. We set $\alpha,~\beta$ (see section {\bf 2.4.2}) fixed to $0$ or $\pi$ for simplicity.

2. We assumed only type I seesaw. Minimal SO(10) model allows both type I and Type II. 

The general case was discussed in \cite{Babu:2005ia}. Their data fitting was improved but not in very good agreement yet. 
More important is that it suffers from some internal problems like the instability of gauge coupling unification due to intermediate energy scales and the no-go theorem on the SUSY breaking etc.

3. If we relax the \lq\lq minimal\rq\rq condition but preserving renormalizability, we may add ${\bf 120}$-plet Higgs. This may improve the conflict between the gauge couplin unification and several intermediate energy scales. Unfortunately, no one has succeeded in it explicitly.

Both on the detailed mismaches of data fitting and on the internal consistency like the no-go theorem of $U(1)_R$ symmetry, there are several loopholes, making us stay in 4D. However, it seems a natural choice for us to go to extra dimensions. This is rather reasonable since our bottom-up approach should have a harmonic encounter with the top-down approach at GUT scales.
There it is very natural for GUT to have some remnants of extra dimensions.
Thus we have proceeded to 5D SO(10) orbifold GUT.

Dimension five may not be our final goal but it must provide essential merits for extra dimensions if it is real.

We have repeated a comprehensive analyses in 5D which was done in 4D and showed how they improve the situations compared with those in 4D.
The problems and deficits discussed in section {\bf 2.8} are remedied in 5D. 

4. Data fitting is expected to be improved since $M_R$ is free from $M_{126}$ unlike the original SO(10) case.

5. Superpotential \bref{SP2} breaks the PS symmetry to the SM group without any intermediate energy scale and does $U(1)_R$ consistently.

Though we have accepted unrenormalizable terms but they are strictly guided by the renormalizable action and inherit the merits of the renormalizable model.

The recent LHC discoveries of a Higgs-like boson around 125 GeV and of the null result of SUSY particle searches etc. give very important constraints on the SO(10) GUT model in 4D and 5D, which forces us to make some modifications to them. 
Neverthless we have cited the old results some of which are not compatible with the LHC results. This is because it is very important to first clearify the discrepancies between the theoritical predictions and observations.
On going and near future experiments of accelerators and cosmological observations must drive GUT to precision science.

\section*{Acknowledgments}
We are greatly indebted to all our colleagues, especially N.Okada, S.Meljanac, A.Ilacovac, T.Kikuchi.
This work is supported in part by 
 the Grant-in-Aid for Scientific Research from the Ministry 
 of Education, Science and Culture of Japan \#20540282. 



\newpage

\begin{appendix}

\chapter{SO(10) Group Properties}
\section{Spinorial representation}
This section is based on \cite{Mohapatra}

SO(2N) is expanded in SU(N) basis, $\chi_i~~(i=1,...,N$), constitutes
Clifford algebra,

\bea
\{\chi_i,\chi_j^\dagger\}&=&\delta_{ij}\nonumber\\
\{\chi_i,\chi_j\}&=&\{\chi_i^\dagger,\chi_j^\dagger\}=0
\eea
Now let us define the 2N operators which constitutes vector basis,
\bea
\Gamma_{2j-1}&=&-i(\chi_j-\chi_j^\dagger)\nonumber\\
\Gamma_{2j}&=&(\chi_j+\chi_j^\dagger)~~~j=1,2,...,N
\eea
$\Gamma_\mu$ satisfies
\be
\{\Gamma_\mu,~\Gamma_\nu\}=2\delta_{\mu\nu}~~\mu,~\nu=1,...,2N
\ee
Then we can construct a representation of SO(2N), where
\be
\Sigma_{\mu\nu}\equiv \frac{1}{2i}[\Gamma_\mu,~\Gamma_\nu]
\ee
are the {\bf 45} dimensional generators. The spinor representation in SO(2N) is $2^N$ dimensional and is split into $2^{N-1}$ dimensional representation under a chiral projection operator,
\be
\Gamma_0=i^N\Gamma_1\Gamma_2...\Gamma_{2N}
\ee
For the case of SO(10),
\bea
|{\bf 16}_-\ra &=&|{\bf 16}_L\ra=\chi_1^\dagger\chi_2^\dagger\chi_3^\dagger\chi_4^\dagger\chi_5^\dagger|0\ra {\bf 1}+\frac{1}{12}\epsilon^{ijklm}\chi_k^\dagger \chi_l^\dagger\chi_m^\dagger|0\ra{\bf 10}_{ij}+\chi_i^\dagger|0\ra\overline{{\bf 5}}^i\nonumber\\
|{\bf 16}_+\ra&=&|{\bf 16}_R\ra=|0\ra {\bf 1}+\frac{1}{2}\chi_i^\dagger \chi_j^\dagger|0\ra\overline{{\bf 10}}^{ij}+\frac{1}{24}\epsilon^{ijklm}\chi_j^\dagger\chi_k^\dagger\chi_l^\dagger\chi_m^\dagger|0\ra{\bf 5}_i
\eea
The Yukawa coupling of chiral Fermions with Higgs scalars is given by
\be
\overline{\Psi}\Gamma_\mu...\Gamma_\rho\Psi \Phi_{\mu...\rho}
\ee
Here $\Phi_{\mu...\rho}$ is Higgs scalar in tensorial representation which is discussed in the next section.

\section{Tensorial representation}
The arguments of this appendix is based on \cite{fuku1}.
Our formulation is in tensorial representation.

Here we first introduce $Y$ diagonal basis: 
$1 = 1^{\prime} + 2^{\prime} i$, $2 = 1^{\prime} - 2^{\prime} i$, 
$3 = 3^{\prime} + 4^{\prime} i$, $4 = 3^{\prime} - 4^{\prime} i$, 
$5 = 5^{\prime} + 6^{\prime} i$, $6 = 5^{\prime} - 6^{\prime} i$, 
$7 = 7^{\prime} + 8^{\prime} i$, $8 = 7^{\prime} - 8^{\prime} i$, 
$9 = 9^{\prime} + 0^{\prime} i$, $0 = 9^{\prime} - 0^{\prime} i$, 
up to the normalization factor 
$\frac{1}{\sqrt{2}}$.  It is more convenient since 
$\left(1, 3, 5, 7, 9 \right)$ 
transforms as ${\bf 5}$-plet and 
$\left(2, 4, 6, 8, 0 \right)$ 
transforms as ${\bf \overline{5}}$-plet under $SU(5) \times U(1)_{X}$
(for that reason $Y$ diagonal basis could also be called $SU(5)$ basis).  
Consequently, 
$\left(1, 3 \right)$ and 
$\left(2, 4 \right)$ are $SU(2)_{L}$ doublets with definite 
hypercharges $Y=\frac{1}{2}$ and $Y=-\frac{1}{2}$, respectively.  
Similarly, 
$\left(5, 7, 9 \right)$ and 
$\left(6, 8, 0 \right)$ transform under $SU(3)_{C}$ as 
${\bf 3}$ and ${\bf \overline{3}}$ with definite hypercharges 
$Y=-\frac{1}{3}$ and $Y=\frac{1}{3}$, respectively.

Note that under the complex conjugation ($c.c.$), 
$\overline{1} = 2$, 
$\overline{3} = 4$, 
$\overline{5} = 6$, 
$\overline{7} = 8$, 
$\overline{9} = 0$, and vice versa.  
The $SO(10)$ invariants are build in such a way that an index $a$ 
is contracted (summed) with the corresponding $c.c.$ index $\overline{a}$, 
for example, $T_{\cdots a \cdots} T_{\cdots \overline{a} \cdots}$. 
 
The basis in $A = {\bf 45}$, $D = {\bf 120}$, $\Phi = {\bf 210}$ and 
$\Delta + \overline{\Delta} = {\bf 126 + \overline{126}}$ dimensional spaces 
are defined by totally antisymmetric (unit) tensors 
$(a^{\prime} b^{\prime})$, 
$(a^{\prime} b^{\prime} c^{\prime})$, 
$(a^{\prime} b^{\prime} c^{\prime} d^{\prime})$ and 
$(a^{\prime} b^{\prime} c^{\prime} d^{\prime} e^{\prime})$, 
respectively,  
and similarly in $a$, $b$, $c$, $d$, $e$ indices in $Y$ diagonal basis. 
The states of the $\Delta$ and $\overline{\Delta}$
have additional properties,
\bea
i \varepsilon_{\bar{a}_1\bar{a}_2\bar{a}_3\bar{a}_4\bar{a}_5
 \bar{a}_6\bar{a}_7\bar{a}_8\bar{a}_9\bar{a}_{10}}
 \overline{\Delta}_{a_6a_7a_8a_9a_{10}} &=&
 \overline{\Delta}_{\bar{a}_1\bar{a}_2\bar{a}_3\bar{a}_4\bar{a}_5},
\nonumber \\
i \varepsilon_{\bar{a}_1\bar{a}_2\bar{a}_3\bar{a}_4\bar{a}_5
 \bar{a}_6\bar{a}_7\bar{a}_8\bar{a}_9\bar{a}_{10}}
 \Delta_{a_6a_7a_8a_9a_{10}} &=&
 -\Delta_{\bar{a}_1\bar{a}_2\bar{a}_3\bar{a}_4\bar{a}_5},
\eea
that allow one to project out the $\Delta$ and $\overline{\Delta}$ states, respectively,
from the $256$ antisymmetric states $(abcde)$.
The explicit expressions for antisymmetric tensors are, for example, 
\bea
(ab) &=& ab - ba,
\nonumber\\
(abc) &=& abc +cab + bca - bac - acb - cba
\eea
{\it etc}.  
Important relations are 
\bea
(12) &=& -i (1^{\prime} 2^{\prime}),
\nonumber\\
(34) &=& -i (3^{\prime} 4^{\prime}),
\nonumber\\
(56) &=& -i (5^{\prime} 6^{\prime}),
\nonumber\\
(78) &=& -i (7^{\prime} 8^{\prime}),
\nonumber\\
(90) &=& -i (9^{\prime} 0^{\prime})
\eea
Symmetric $E = {\bf 54}$ dimensional space is spanned by traceless symmetric states
$\left\{a'b' \right\} \equiv a'b' + b'a' \,\, 
\left(a^{\prime}, b^{\prime} = 
1^{\prime}, 2^{\prime}, \cdots, 9^{\prime}, 0^{\prime} \right)$ 
and $\sum_{a^{\prime}} c_{a^{\prime}}\, \{a^{\prime} a^{\prime}\}$ 
with $\sum_{a^{\prime}} c_{a^{\prime}} \equiv 0$.  
Also, important relations are 
\bea
\left\{12 \right\} = 
   1^{\prime} 1^{\prime} 
 + 2^{\prime} 2^{\prime},
\nonumber\\
\left\{34 \right\} = 
   3^{\prime} 3^{\prime} 
 + 4^{\prime} 4^{\prime},
\nonumber\\
\left\{56 \right\} =
   5^{\prime} 5^{\prime} 
 + 6^{\prime} 6^{\prime},
\nonumber\\
\left\{78 \right\} =
   7^{\prime} 7^{\prime} 
 + 8^{\prime} 8^{\prime},
\nonumber\\
\left\{90 \right\} =
   9^{\prime} 9^{\prime} 
 + 0^{\prime} 0^{\prime}.
\eea
SO(10) invariants are defined in the fundamental $SO(10)$ basis 
$1^{\prime},2^{\prime}, \cdots, 9^{\prime},0^{\prime}$ and in the $Y$ diagonal 
basis $1, 2, \cdots, 9, 0$. We give some typical examples:
\begin{eqnarray}
\Phi^2
&\equiv& 
\Phi_{a'b'c'd'} \Phi_{a'b'c'd'}
\ =\ \Phi_{abcd}\Phi_{\overline{a}\overline{b}\overline{c}\overline{d}}
\nonumber\\
\Phi^3
&\equiv& \Phi_{a'b'c'd'} \Phi_{a'b'e'f'} \Phi_{c'd'e'f'} 
\ =\ \Phi_{\overline{a}\overline{b}\overline{c}\overline{d}}
\Phi_{abef} \Phi_{cd\overline{e}\overline{f}},\\
\Phi \overline{\Delta} \Delta
&\equiv& \Phi_{a'b'c'd'} \overline{\Delta}_{a'b'e'f'g'} \Delta_{c'd'e'f'g'}
\ =\ \Phi_{\overline{a}\overline{b}\overline{c}\overline{d}} 
\overline{\Delta}_{abefg} \Delta_{cd\overline{e}\overline{f}\overline{g}},
~~etc.\nonumber
\end{eqnarray}
Here $a^{\prime}, b^{\prime}, c^{\prime}, \cdots$ 
$\left(a,b,c, \cdots \right)$ run over all the $SO(10)$ vector 
($Y$ diagonal) indices. 

Now, the Higgs fields $A$, $E$, $\Delta$, $\overline{\Delta}$, and $\Phi$ 
contain 8 directions of singlets under the $G_{321}$ subgroup 
(see Appendix of \cite{fuku1}). The 
corresponding vacuum expectation values (VEVs) are defined by: 
\bea
\langle A \rangle &=& \sum_{i=1}^2 A_i \, \widehat{A}_i,
\\
\langle E \rangle &=& E \, \widehat{E}, 
\\
\langle \Delta \rangle &=& v_R \, \widehat{v_R}, 
\\
\langle \, \overline{\Delta} \, \rangle &=& 
\overline{v_R} \, \widehat{\overline{v_R}}, 
\\
\langle \Phi \rangle &=& \sum_{i=1}^3 \Phi_i \, \widehat{\Phi}_i,  
\eea
where unit directions $\widehat{A}_i \,\, (i=1,2)$, $\widehat{E}$, 
$\widehat{v_R}$, $\widehat{{\overline{v_R}}}$ and 
$\widehat{\Phi}_i \,\, (i=1,2,3)$ in the $Y$ diagonal basis are: 
\bea
\label{ud1}
\widehat{A}_1 
&=& \widehat{A}^{(1,1,0)}_{(1,1,3)}\ =\ \frac{i}{2}(12+34), 
\label{singlet1}
\\
\label{ud2}
\widehat{A}_2 
&=& \widehat{A}^{(1,1,0)}_{(15,1,1)}\ =\ \frac{i}{\sqrt{6}}(56+78+90), 
\\
\label{ud3}
\widehat{E} 
&=& \widehat{E}^{(1,1,0)}_{(1,1,1)}\ =\ 
\frac{1}{\sqrt{60}}\{{\it 3\times}[12+34]-{\it 2\times}[56+78+90]\}, 
\\
\label{ud4}
\widehat{v_R} 
&=& \widehat{\Delta}^{(1,1,0)}_{(\overline{10},1,3)}\ =\ 
\frac{1}{\sqrt{120}}(24680), 
\\
\label{ud5}
\widehat{\overline{v_R}} 
&=& \widehat{\overline{\Delta}}^{(1,1,0)}_{(10,1,3)}\ =\ 
\frac{1}{\sqrt{120}}(13579), 
\\
\label{ud6}
\widehat{\Phi}_1 &=& \widehat{\Phi}^{(1,1,0)}_{(1,1,1)}\ =\ 
-\frac{1}{\sqrt{24}}(1234), 
\\
\label{ud7}
\widehat{\Phi}_2 &=& \widehat{\Phi}^{(1,1,0)}_{(15,1,1)}\ =\ 
-\frac{1}{\sqrt{72}}(5678+5690+7890), 
\\
\label{ud8}
\widehat{\Phi}_3 &=& \widehat{\Phi}^{(1,1,0)}_{(15,1,3)}\ =\ 
-\frac{1}{12}([12+34][56+78+90]).  
\label{singlet8}
\eea
The upper and the lower indices indicate the 
$SU(3)_C \times SU(2)_L \times U(1)_Y$, 
$SU(4) \times SU(2)_L \times SU(2)_R$ quantum numbers, respectively 
in the case of double indices. A word about notation: the square brackets
are used for grouping of indices. This grouping of indices 
is used to emphasize the $SU(2)_L$ and $SU(3)_C$ structures within the state vectors.
The square brackets satisfy usual distributive law with respect
to summation of indices and tensor product of indices, e.g.
\bea
&&([12+34][56+78+90])
\nonumber\\ 
&&=\ (1256+1278+1290+3456+3478+3490)
\nonumber\\ 
&&=\ (1256)+(1278)+(1290)+(3456)+(3478)+(3490),
\nonumber\\
&&([1,3][5[78+90],7[56+90],9[56+78]])
\nonumber\\
&&=\ (1578+1590,1756+1790,1956+1978,
\nonumber\\
&&\qquad  3578+3590,3756+3790,3956+3978)
\nonumber\\
&&=\ (1578,1756,1956,3578,3756,3956)
\nonumber\\
&&+\ \ (1590,1790,1978,3590,3790,3978).
\eea

The unit directions appearing in VEVs satisfy the following orthonormality relations
\bea
\widehat{A_i} \cdot \widehat{A_j} &=& 
\delta_{ij} \,\, \left(i,j = 1,2 \right),  
\nonumber\\
\widehat{E}^2 &=& 1, 
\nonumber\\
\widehat{v_R} \cdot \widehat{v_R} &=& 
\widehat{\overline{v_R}} \cdot \widehat{\overline{v_R}} = 0, 
\nonumber\\
\widehat{v_R} \cdot \widehat{\overline{v_R}} &=& 1,
\nonumber\\
\widehat{\Phi_i} \cdot \widehat{\Phi_j} &=& 
\delta_{ij} \,\, \left(i,j = 1,2,3 \right).  
\eea
Due to the D-flatness condition the absolute values of the VEVs, 
$v_R$ and ${\overline{v_R}}$ are equal, 
\be
|v_R| = |\overline{v_R}|.
\ee
More details should be reffered to the original paper \cite{fuku1}.
\end{appendix}
\end{document}